\documentclass[12pt]{article}

\usepackage{arxiv}

\usepackage[utf8]{inputenc} % allow utf-8 input
\usepackage[T1]{fontenc}    % use 8-bit T1 fonts
\usepackage{hyperref}       % hyperlinks
\usepackage{url}            % simple URL typesetting
\usepackage{booktabs}       % professional-quality tables
\usepackage{amsfonts}       % blackboard math symbols
\usepackage{nicefrac}       % compact symbols for 1/2, etc.
\usepackage{microtype}      % microtypography
\usepackage{lipsum}		% Can be removed after putting your text content
\usepackage{graphicx}
\usepackage{doi}
\usepackage{scalerel,stackengine}
\usepackage{graphicx,multicol}
\usepackage{multirow} 
\usepackage{indentfirst,csquotes}
 \usepackage{array} 
\usepackage{abstract}
\usepackage{algorithm}
\usepackage{algpseudocode}

\usepackage[caption=false]{subfig}
\usepackage{amssymb,amsthm,amsmath}
\usepackage{xcolor,paralist,hyperref,fancyhdr,etoolbox}
\newtheorem{theorem}{Theorem}[]
\newtheorem{definition}[theorem]{Definition}

\newtheorem{lemma}[]{Lemma}

\newtheorem{cor}{\bf Corollary}%[section]
\usepackage[round]{natbib} 
\newtheorem{lem}{\bf Lemma}%[section]
\newtheorem{rmk}{\bf Remark}%[section]
\newtheorem{thm}{Theorem}%[chapter]

\usepackage[utf8]{inputenc}
\numberwithin{equation}{section}

\usepackage{hyperref}

\hypersetup{ colorlinks=true, linkcolor=black, filecolor=black, urlcolor=black,citecolor = blue }

\usepackage{lipsum}
\usepackage{multicol}

\usepackage{listings}
\title{Intrinsic Geometry-Based Angular Covariance:\\ A Novel Framework for Nonparametric\\ Changepoint Detection in Meteorological Data}

\author{ { Surojit Biswas}\thanks{ webpage: https://sites.google.com/view/surojitbiswas/home?authuser=0} \\
	Department of Mathematics\\  IIT Kharagpur, India-$721302$ \\
	\texttt{surojit23@iitkgp.ac.in} \\
	\And
    {Buddhananda Banerjee}\thanks{ webpage: https://sites.google.com/site/buddhanandastat/} \\
	Department of Mathematics\\  IIT Kharagpur, India-$721302$ \\
	\texttt{bbanerjee@maths.iitkgp.ac.in } \\
	 \AND
     {Arnab Kumar Laha} \\
	 Interdisciplinary Statistical
Research Unit\\  ISI Kolkata, India-700108 \\
	\texttt{arnablaha@isical.ac.in} \\
}

\date{}

\renewcommand{\shorttitle}{Toroidal and Spherical Changepoint detection in Cyclone Data}

\hypersetup{
pdftitle={A template for the arxiv style},
pdfsubject={q-bio.NC, q-bio.QM},
pdfauthor={David S.~Hippocampus, Elias D.~Striatum},
pdfkeywords={First keyword, Second keyword, More},
}

\begin{document}
\maketitle
%%% For table of content enable it
% \tableofcontents{}
% \bigskip
% \bigskip
% \vspace{0.5cm}
\begin{abstract}

In many temporal datasets, the parameters of the underlying distribution may change abruptly at unknown times. Detecting such changepoints is crucial for numerous applications. Although such a problem has been extensively studied for linear data, there has been notably less research on bivariate angular data. To the best of our knowledge, this paper presents the first attempt to address the changepoint detection problem for the mean direction of toroidal and spherical data.
By defining the ``square of an angle'' through intrinsic geometry, we construct a curved dispersion matrix for bivariate angular data, analogous to the linear dispersion matrix in Euclidean space. Using the analogous measure of the ``Mahalanobis distance,'' we develop two new non-parametric tests to identify changes in the mean direction parameters for toroidal and spherical distributions. The pivotal distributions of the test statistics are shown to follow the Kolmogorov distribution under the null hypothesis. Under the alternative hypothesis, we establish the consistency of the proposed tests. We also apply the proposed methods to detect changes in mean direction for hourly wind-wave direction (toroidal) measurements and the path (spherical) of the cyclonic storm ``Biporjoy,'' which occurred between  6th and 19th June 2023 over the Arabian Sea, western coast of India. 
% This study represents a novel application of torus-to-torus regression in meteorological analysis, highlighting its potential for broader use in directional data modeling.  
% To the best of our knowledge, this paper provides a mathematical foundation for modeling the regression between bivariate angular predictors and bivariate angular responses, for the first time. Built upon a modified M\"{o}bius transformation, the proposed model makes use of the intrinsic geometry of the torus. We propose a new loss function grounded in differential geometry to enable effective semi-parametric estimation without assuming any specific angular error distribution. The practical utility of the model is demonstrated through an application to wind and wave direction data observed during two major cyclonic events Biparjoy and Amphan, affecting the coastlines of India. 
\end{abstract}

% keywords can be removed

\section{Introduction}
The presence of bivariate angular or directional data is very common in different disciplines of sciences, for example dihedral (torsion)  angles in protein structures (bioinformatics), wind directions and  sea wave directions (meteorology), the path of a cyclone (climatology),  daily occurrence time of maximum and minimum share price of a stock (finance), etc.
Such data refers to measurements that exhibit a circular or periodic nature. After a suitably chosen location of the origin, the data can be represented on a torus ($\mathbb{S}_1 \times \mathbb{S}_1 $) or the sphere $\mathbb{S}_2$ depending on the range of the data. For a comprehensive exploration of bivariate circular data, refer to \cite{mardia2000directional} and \cite{ley2017modern}.
 
Change point analysis is a key statistical technique used to detect unexpected shifts or changes in a data sequence over time. These changes can occur due to variations in the parameters within the same distribution family or a complete switch to a different distribution family. The presence of change points can significantly disrupt standard statistical analyses. Therefore, the main goal of change point analysis is to conduct a statistical test to determine if a change point exists in the dataset. A substantial amount of research in change point analysis has been carried out for real-valued random variables   \cite[see][]{Hovarth_1999,Antoch_1997,Cobb_1978, Davis_1995}, vector-valued random variable \cite[see][]{Kirch_2014,Kokoszka_2000, shao2010testing, anastasiou2023generalized}, and functional valued random variable \cite[see][]{horvath2012inference, banerjee2018more, 
Horman_2010, banerjee2020data}. In the context of angular data, there has been limited exploration of the change point problem. The change point in angular data may occur in the mean direction, concentration, or both. For the first time, \cite{lombard1986change} introduced a pioneering rank-based test to detect change points in the change in location, and change in concentration parameter for angular data. Following this work, \cite{grabovsky2001change} put forth a modified CUSUM procedure for testing the change in concentration parameter of the angular distribution.
 \cite{ghosh1999change},  proposed a likelihood-based approach for addressing change-point detection in the mean direction for the von Mises distribution only. Additionally, \cite{sengupta2008likelihood} introduced a novel likelihood-based method, referred to as the likelihood integrated method.\\
 % Recently, \cite{biswas2024changepoint} has introduced a new method driven by the intrinsic geometry of curved torus for changepoint detection in angular data. They have implemented the method to medical science, engineering, and meteorological datasets.

\textbf{Our contribution:}
This paper investigates the problem of changepoint detection in toroidal and spherical data, representing important forms of bivariate angular observations. To the best of our knowledge, this is the first systematic study addressing changepoint detection for the mean direction parameter in such settings. Our key contributions are as follows:
\begin{enumerate}
    \item  Extending the notion of the ``square of an angle" for toroidal data introduced by \citet{biswas2025semi}, we introduce an analogous concept for spherical data, thus providing a unified framework for bivariate angular variables.

     \item   We define the ``curved dispersion matrix'' for toroidal and spherical random variables, serving as a natural analog of the classical dispersion matrix for bivariate linear data.
     
     \item Building on this framework, and an analogous measure of Mahalanobis distance, we propose two novel non-parametric tests to detect changepoints in the mean direction parameter for toroidal and spherical data.

     \item The pivotal distributions of the proposed test statistics are shown to follow the Kolmogorov distribution under the null hypothesis, while their consistency is theoretically established under the alternative hypothesis.

     \item    To demonstrate the practical applicability of the proposed tests, we have implemented the proposed tests on the cyclone data.
    
\end{enumerate}

 \textbf{Challenges with Cyclone data:}
 During a cyclone, the relationship between wave direction and wind direction is dynamic and complex. Initially, waves align with prevailing winds as intense gusts transfer energy to the ocean surface. However, as the cyclone evolves, rotating wind fields lead to continuous shifts in wind direction around the eye. This generates waves that initially remain aligned with the wind but later travel outward independently. The cyclone’s forward motion, size, and coastal topography further influence wave behavior. As waves move away from the storm center, they are shaped by additional environmental factors such as ocean currents and atmospheric conditions.

  The path of the cyclone may change multiple times due to Earth's rotation and the surrounding high and low-pressure systems. When approaching land, cyclones bring heavy rainfall, high winds, and storm surges, causing severe damage. As they move over land or cooler waters, they gradually weaken and dissipate. Each cyclone follows a unique path determined by complex meteorological interactions. Wind and wave directions can be modeled as toroidal data, while the cyclone's trajectory, defined by latitude and longitude, can be represented on the sphere. 

  Apart from changes in wind speed, the occurrence of a cyclone often introduces changepoints in both wind-wave directions and the cyclone’s trajectory. Detecting these changepoints is challenging due to the bivariate directional nature of the data, which lies on curved manifolds rather than in linear space. Traditional linear covariance structures are inadequate for such data because they fail to account for the underlying topology. To address this, we have developed a novel notion of covariance for directional data using tools from differential geometry. This framework enables a moment-based, non-parametric analysis that is better suited to capturing the intrinsic geometry of directional observations.
For example, the track and wind-wave interactions of Super Cyclone "Biporjoy," which struck the western coast of India, offer an interesting case study for such modeling approaches.

\textbf{Data of Biporjoy cyclone:} 
\label{biporjoy section}
The Extremely Severe Cyclonic Storm ``Biparjoy” over the east-central Arabian Sea occurred from 6th June to 19th June 2023, severely affecting some states of western India. According to the report by the Regional Specialized Meteorological Center - tropical cyclones, New Delhi India Meteorological Department (IMD)  this cyclone was longest duration cyclone since 1977. Cyclone Biparjoy, a very severe cyclonic storm in June 2023, exemplified the intricate interplay between wind direction and wave direction during its course over the Arabian Sea. Originating from a low-pressure area, Biparjoy intensified and reached peak wind speeds of 195 km/h (121 mph), which played a crucial role in wave formation. These intense winds imparted substantial energy to the surface of the ocean, producing large and powerful waves, resulting in severe coastal flooding and erosion, particularly along the Gujarat coast. This surge, along with the powerful waves, caused extensive damage to coastal infrastructure and ecosystems, highlighting the storm's destructive potential.

An upper air cyclonic circulation formed over the Southeast Arabian Sea and developed into a depression early on June 6. It moved northwards, intensifying into a deep depression and then into Cyclonic Storm ``Biparjoy" in the adjoining Southeast Arabian Sea. Continuing its northward trajectory, it intensified into a Severe Cyclonic Storm (CS) over the east-central Arabian Sea and further into a Very Severe Cyclonic Storm (VSCS) in the same region. From June 7 to June 11, Biparjoy followed a recurving path, moving gradually north-northwestwards, then north-northeastwards, and finally northwards. As it moved northwards, it intensified into an Extremely Severe Cyclonic Storm (ESCS) over the east-central Arabian Sea. It then shifted north-northeastwards briefly before returning to a northward path, maintaining its intensity as an ESCS. Subsequently, it moved north-northwestwards and weakened into a VSCS over the northeast and adjoining east-central Arabian Sea. Continuing its north-northwestward, then northward, and finally northeastward movement, the storm gradually weakened. It crossed the Saurashtra and Kutch regions of India and the adjoining Pakistan coasts between Mandvi (Gujarat) and Karachi (Pakistan), near latitude $23.28^{\circ}N$ and longitude $68.56^{\circ}E$. After landfall, Biparjoy moved east-northeastwards, weakening into a cyclonic storm over Saurashtra and Kutch. It then moved northeastwards and weakened into a deep depression over Southeast Pakistan and adjoining Southwest Rajasthan and Kutch. Continuing its east-northeastward movement, it further weakened into a depression over South Rajasthan and adjoining north Gujarat and eventually into a well-marked low-pressure area over central Northeast Rajasthan and its surroundings by the morning of June 19.

This article is structured as follows. First, Section-\ref{int_geo_torus} discusses the area decomposition of a curved torus, while Section-\ref{int_geo_sphere} extends this discussion to the sphere. In Section-\ref{curved_disp_matrix} we revisit the ``square of an angle'' for torus and extend it for sphere, based on this, we define the ``curved variance,'' ``curved covariance,'' and the ``curved dispersion matrix.'' Using these notions, we propose two non-parametric tests for changepoint detection in the mean direction for toroidal and spherical data, which are presented in Section-\ref{changepoint_mean_torus} and Section-\ref{changepoint_mean_sphere}, respectively. Section-\ref{simulation} contains an extensive simulation study evaluating the performance of these tests when applied to data generated from the von Mises sine model (for toroidal data) and the Fisher distribution (for spherical data). In Section-\ref{data_analysis_biporjoy}, we apply the proposed tests to data from the extremely severe cyclonic storm ``Biparjoy.'' The article concludes with Section-\ref{conclusion}.
 % followed by appendices that include essential differential geometry tools in \ref{int_geo_general}, the proof of Lemma \ref{est_cd_matrix_lemma} in \ref{proof_lemma_covariance}, and the proof of Corollary \ref{consistency_corr_torus} in \ref{consistency_proof_torus}.

\begin{figure}[b]
    \centering
    \subfloat[ ]{%
        {\includegraphics[trim= 0 0 0 0, clip,width=0.6\textwidth, height=0.35\textwidth]{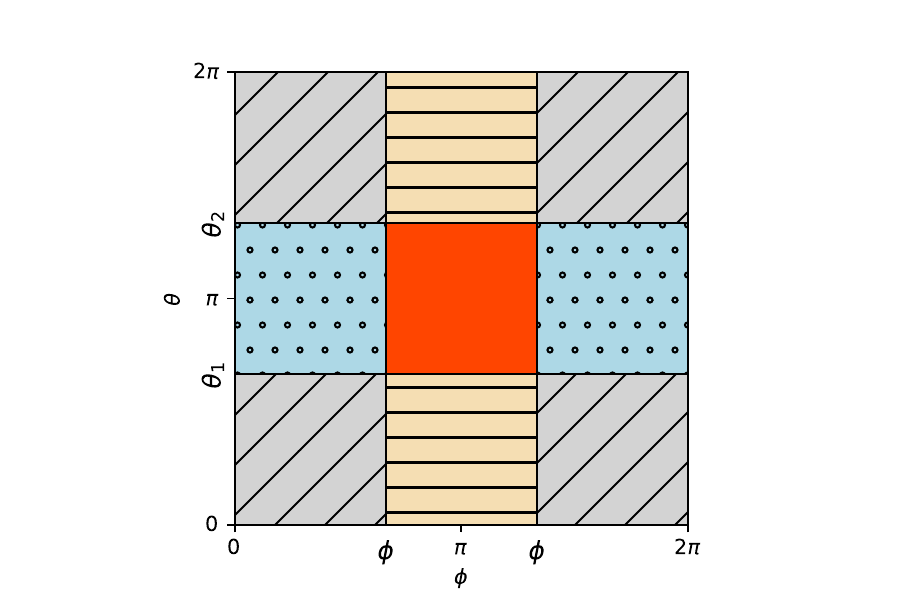}}}
          \subfloat[]{%
        {\includegraphics[trim= 80 80 80 80, clip, width=0.4\textwidth, height=0.38\textwidth]{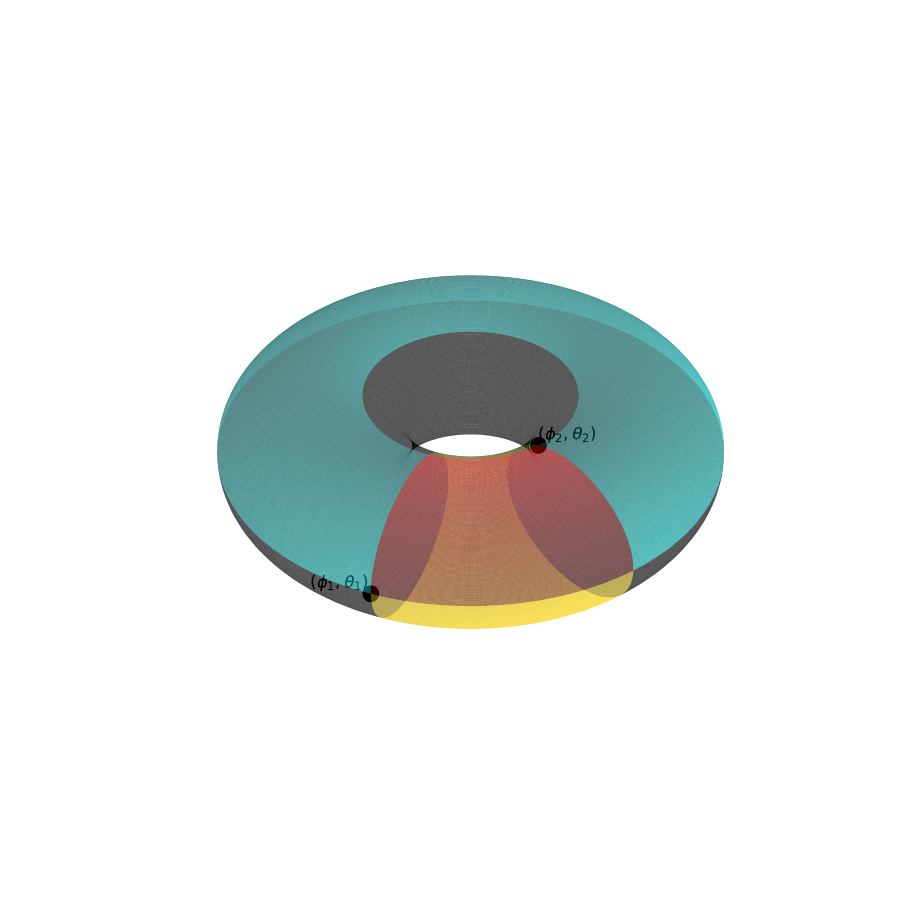}}}\hspace{5pt}
    \caption{Area between $(\phi_1, \theta_1)$, and $(\phi_2, \theta_2)$ (a) flat torus, (b)  curved torus. }
    \label{torus_area_plot}
\end{figure}

 \section{Intrinsic geometry of torus and sphere}
In this section, we discuss the area decompositions of the curved torus and sphere using fundamental intrinsic geometric tools from Riemannian geometry. Additionally, we introduce two useful definitions of proportioned area for the curved torus and sphere.
\subsection{Intrinsic geometry of torus}
\label{int_geo_torus}
The curved torus is defined by the parametric equation 
\begin{equation}
  X(\phi,\theta)=\{  (R+r\cos{\theta})\cos{\phi}, (R+r\cos{\theta})\sin{\phi}, r\sin{\theta} \}\subset \mathbb{R}^3, 
  \label{torus para equn}
\end{equation}
with the parameter space $\{ 
 (\phi,\theta):0<\phi,\theta<2\pi\}= \mathbb{S}_1 \times \mathbb{S}_1,$ known as 2-torus. Here, $r, R$  are the vertical and horizontal radii, respectively.

 Now, the partial derivatives of $X$ with respect to $\phi$, and $\theta$ are 
$$ X_{\phi}=\{ -(R+r\cos{\theta})\sin{\phi}, (R+r\cos{\theta})\cos{\phi}, 0 \}$$ and
$$X_{\theta}=\{  -r\sin{\theta} \cos{\phi}, -r\sin{\theta}\sin{\phi}, r\cos{\theta}  \}$$ respectively. Hence, the coefficients of the first fundamental form are
 \begin{equation}
 \begin{aligned}
     E=\langle X_{\phi},X_{\phi}\rangle &= (R+r\cos{\theta})^2\\
    F=\langle X_{\phi},X_{\theta}\rangle &= 0 \\
     G=\langle X_{\theta},X_{\theta}\rangle &= r^2
 \end{aligned}
 \label{fff cof torus}
 \end{equation}
 Now, following the calculation, the area element of the curved torus (Equation-\ref{torus para equn}) can be calculated as
\begin{equation}
    dA^{(\mathcal{T})}=r(R+r\cos{\theta})~d\theta~d\phi.
    \label{torus area element}
\end{equation}
One can see \cite{biswas2025semi} for the detailed calculation of the above area element.
\subsection{Area Decomposition of Curved Torus}
 Let $ \phi$ and $ \theta$ be the horizontal and vertical angles of a torus, respectively, where $ \phi$ and $ \theta$ are in the range of $ [0,2\pi)$. To start, we establish the area bounded by the coordinates $(0,0)$ and $(\phi,\theta)$ on the surface of a curved torus.  Consider two points on a flat torus, denoted as $(0,0)$ and $(\phi,\theta)$. The flat torus is defined by the interval $[0,2\pi)$ in both the horizontal and vertical directions. The area between these two points, when mapped onto the surface of the curved torus with horizontal radius $R$ and vertical radius $r$, can be calculated using Equation-\ref{torus area element}.  It is important to observe that when considering two locations $(0,0)$ and $(\phi,\theta)$ on a flat torus, the surface of the curved torus is divided into four subsets that are both mutually exclusive and exhaustive. The mapping above decomposition $\mathbb{T}_1:=[0, \phi]\times [0,\theta]$, $\mathbb{T}_2:=[\phi,2\pi]\times [0,\theta] $,  $\mathbb{T}_3:=[0,\phi]\times [\theta,2\pi]$ and $\mathbb{T}_4:=(\phi,2\pi] \times [\theta,2\pi]$ are represented on the surface of a curve torus (Equation-\ref{torus para equn}) in Figure-\ref{torus_area_plot}. The corresponding areas on the surface of the curved torus are denoted as $A_1$, $A_2$, $A_3$, and $A_4$, respectively. \cite{biswas2025semi} have calculated of the areas $A_i=\displaystyle\iint_{\mathbb{T}_i}dA(s,t) ~ \text{for} ~i=1,2,3,4$ using $dA (\phi, \theta)=rR\left(1+\frac{r}{R}\cos{\theta}\right) d\phi ~d\theta$ from Equation-\ref{torus area element} .

Analogous to the notion of circular distance - which is the length of the smaller arc between two angles, and the notion of geodesic distance on a surface - which is the length of the shortest path joining two points on the surface, \cite{biswas2025semi} have defined the \textit{proportionate area} included between these points $(0,0)$, $(\phi,\theta)$ as given   Definition-\ref{torus area of a segemnt}. 
\begin{definition}
	The \textit{proportionate area included between the   $(0,0)$, and $(\phi,\theta)$ } is defined as 
	
	$$A_T\left[(0,0),(\phi,\theta)\right]=\frac{\min \{A_1,A_2,A_3,A_4\}}{4\pi^2rR}.
	\label{torus area of a segemnt}
	$$
\end{definition}

\subsection{Intrinsic geometry of sphere}
\label{int_geo_sphere}

The parametric equation of the sphere is given by
\begin{equation}
  X(\phi,\theta)=\{  r\sin{\theta}\cos{\phi}, r\sin{\theta}\sin{\phi}, r\cos{\theta} \}\subset \mathbb{R}^3, 
  \label{sphere para equn}
\end{equation}
with the parameter space $\{ 
 (\phi,\theta):0\leq\phi<2\pi,~~ 0\leq\theta<\pi\}$.
 Now, the partial derivatives of $X$ with respect to $\phi$, and $\theta$ are 
$$ X_{\phi}=\{ -r\sin{\theta}\sin{\phi}, r\sin{\theta}\cos{\phi}, 0 \}$$ and
$$X_{\theta}=\{  r\cos{\theta} \cos{\phi}, r\cos{\theta}\sin{\phi}, -r\sin{\theta}  \}$$ respectively. Hence, the coefficients of the first fundamental form are
 \begin{equation}
 \begin{aligned}
     E=\langle X_{\phi},X_{\phi}\rangle &= r^2\sin^2{\theta}\\
    F=\langle X_{\phi},X_{\theta}\rangle &= 0 \\
     G=\langle X_{\theta},X_{\theta}\rangle &= r^2
 \end{aligned}
 \label{fff cof sphere}
 \end{equation}
Again, following the calculation,  the area element of the sphere (Equation-\ref{sphere para equn}) can be derived as
\begin{equation}
    dA^{(\mathcal{S})}=r^2\sin{\theta}~d\theta~d\phi.
    \label{sphere area element}
\end{equation}

\subsection{Area Decomposition of Sphere} 
\begin{figure}[t]
    \centering
            \subfloat[ ]{%
        {\includegraphics[trim= 15 20 20 20, clip, width=0.4\textwidth, height=0.3\textwidth]{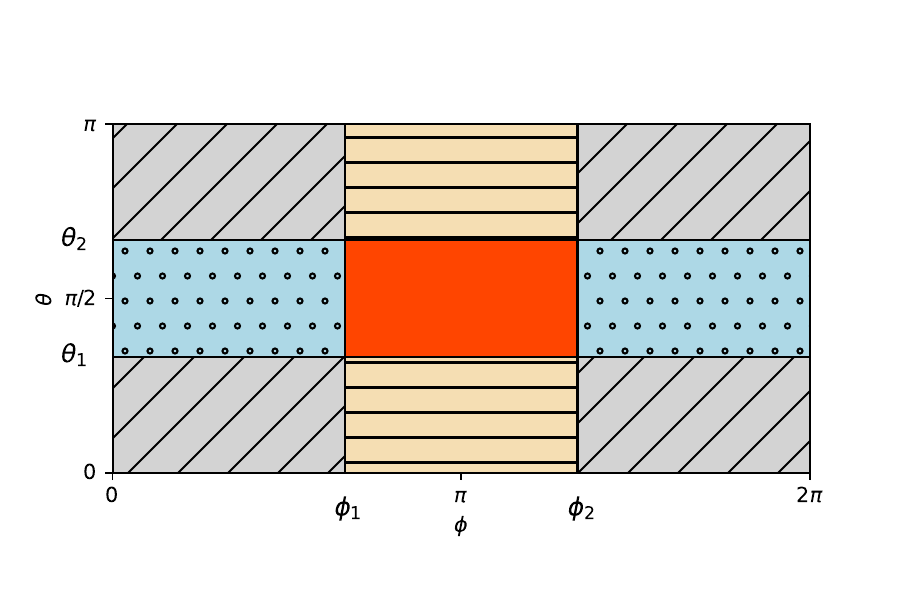}}}
          \subfloat[]{%
        {\includegraphics[trim= 100 100 100 100, clip, width=0.4\textwidth, height=0.35\textwidth]{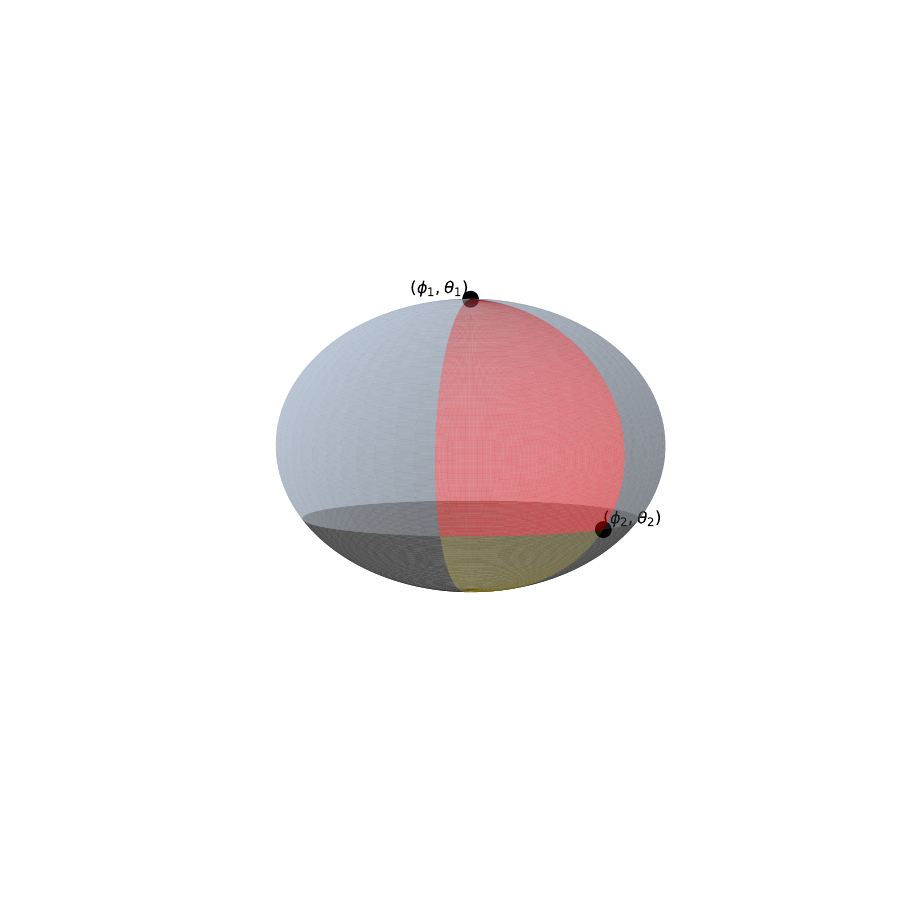}}}

    \caption{Area between $(\phi_1, \theta_1)$, and $(\phi_2, \theta_2)$  (a) flat sphere, (b)  sphere. }
    \label{sphere_area_plot}
\end{figure}

Let  $ \phi \in [0,2\pi)$, $ \theta \in [0,\pi)$ denote the horizontal and vertical angles of a sphere, respectively. We begin by defining the area between two points on the surface of the sphere.  Let $(\phi_1,\theta_1)$ and $(\phi_2,\theta_2)$ be two points on the sphere $ [0,2\pi) \times [0,\pi)$, then the \textit{proportionate area included between these two diagonally opposite points} when mapped on the surface of the sphere with radius $r$ can be computed by the following method using Equation-\ref{sphere area element} considering $0<\phi_1<\phi_2<2\pi$ and $0<\theta_1<\theta_2<\pi.$  Note that for two such diagonally opposite points  $(\phi_1,\theta_1)$ and $(\phi_2,\theta_2)$ on flat sphere, the surface on the sphere get partitioned into four mutually exclusive and exhaustive subsets as images (using Equation-\ref{sphere para equn}) of the following sets  $\mathbb{S}_1:=[\phi_1, \phi_2]\times [\theta_1,\theta_2]$, $\mathbb{S}_2:=([\phi_2,2\pi]\cup [0,\phi_1])\times [\theta_1,\theta_2] $,  $\mathbb{S}_3:=[\phi_1,\phi_2]\times([\theta_2,\pi]\cup [0,\theta_1])$ and $\mathbb{S}_4:=([\phi_2,2\pi]\cup [0,\phi_1])\times([\theta_2,\pi]\cup [0,\theta_1]).$ Let us call these areas  $A^{\mathcal{S}}_1$, $A^{\mathcal{S}}_2$, $A^{\mathcal{S}}_3$ and $A^{\mathcal{S}}_4$ respectively.  A diagrammatic representation of this decomposition is given in Figure-\ref{sphere_area_plot}.  We now provide details of the computation of the areas $A^{\mathcal{S}}_1$, $A^{\mathcal{S}}_2$, $A^{\mathcal{S}}_3$ and $A^{\mathcal{S}}_4$. 

\begin{itemize}
     \item Case-1: Using   Equation-\ref{sphere area element} on ${S_1}$ we get 
 \begingroup
\allowdisplaybreaks
\begin{align}
A^{\mathcal{S}}_1&=\displaystyle\iint_{S_1} dA^{\mathcal{S}}=\int_{\phi_1}^{\phi_2}\int_{\theta_1}^{\theta_2}   dA^{\mathcal{S}} 
= r\int_{\phi_1}^{\phi_2}d\phi~\int_{\theta_1}^{\theta_2}  \sin{\theta}~d\theta. \nonumber\\
&= r^2(\phi_2-\phi_1) \left( \cos{\theta_1}-\cos{\theta_2}\right).
 \label{area first segment sphere}  %labeling for ith cell
\end{align}%
\endgroup
\item Case-2:
Using   Equation-Equation-\ref{sphere area element} on ${S_2}$ we get
\begingroup
\allowdisplaybreaks
\begin{align}
A^{\mathcal{S}}_2&=\displaystyle\iint_{S_2} dA^{\mathcal{S}}=\int_{\phi_2}^{2\pi}\int_{\theta_1}^{\theta_2}   dA^{\mathcal{S}}+ \int_{0}^{\phi_1}\int_{\theta_1}^{\theta_2}   dA^{\mathcal{S}} \nonumber\\
&=r^2[2\pi-(\phi_2-\phi_1)]\left( \cos{\theta_1}-\cos{\theta_2}\right)
 \label{area second segment sphere} %labeling for ith cell
\end{align}%
\endgroup
    \item Case-3:
Using   Equation-Equation-\ref{sphere area element} on ${S_3}$ we get
    \begingroup
\allowdisplaybreaks
\begin{align}
A^{\mathcal{S}}_3&=\displaystyle\iint_{S_3} dA^{\mathcal{S}}=\int_{\phi_1}^{\phi_2}\int_{\theta_2}^{\pi}   dA^{\mathcal{S}}+\int_{\phi_1}^{\phi_2}\int_{0}^{\theta_1}   dA^{\mathcal{S}} \nonumber\\
&= r^2(\phi_2-\phi_1)\left[2+ (\cos{\theta_2}-\cos{\theta_1} )\right]
 \label{area third segment sphere} %labeling for ith cell
\end{align}%
\endgroup
 \item Case-4: 
Using   Equation-Equation-\ref{sphere area element} on ${S_4}$ we get
\begingroup
\allowdisplaybreaks
\begin{align}
A^{\mathcal{S}}_{4}&=\displaystyle\iint_{S_4} dA^{\mathcal{S}}=\int_{\phi_2}^{2\pi}\int_{\theta_2}^{\pi}  dA^{\mathcal{S}} +\int_{\phi_2}^{2\pi}\int_{0}^{\theta_1}  dA^{\mathcal{S}}
+\int_{0}^{\phi_1}\int_{\theta_2}^{\pi}
dA^{\mathcal{S}}+\nonumber\\ &\int_{0}^{\phi_1}\int_{0}^{\theta_1}dA^{\mathcal{S}}=r^2 \left[2\pi-(\phi_2-\phi_1)\right]\left[2+ (\cos{\theta_2}-\cos{\theta_1} )\right].
 \label{area fourth segment sphere} %labeling for ith cell
\end{align}%
\endgroup
\end{itemize}
Now, below, we define the \textit{proportionate area included between these two diagonally opposite points $(\phi_1,\theta_1)$, $(\phi_2,\theta_2)$} as given below. It may be noted that,  since $r$ is arbitrary and $\min \{A^{\mathcal{S}}_1,A^{\mathcal{S}}_2,A^{\mathcal{S}}_3,A^{\mathcal{S}}_4\}$ is dependent on $r$,  we divide it by total surface area of the sphere, $4\pi r^2$ which is the total area of the torus to remove this dependency.

\begin{definition}
        The \textit{proportionate area included between these two diagonally opposite points $(\phi_1,\theta_1)$, $(\phi_2,\theta_2)$ } is defined as 
        %to be the proportionate area between the  two points $(\phi_1,\theta_1)$, $(\phi_2,\theta_2)$ on the surface of the curved torus
   $$A_S\left[(\phi_1,\theta_1),(\phi_2,\theta_2)\right]=\frac{\min \{A^{\mathcal{S}}_1,A^{\mathcal{S}}_2,A^{\mathcal{S}}_3,A^{\mathcal{S}}_4\}}{4\pi r^2}.
    \label{sphere area of a segemnt}
$$
\end{definition}

\section{Curved dispersion matrix}
\label{curved_disp_matrix} 
In this section, we define   ``curved variance'' and ``curved co-variance'' for the toroidal and spherical data as follows.

Following the  Definition-\ref{torus area of a segemnt} and \ref{sphere area of a segemnt}, we obtain the proportionate  area between  $(0,0)$ to any arbitrary point $(\phi,\theta)$ on the curved torus or sphere, respectively, as:
\begin{equation}
A^{(0)}(\phi,\theta)= \displaystyle 
    \begin{cases}
        A_T\left[(0,0),(\phi,\theta)\right]& \text{for the curved torus.} \\
         A_S\left[(0,0),(\phi,\theta)\right] & \text{for the sphere.}
    \end{cases}
    \label{zero centered torus area of a segemnt}
\end{equation}
Now, we apply this notion to angular data for defining the square of an angle as well as the variance of an angular random variable as follows.

\begin{definition} \textit{The square of an angle} $\theta$ is defined as

\begin{equation}
A_C^{(0)}(\theta)= A^{(0)}(\theta,\theta)=\displaystyle 
    \begin{cases}
        A_T\left[(0,0),(\theta,\theta)\right]& \text{for the curved torus}.\\
        A_S\left[(0,0),(\theta,\theta)\right] & \text{for the sphere.}
    \end{cases}
      \label{zero centered circle area of a segemnt}
\end{equation}
\label{square_of_an_angle}
\end{definition}

\begin{definition}
      Let $\Theta$ be a zero-centered circular random variable with probability density function $f(\theta)$ on the unit circle $\mathbb{S}_1$. Then, the \textit{curved-variance} of the random variable $\Theta$ is 
    $$CVar(\Theta)=E_{f}\left[ A_C^{(0)}(\Theta) \right].$$
 If the circular mean of $\Theta \mbox{~is~}\mu \neq 0$ then  $\Theta$ can be replaced by $\Theta'= \left[(\Theta-\mu) \mod k\pi\right]$, where $k=2$ when $\Theta \in [0,2\pi)$, and  $k=1$ when $\Theta \in [0,\pi).$
\label{curved variance}
\end{definition}
If a random sample $\theta_1, \ldots, \theta_n$ is given then using weak law of large number (WLLN), $CVar(\Theta)$ can be consistently estimated as $ \widetilde{CVar(\Theta)}= \displaystyle \frac{1}{n}\sum_{i=1}^{n}A_C^{(0)}\left[(\theta_i-{\mu}) \mod k\pi\right],$ when $\mu$ is known.
 %Given a random sample $\theta_1,\theta_2\ldots,\theta_n$ we can consistently estimate ${CVar(\Theta)}$ by its sample analogue will be  $ \widehat{CVar(\Theta)}= \displaystyle \frac{1}{n}\sum_{i=1}^{n}A_C^{(0)}\left[(\theta_i-{\mu}) \mod 2\pi\right],$  using law of large number when  ${\mu}$ is know. 
 When $\mu$ is unknown we can use the plug-in estimator $$ \widehat{CVar(\Theta)}= \displaystyle \frac{1}{n}\sum_{i=1}^{n}A_C^{(0)}\left[(\theta_i-\widehat{\mu}) \mod k\pi\right],$$ where $\widehat{\mu}$ is the estimated circular mean of the data. Continuing the analogy we define the curved co-variance in the following section.

  % \begin{rmk}
  %     $A_1+A_2+A_3+A_4=4\pi^2rR$ is the total surface area of the curved torus.
  % \end{rmk}

\subsection{Curved co-variance}
In this section, we will define a measure similar to covariance in linear data for angular random variables, termed ``area covariance" (ACov), using Equation \ref{zero centered torus area of a segemnt}. Now, without loss of generality, we define the sign of the  circular random variables, $\phi, \theta \in [0, 2\pi)$ as   
\begin{eqnarray}
    sgn(\phi)&=& 2\left( \delta_{(\phi<\pi)}-0.5 \right) \in \{ 
 -1,1\}.
    \label{sign_function}
\end{eqnarray}
Similarly, it will hold for $\theta$ as well.

\begin{definition}
    Let $\Theta$, $\Phi$ be two zero-centered circular random variables with the joint probability density function $f(\phi,\theta)$. Then, using Equation-\ref{zero centered torus area of a segemnt}, the ``area covariance''  is defined as: 
$$ ACov(\Phi,\Theta)=\displaystyle E_{f(\phi,\theta)}\left[sgn(\phi)~~sgn(\theta) \sqrt{ A^{(0)}_C(\phi) \cdot A^{(0)}_C(\theta)}
 \right]. $$

\label{area covariance}
\end{definition}
If a random sample $\{(\phi_1,\theta_1), \ldots, (\phi_n,\theta_n)\}$ is given then using WLLN, $ACov(\Phi,\Theta)$ can be consistently estimated as
\begin{eqnarray}
\widetilde{ACov(\Phi,\Theta)} &=& \frac{1}{n} \sum_{i=1}^{n} sgn\left\{ (\phi_i - {\mu_\phi}) \mod 2\pi \right\}  sgn\left\{ (\theta_i - {\mu_\theta}) \mod k\pi \right\} \times  \nonumber \\ & & 
\Bigg[  A^{(0)}_C\left\{ (\phi_i - {\mu_\phi}) \mod 2\pi \right\} 
~ A^{(0)}_C\left\{ (\theta_i - {\mu_\theta}) \mod k\pi \right\} \Bigg]^{1/2},\nonumber
\end{eqnarray}
when $\mu_\phi, \mu_\theta$ are known mean directions.
 When $\mu_\phi, \mu_\theta$ are unknown, we can use the plug-in estimator
\begin{eqnarray}
\widehat{ACov(\Phi,\Theta)} &=& \frac{1}{n} \sum_{i=1}^{n} sgn\left\{ (\phi_i - {\hat{\mu}_\phi}) \mod 2\pi \right\}  sgn\left\{ (\theta_i - {\hat{\mu}_\theta}) \mod k\pi \right\} \times  \nonumber \\ & & \quad  \Bigg[  A^{(0)}_C\left\{ (\phi_i - {\hat{\mu}_\phi}) \mod 2\pi \right\}
 A^{(0)}_C\left\{ (\theta_i - {\hat{\mu}_\theta}) \mod k\pi \right\} \Bigg]^{1/2},\nonumber
\end{eqnarray}
 where $\widehat{\mu}_\phi, \widehat{\mu}_\theta$ are the estimated circular mean directions of the data. $k$ can be suitably chosen based on the range of the angular data as introduced in Definition-\ref{curved variance}. Note that if $k=1$, that is for spherical data, the sign corresponding to the vertical angle $\theta$ is not necessarily to be considered.

Without loss of generality assuming a marginal probability density function be denoted as $f(\cdot)$,  we obtain   from Definition-\ref{area covariance} and Equation-\ref{zero centered circle area of a segemnt}
\begin{equation}
     ACov(\cdot,\cdot)=\displaystyle E_{f(\cdot)}\left[  A^{(0)}(\cdot,\cdot)
 \right]= E_{f(\cdot)}\left[  A_{C}^{(0)}(\cdot)\right]=CVar(\cdot).
 \label{cov_var_theta}
\end{equation}
for any of the angular random variables $\Phi$ and $\Theta.$ 
Here, we can find $f(\theta)$ from the joint probability distribution function $f(\phi,\theta)$ as 
$f(\theta)=\displaystyle \int_{\phi=0}^{2\pi}f(\phi,\theta)~d\phi$. Using the expression of $f(\theta)$ in the Equation-\ref{cov_var_theta} we have

\begin{equation}
   CVar(\Theta)= E_{f(\phi,\theta)}\left[  A^{(0)}(\theta,\theta)
 \right]=  E_{f(\theta)}\left[  A^{(0)}(\theta,\theta)
 \right].
 \label{a11}
\end{equation}

Similarly, we can say that 
\begin{equation}
   CVar(\Phi)= E_{f(\phi,\theta)}\left[  A^{(0)}(\phi,\phi)
 \right]=E_{f(\phi)}\left[  A^{(0)}(\phi,\phi)
 \right].
 \label{a22}
\end{equation}

  \begin{lemma}
     Let, $a_{11}=CVar(\Phi), a_{22}=CVar(\Theta)$, and $a_{12}=a_{21}=ACov(\Phi,\Theta)$, then the curve dispersion (CD) matrix $\Sigma_{A}$ defined by 

$$\Sigma_{A}=\begin{pmatrix}
a_{11} & a_{12} \\
a_{21} & a_{22}
\end{pmatrix}$$ is a symmetric and positive semi-definite matrix.
 \label{cd_matrix}
\end{lemma}

\begin{proof}
  By construction, the matrix, $\Sigma_{A}$ is symmetric. 
   Now consider

     \begingroup
\allowdisplaybreaks
\begin{align}
a_{11} \cdot a_{22}&=  E_{f(\phi,\theta)}\left[  A^{(0)}(\phi,\phi)
 \right] \cdot E_{f(\phi,\theta)}\left[  A^{(0)}(\theta,\theta)
 \right] \nonumber \\
&=  E_{f(\phi,\theta)}\left[ \left( \sqrt{A^{(0)}(\phi,\phi)} \right)^2
 \right] \cdot E_{f(\phi,\theta)}\left[ \left( \sqrt{A^{(0)}(\theta,\theta)} \right)^2
 \right] \nonumber \\
&\geq \left(   E_{f(\phi,\theta)}\left[ \sqrt{A^{(0)}(\phi,\phi) \cdot A^{(0)}(\theta,\theta)} 
 \right]\right)^2  \geq [ACov(\Phi,\Theta)]^2 \nonumber \\
a_{11} \cdot a_{22}&\geq a_{12}^2 \mbox{ by Cauchy-Schwarz inequality}
 \label{covariance matrix definiteness}
\end{align}%
\endgroup
 This implies that $|\Sigma_{A}|\geq 0$ and $trace(\Sigma_{A})>0$.  Hence, the eigenvalues of $\Sigma_{A}$ are non-negative.  As a consequence  $\Sigma_{A}$ is positive semi-definite.
   \end{proof}

  \begin{rmk}
     The proportionate area $A_T\left[(\phi_1,\theta_1) (\phi_2,\theta_2)\right] \in [0, \frac{1}{4}]$. It may be noted that $A_T\left[(\phi_1,\theta_1) (\phi_2,\theta_2)\right]$ only depends only on  $\frac{r}{R}$ where, $0<\frac{r}{R}\leq 1$. 
     \label{remark_bounded-area}
  \end{rmk}

  \begin{rmk}
  $CVar$ does not depend on the (known) mean direction. 
  \end{rmk}
  %i.e.,  $CVar(\Theta)=CVar[(\Theta+\delta) \mod 2\pi] $, for any $\delta \in [0,2\pi).$  
  %Although the $CVar$ is invariant with respect to the mean direction of the circular data, 
  %the measurement will vary with the choice of zero-point on the surface of the curved torus because of its non-constant curvature. 
  \begin{rmk}
  As a natural choice, put $(\phi, \theta)=(0,0)$ in  Equation-\ref{torus para equn}, the zero-point on the curved torus is assumed to be $(R+r,0,0)$, and counter-clock-wise rotation is considered to be conventional.
  \end{rmk}

\begin{rmk}
For univariate linear random variable $X$ with expectation $\eta$ it is well-known that 
$$Var(X)=E(X-\eta)^2=EX^2-\eta^2.$$
% On the other hand, the dispersion matrix of multivariate random vector $\mathbf{X}$ with mean vector $\boldsymbol{\mu}$ is 
% $$D(\mathbf{X})=E[(\mathbf{X}-\boldsymbol{\mu})(\mathbf{X}-\boldsymbol{\mu})^T]=E\mathbf{X}\mathbf{X}^T-\boldsymbol{\mu}\boldsymbol{\mu}^T.$$
% Similarly for the Hilbert-valued functional data $\mathbf{X}(t)$ with mean function $\boldsymbol{\mu}(t)$ the covariance operator is 
% $$C(\mathbf{X})=E[(\mathbf{X}-\boldsymbol{\mu})\otimes(\mathbf{X}-\boldsymbol{\mu})]=E\mathbf{X}\otimes\mathbf{X}-\boldsymbol{\mu}\otimes\boldsymbol{\mu}.$$
% The above relations hold because of the commutativity of the product in real numbers. However, the notion of the product of two angles is not defined similarly.
%Hence, we have introduced the square of an angle in Equation-\ref{zero centered circle area of a segemnt}, which leads to the Definition-\ref{curved variance} of Curved-Variance. 
Though Definition-\ref{curved variance} is a generalization of the definition
$Var(X)=E(X-\eta)^2$, but the simplification  $Var(X)=EX^2-\eta^2$ is not generalizable to the case of angular data. i.e. $ E_f(A_C^{(0)}([(\Theta-\mu) \mod 2\pi])) \neq E_f(A_C^{(0)}(\Theta))-A_C^{(0)}(\mu)$ in general, where $\Theta$ is a circular random variable with circular mean $\mu$. %as defined in Definition-\ref{curved variance}. 
Consider a circular random variable $\Theta$ with probability density function $f(\theta)$ and mean direction at $\pi$ as a counter-example. Note that 
$$ A_C^{(0)}(\theta)-A_C^{(0)}(\pi)=A_C^{(0)}(\theta)-\frac{1}{4}\leq0  \mbox{~~for all } \theta \in [0,2\pi).$$ As a consequence $ \displaystyle E_f\left[ A_C^{(0)}(\Theta)\right]-\frac{1}{4}\le 0$.  Hence, $ \displaystyle E_f\left[ A_C^{(0)}(\Theta)\right]-\frac{1}{4}<0$ unless $\Theta$ is degenerate at $\pi$. Thus we see that  $CVar(\Theta)= E_f(A_C^{(0)}([(\Theta-\pi) \mod 2\pi])) \neq E_f(A_C^{(0)}(\Theta))-A_C^{(0)}(\pi)$ when $\Theta$ is not degenerate at $\pi$.  
The definition of ${CVar}$ considers the non-constant curvature through the area element of the surface of the torus. A similar approach, when applied to linear univariate data, would yield the usual definition of variance of linear univariate data since the curvature is constant. 
\end{rmk}

\begin{rmk}
     When we consider the sphere with the radius, $r$, the distribution $f(\phi,\theta)$ is a spherical distribution, and hence the curved variance and curved co-variance will be calculated using the formulas defined for the sphere.
 \end{rmk}

 \begin{lemma}
     Let, $\widehat{a}_{11}=\widehat{CVar}(\Phi), \widehat{a}_{22}=\widehat{CVar}(\Theta)$, and $\widehat{a}_{12}=\widehat{a}_{21}=\widehat{ACov}(\Phi,\Theta)$, then the estimated curve dispersion (CD) matrix $\widehat{\Sigma}_{A}$ defined by 

$$\widehat{\Sigma}_{A}=\begin{pmatrix}
\widehat{a}_{11} & \widehat{a}_{12} \\
\widehat{a}_{21} & \widehat{a}_{22}
\end{pmatrix}$$ converges in probability to the curve dispersion (CD) matrix $\Sigma_{A}$, defined in Lemma-\ref{cd_matrix} as the sample sizes increase to infinity.
 \label{est_cd_matrix_lemma}
\end{lemma}
\begin{proof}
 Let $\Phi$ be a zero-centered circular random variable with probability density function $f(\phi)$ on the unit circle $\mathbb{S}_1$. Also, assume that $\phi_1,\phi_2,\cdots,\phi_n$ are random samples from some circular distribution, with true and estimated mean directions be $\mu_{\phi}$, and $\hat{\mu}_\phi$, respectively. 
 Now, let $\delta_n=[(\hat{\mu}_\phi-\mu_\phi)\mod 2\pi]\xrightarrow{p} 0$ as $n\xrightarrow{} \infty$, which is a valid assumption, for more details \cite[see][Sec. 4.8, p.76]{mardia2000directional}. Again, suppose
 $\hat{a}_i=A_C^{(0)}[(\phi_i-\hat{\mu}_\phi) \mod 2\pi] \mbox{~~and~~} \tilde{a}_i=A_C^{(0)}[(\phi_i-\mu_\phi) \mod 2\pi].$
Consider,
\begin{eqnarray}
    \hat{a}_i &=& A_C^{(0)}[(\phi_i-\hat{\mu}_\phi) \mod 2\pi]\nonumber\\
    &=& A_C^{(0)}[(\phi_i-\mu_\phi+\mu_\phi-\hat{\mu}_\phi) \mod 2\pi]\nonumber\\
    &=& A_C^{(0)}[((\phi_i-\mu_\phi)\mod 2\pi+\delta_n) \mod 2\pi], \nonumber\\
    &=& A_C^{(0)}[(\phi_i-\mu_\phi)\mod 2\pi]+\delta_n \frac{d}{d\phi_i}A_C^{(0)}[(\phi_i-\mu_\phi)\mod 2\pi]+ \nonumber\\
    &&\frac{\delta_n ^2}{2!} \frac{d^2}{d\phi_i^{2}}A_C^{(0)}[(\phi_i-\mu_\phi)\mod 2\pi]+\cdots \cdots\cdots \nonumber\\
    &=& \tilde{a}_i+\delta_n \tilde{a}_i^{'}+ \frac{\delta_n ^2}{2!} \tilde{a}_i^{''}+\cdots \cdots\cdots
    \label{taylor_exp}
\end{eqnarray}
Now, to prove this lemma, it suffices to establish that  
\[
\hat{a}_{11} \xrightarrow{p} a_{11}, \quad \hat{a}_{22} \xrightarrow{p} a_{22}, \quad \text{and} \quad \hat{a}_{12} \xrightarrow{p} a_{12} \quad \text{as} \quad n\rightarrow \infty.
\]
Considering up-to first order of the Equation-\ref{taylor_exp}, 
\begin{eqnarray}
    |\hat{a}_{11}-a_{11}| &=&\left|\frac{1}{n} \sum_{i=1}^n \hat{a}_i- E_{f(\phi)}\left[ A_C^{(0)}(\Phi)\right] \right| \nonumber\\
    &=&\left|\frac{1}{n} \sum_{i=1}^n \hat{a}_i-\frac{1}{n} \sum_{i=1}^n \tilde{a}_i +\frac{1}{n} \sum_{i=1}^n \tilde{a}_i  - E_{f(\phi)}\left[ A_C^{(0)}(\Phi)\right] \right|\nonumber\\
    &\leq&  \left|\frac{1}{n} \sum_{i=1}^n \hat{a}_i-\frac{1}{n} \sum_{i=1}^n \tilde{a}_i \right| +\left| \frac{1}{n} \sum_{i=1}^n \tilde{a}_i  - E_{f(\phi)}\left[ A_C^{(0)}(\Phi)\right] \right| \nonumber\\
    &=&  \left|\frac{1}{n} \sum_{i=1}^n (\tilde{a}_i+\delta_n  \tilde{a}_i^{'})-\frac{1}{n} \sum_{i=1}^n \tilde{a}_i \right| +\left| \frac{1}{n} \sum_{i=1}^n \tilde{a}_i  - E_{f(\phi)}\left[ A_C^{(0)}(\Phi)\right] \right| \nonumber\\
    &\leq& |\delta_n| \left( \frac{1}{n}  \sum_{i=1}^n \left|\tilde{a}_i^{'}\right| \right) +\left| \frac{1}{n} \sum_{i=1}^n \tilde{a}_i  - E_{f(\phi)}\left[ A_C^{(0)}(\Phi)\right] \right|.
    \label{est_true_cvar_convergence}
\end{eqnarray}
Now, from the Remark 1 in the main text, we observe that each \(\tilde{a}_i \leq \frac{1}{4} \), and \( |\tilde{a}^{'}_i| \leq M_{\tilde{a}_i^{'}}\) which is a positive constant. Therefore,  $\frac{1}{n} \sum_{i=1}^n \left|\tilde{a}_i^{'}\right|\leq M_{\tilde{a}_i^{'}}$ is a finite quantity.
Hence, we have
\begin{eqnarray}
   |\delta_n|  \left( \frac{1}{n}  \sum_{i=1}^n \left|\tilde{a}_i^{'}\right| \right) &=& O \left(\frac{1}{n} \right) \text{~~~and~~~}  \\
   \left| \frac{1}{n} \sum_{i=1}^n \tilde{a}_i  - E_{f(\phi)}\left[ A_C^{(0)}(\Phi)\right] \right|&=&O \left(\frac{1}{n} \right).
\end{eqnarray}
As a consequence, we have 
$
\hat{a}_{11} \xrightarrow{p} a_{11} \text{~~as~~} n\xrightarrow{} \infty.
$ By employing a similar argument, we can show that  
$\hat{a}_{22} \xrightarrow{p} a_{22} \quad \text{as} \quad n\xrightarrow{} \infty.$
Since we have both \(\hat{a}_{11} \xrightarrow{p} a_{11}\) and \(\hat{a}_{22} \xrightarrow{p} a_{22}\) as \(n\xrightarrow{} \infty\), it immediately follows that  
$
\hat{a}_{12} \xrightarrow{p} a_{12}  \text{~~as~~}  n\xrightarrow{} \infty.$ This completes the proof.
\end{proof}

\section{Change in mean direction of toroidal data}
\label{changepoint_mean_torus}

%\subsection{Changepoint detection in the mean direction}  
\label{section_cp_torus_mean}

\subsection{Single changepoint}

Let $\Psi_i=(\phi_i,\theta_i) \in [0,2\pi) \times [0,2\pi), \mbox{~~for~~} i=1, \ldots,n$ be independent angular random vectors. We are interested in addressing  the following testing problem : 

\begin{eqnarray}
     H_{0t} &:& \Psi_{i}\overset{\mathrm{i.i.d.}}{\scalebox{1.5}{$\sim$}} F(\psi;\boldsymbol{\xi}_1,\boldsymbol{\eta}) 
     \mbox{~~for all~~} i =1,2,\ldots n, \nonumber\\
   H_{1t}&:&   \begin{cases}
               \Psi_{i} \overset{\mathrm{i.i.d.}}{\scalebox{1.5}{$\sim$}}   F(\psi;\boldsymbol{\xi}_1,\boldsymbol{\eta}) & \text{, } 1 \leq i \leq k^*\\
         \Psi_{i}\overset{\mathrm{i.i.d.}}{\scalebox{1.5}{$\sim$}}F(\psi;\boldsymbol{\xi}_2,\boldsymbol{\eta})   & \text{, } (k^*+1) \leq i \leq n,
    \end{cases}
    \label{torus mean test}
\end{eqnarray}
where $\boldsymbol{\xi}_1,\boldsymbol{\xi}_2 $ are  suitable  vector-valued  parameters representing the location (mean directions) of the distributions and $\boldsymbol{\xi}_1 \neq \boldsymbol{\xi}_2,$ 
under the alternative hypothesis $H_{1t}$. In both hypotheses, it is assumed that the concentration cum shape-parameter vector $\boldsymbol{\eta}$ remains unchanged for the entire sequence. Here we assume that  $\boldsymbol{\eta}$ is unrelated to $\boldsymbol{\xi}_1$ and $\boldsymbol{\xi}_2 $. 
We consider the corresponding  mean shifted angles for $$\phi_i^c=[(\phi_i-\widehat{\mu}_\phi) \mod 2\pi] \mbox{~and~}  \theta_i^c=[(\theta_i-\widehat{\mu}_\theta) \mod 2\pi], $$ where   $\widehat{\mu}_\phi$, $\widehat{\mu}_\theta$ are the estimated circular mean direction of 
$\phi_i, \theta_i,$ respectively, $ \text{~for~} i=1, \cdots,n$, on the surface of a curved torus. Now onwards  $\widehat{\mu}_\phi$, $\widehat{\mu}_\theta$ are treated to be constants. This is reasonable particularly when the sample size is large.
% Without loss of generality, the sign of the  circular random variables, $\phi, \theta \in [0, 2\pi)$ can be defined as   
% \begin{eqnarray}
%     sgn(\phi)&=& 2\left( \delta_{(\phi<\pi)}-0.5 \right) \in \{ 
%  -1,1\}.
%     \label{sign_function}
% \end{eqnarray}
% Similarly, it will hold for $\theta$ as well.
Using the Definition-\ref{square_of_an_angle} (for torus), we get the corresponding square areas as
$$\widehat{a}_i=sgn^2(\phi_i^c)A_C^{(0)}(\phi_i^c) \mbox{~~and~~} \widehat{b}_i=sgn^2(\theta_i^c)A_C^{(0)}(\theta_i^c),$$ respectively, together with $\widehat{(ab)}_i=sgn(\phi_i^c)~~sgn(\theta_i^c) \sqrt{A_C^{(0)}(\phi_i^c) \cdot A_C^{(0)}(\theta_i^c)}$. Hence, calculate the curved variance and curved co-variance as
\begin{eqnarray}
    \widehat{cv}_\phi&=&\frac{1}{n}\sum_{i=1}^n~ \widehat{a}_i,\hspace{0.4cm} \widehat{cv}_\theta=\frac{1}{n}\sum_{i=1}^n~ \widehat{b}_i,~\mbox{and~}
    \widehat{ccv}_{\phi\theta}=\frac{1}{n}\sum_{i=1}^n~ \widehat{(ab)}_i.
\end{eqnarray}
Now, we obtain  the estimated  \textit{curved dispersion matrix} as 
\begin{eqnarray}
   \widehat{ \Sigma}_{(t)}= \begin{bmatrix}
\widehat{cv}_\phi & \widehat{ccv}_{\phi\theta}  \\
\widehat{ccv}_{\phi\theta} & \widehat{cv}_\theta
\end{bmatrix}
\label{cvar_matrix torus}
\end{eqnarray}
which converges to its theoretical analog $ \Sigma_{(t)}$ with probability $1$ following the Lemma-\ref{est_cd_matrix_lemma}. Hence, a similar convergence holds for the inverse given by
\begin{eqnarray}
    \widehat{\Sigma}_{(t)}^{-1}= \frac{1}{\widehat{cv}_\phi \cdot
\widehat{cv}_\theta-\widehat{ccv}_{\phi\theta}^2} \begin{bmatrix}
\widehat{cv}_\theta & -\widehat{ccv}_{\phi\theta}  \\
-\widehat{ccv}_{\phi\theta} & \widehat{cv}_\phi
\end{bmatrix}~~ \mbox{and denoted as} \begin{bmatrix}
\iota \widehat{cv}_\phi & \iota \widehat{ccv}_{\phi\theta}  \\
\iota \widehat{ccv}_{\phi\theta} & \iota  \widehat{cv}_{\theta}
\end{bmatrix}.
\label{inv_matrix torus}
\end{eqnarray}
% We consider the corresponding mean $(\mu_\phi,\mu_\theta )$  of the initial $\lambda$\%  data under the assumption that the changepoint may occur in the rest $(100-\lambda)\%$  of the data. Hence we obtain  $$\phi_i^m=[(\phi_i-\mu_\phi) \mod 2\pi] \mbox{~and~}  \theta_i^m=[(\theta_i-\mu_\theta) \mod 2\pi], $$ where   $\mu_\phi$, $\mu_\theta$ are the circular mean direction before changepoint of 
% $\phi_i, \theta_i,$ respectively, $ \text{~for~} i=1, \cdots,n$, on the surface of a curved torus.
Now, let us consider ${a}_i=sgn^2(\phi_i) A_C^{(0)}(\phi_i)$, and ${b}_i=sgn^2(\theta_i) A_C^{(0)}(\theta_i)$, respectively, together with ${(ab)}_i= sgn(\phi_i)~~sgn(\theta_i) 
 \sqrt{A_C^{(0)}(\phi_i) \cdot A_C^{(0)}(\theta_i)}$. We calculate an expression analogous to the quadratic form associated with the \textit{Mahalanobis distance } using the matrix $\widehat{\Sigma}_{(t)}^{-1}$ and the vector 
 $\left( sgn(\phi_i)
 \sqrt{A_C^{(0)}(\phi_i)}, sgn(\theta_i) 
 \sqrt{A_C^{(0)}(\theta_i)} \right)^{T}$ to obtain
\begin{eqnarray}
    Q_i&=& \iota \widehat{cv}_\phi \cdot {a}_i +\iota \widehat{cv}_\theta \cdot {b}_i +2\cdot \iota \widehat{ccv}_{\phi\theta} \cdot {(ab)}_i 
\label{qform_nc}
\end{eqnarray}
$  \mbox{~for~} i=1,2,\ldots n$. Now, consider the estimated variance of the sequence, $\{ Q\}$ as
$\widehat{S}_{Q}^2{=}\frac{1}{n-1} \sum_{i=1}^{n}\left(Q_i-\Bar{Q}\right)^2,$ where $n~\bar{Q}=\displaystyle \sum_{i=1}^{n}Q_i,$ and we define a  CUSUM process
for fixed $\widehat{S}_{Q}^2$  as
\begin{equation}
       U_Q(k)=\frac{1}{\sqrt{n~\widehat{S}_{Q}^2}}  \sum_{i=1}^{k} \left(Q_i-\bar{Q}  \right)\mbox{~~for all~~} k=1,\ldots,n 
       \label{torus cusum process}
\end{equation}
to obtain the test statistic
\begin{equation}
    \mathcal{M}_n=\displaystyle \max_{1 \leq k < n} |U_Q(k)|.
    \label{torus_test_statistic}
\end{equation}
We then suggest an estimate of the change point location $\widehat{k}^{*}$ ,
based on the above test statistic as 
\begin{equation}
\widehat{k}^{*}=\displaystyle \arg \max_{1 \leq k < n} |U_Q(k)|.
    \label{est_loc}
\end{equation}
Hence, we reject the null hypothesis, $H_{0t}$, if $\mathcal{M}_n>k_{\alpha}$, where, $k_{\alpha}$ is the upper $\alpha$ point of the exact (or asymptotic) distribution of $\mathcal{M}_n$ under the null hypothesis. The closed-form distribution of $\mathcal{M}_n$ is not available; hence, we need to take recourse to simulation to obtain the cut-off value $k_{\alpha}$. When $n$ is large, the limiting distribution of $\mathcal{M}_n$ can be derived as follows.
Let us consider $u \in (0,1)$, and denote $k= \lfloor{nu}\rfloor$, which is a standard assumprtion for CUSUM processes  \cite[see][p. 36, Theorem-4.3] {de2009self}. Hence, from Equation-\ref{torus cusum process} we can write
\begin{equation}
    U_Q^{(n)}(u)=U_Q(\lfloor{nu}\rfloor)=\left| \frac{1}{\sqrt{n ~S_{Q}^2}} \left[ \sum_{i=1}^{\lfloor{nu}\rfloor} Q_i-u \sum_{i=1}^n Q_i   \right]\right|.
    \label{concentration asym}
\end{equation}
Then, with the proper embedding of Skorohod topology in $D[0,1]$ \cite[see][Ch. 3]{billingsley2013convergence}, under the null hypothesis, $H_{0}$, and as $n \rightarrow \infty$, the process $U_Q^{(n)}(u)$  converges weakly to $B_{0}(u),$ where $B_{0}(u)$ is the standard Brownian bridge on $[0,1].$  Now, we can compute the upper-$\alpha$ value, $k_{\alpha},$ from the above limiting random variable $K_\infty$ of the test statistic, $\mathcal{M}_n$ (see Lemma-\ref{bbridge_lemma_torus}). The corresponding large sample approximation is discussed in Section-\ref{simulation}.

% \begin{algorithm}
% \caption{Empirical Threshold Calculation from Kolmogorov distribution}
% \label{threshold_estimation}

% \textbf{Input:} Segment length $n$, desired quantile level $q$ (e.g., $q = 0.95$)

% \textbf{Output:} Empirical threshold value $\tau$ for significance testing

% \begin{algorithmic}[1]
% \Function{EstimateThreshold}{$n$, $q$}
%     \State Initialize $\mathcal{L} \gets $ an empty list 
%     \For{$i = 1$ to $5000$}
%         \State Simulate null data: $z_1, \dots, z_n \sim \mathcal{N}(\mu, \sigma^2)$ %(e.g., $\mu=5$, $\sigma=3$)
%         \State Center the data: $z_t' \gets z_t - \bar{z}$
%         \State Compute cumulative sum statistic: $B_k \gets \left| \sum_{t=1}^k z_t' \right| / (\sqrt{n} \cdot \hat{\sigma})$ for $k = 1, \dots, n$
%         \State Record test value: $\mathcal{L}_i \gets \displaystyle \max_k B_k$
%     \EndFor
%     \State Compute threshold: $\tau \gets$ empirical $q$-quantile of $\mathcal{L}$
%     \State \Return $\tau$
% \EndFunction
% \end{algorithmic}
% \end{algorithm}

\newpage

\begin{algorithm}[t]
\caption{Empirical Threshold Calculation from Kolmogorov distribution}
\label{threshold_estimation}
\textbf{Input:} Segment length $n$, desired quantile level $q$ (e.g., $q = 0.95$)
\textbf{Output:} Empirical threshold value $\tau$ for significance testing
\begin{algorithmic}[1]
\Function{EstimateThreshold}{$n$, $q$}
    \State Initialize $\mathcal{L} \gets $ an empty list 
    \For{$i = 1$ to $5000$}
        \State Simulate null data: $z_1, \dots, z_n \sim \mathcal{N}(\mu, \sigma^2)$ %(e.g., $\mu=5$, $\sigma=3$)
        \State Center the data: $z_t' \gets z_t - \bar{z}$
        \State Compute cumulative sum statistic: $B_k \gets \left| \sum_{t=1}^k z_t' \right| / (\sqrt{n} \cdot \widehat{\sigma})$ for $k = 1, \dots, n$
        \State Record test value: $\mathcal{L}_i \gets \displaystyle \max_k B_k$
    \EndFor
    \State Compute threshold: $\tau \gets$ empirical $q$-quantile of $\mathcal{L}$
    \State \Return $\tau$
\EndFunction
\end{algorithmic}
\end{algorithm}
\subsection{Multiple Changepoint}
In many real-world applications, structural changes may occur at several unknown points in a data sequence. We extend the framework of single changepoint detection to accommodate multiple changepoint scenarios, where both the number and locations of changes are unknown. Formally, the problem may be posed as:
$$
\begin{aligned}
    H_0 &: \Psi_i \overset{\text{i.i.d.}}{\sim} F(\psi; \boldsymbol{\xi}_1, \boldsymbol{\eta}) \quad \text{for all } i = 1, 2, \dots, n, \\
    H_{1,m} &: 
    \begin{cases}
        \Psi_i \overset{\text{i.i.d.}}{\sim} F(\psi; \boldsymbol{\xi}_1, \boldsymbol{\eta}) & \text{for } 1 \leq i \leq k^*_1 \\
        \Psi_i \overset{\text{i.i.d.}}{\sim} F(\psi; \boldsymbol{\xi}_2, \boldsymbol{\eta}) & \text{for } k^*_1 + 1 \leq i \leq k^*_2 \\
        \;\vdots \\
        \Psi_i \overset{\text{i.i.d.}}{\sim} F(\psi; \boldsymbol{\xi}_{m+1}, \boldsymbol{\eta}) & \text{for } k^*_m + 1 \leq i \leq n
    \end{cases}
\end{aligned}
$$
where the changepoints $k^*_1 < k^*_2 < \cdots < k^*_m \in (1, n)$ partition the sequence into $m+1$ homogeneous segments. Each segment is assumed to follow a distribution from the same parametric family, but with distinct location parameters $\boldsymbol{\xi}_j$, while sharing common nuisance parameters $\boldsymbol{\eta}$.

To identify multiple changepoints, we adopt a recursive binary segmentation (RBS) approach. This strategy applies the single changepoint detection method recursively to subsets of the data, splitting the sequence each time a significant changepoint is detected. At each step, the detection is localized within a subinterval $[b_k, e_k]$, and a candidate changepoint $\widehat{\gamma}_k$ is estimated as:
$$
\widehat{\gamma}_k = \arg \max_{j \in [b_k, e_k]} \left| U_j \right|,
$$
where $U_j$ is the normalized test statistic derived from the cumulative sum process over the interval $[b_k, e_k]$. If the test statistic at $\widehat{\gamma}_k$ exceeds a predefined significance threshold, the subinterval is divided at $\widehat{\gamma}_k$, and the procedure is recursively applied to each resulting segment.
To avoid over-segmentation, we impose a minimum segment length constraint and apply significance testing at each step. This ensures the method is robust to spurious fluctuations and only detects substantial structural changes. The final set of changepoints is obtained by collecting all significant split points, denoted $\{ \widehat{\gamma}_1, \widehat{\gamma}_2, \ldots, \widehat{\gamma}_s \}$. The detected
changepoints using binary segmentation are not necessarily in
increasing order, when there are more than one changepoint. So, we sort the set of estimated changepoints in increasing order, that is $\{ \widehat{k}_1^*,\widehat{k}_2^*,\ldots,\widehat{k}_s^* \}=\mbox{sort}\{ \widehat{\gamma}_1, \widehat{\gamma}_2, \ldots, \widehat{\gamma}_s \}$.

\small

\begin{algorithm}[h!]
\caption{Recursive Changepoint Detection for Bivariate Angular Data}
\label{algo:recursive_segmentation_algo}
\textbf{Input:} Matrix of angular data $\mathbf{X} = [(\phi_t, \theta_t)]_{t=1}^n$, minimum segment length $h$, significance level $\alpha$.
\textbf{Output:} Estimated set of changepoint locations $\widehat{\mathcal{C}}$.
\begin{algorithmic}[1]
\Function{DetectChangepoints}{$\mathbf{X}_{[s:e]}$, $s$}
    \State $n \gets e - s + 1$
    \If{$n < h$}
        \State \Return $\emptyset$
    \EndIf
    \State Apply test statistic $\mathcal{M}_n$ to $\mathbf{X}_{[s:e]}$ and obtain:
    \State Estimated changepoint location $\widehat{k}$ (relative to segment), and test value $T$
   % \State Compute $p$-value from null distribution of test statistic
    \State Compute threshold value $\tau \gets$ {EstimateThreshold}~{($n$, $1-\alpha$)} from Algorithm-\ref{threshold_estimation}
    \If{$T \le \tau$}
        \State \Return $\emptyset$
    \EndIf
    \State Global changepoint: $k_{\text{global}} \gets s + \widehat{k} - 1$
    \State Initialize changepoint set: $\widehat{\mathcal{C}} \gets \{k_{\text{global}}\}$
    \State Recursively call left segment: 
        $\widehat{\mathcal{C}} \gets \widehat{\mathcal{C}} ~~\cup$ \Call{DetectChangepoints}{$\mathbf{X}_{[s: k_{\text{global}}]}$, $s$}
    \State Recursively call right segment:
        $\widehat{\mathcal{C}} \gets \widehat{\mathcal{C}} ~~\cup$ \Call{DetectChangepoints}{$\mathbf{X}_{[k_{\text{global}}+1: e]}$, $k_{\text{global}}+1$}
    \State \Return $\widehat{\mathcal{C}}$
\EndFunction
\Statex
%\State \Return \Call{DetectChangepoints}{$\mathbf{X}_{[1:n]}$, $1$}
\end{algorithmic}
\end{algorithm}

\section{Theoretical guarantees}

We study the asymptotic properties of the proposed method for detecting
single and multiple  changepoints in toroidal data.

 \begin{lem}
Under the null hypothesis, $H_0$,
    \begin{equation}
   \mathcal{M}_n  \overset{d}{\to} \displaystyle \sup_{0<u<1} |B_{0}(u)|=K_\infty,
    \label{bbridge}
\end{equation}
where $K_\infty$ follows the \textit{Kolmogorov distribution}.
 \label{bbridge_lemma_torus}
\end{lem}

\begin{proof}
Let $q_i=Q_i-E(Q_i)$, then clearly, $E(q_i)=0$, and $S_{q}^2=\text{Var~}(q_i)<\infty,$ for $i=1,\cdots,n.$ The estimated variance for $q_i$'s is 
$$S_{q}^2=Var(q_1)\widehat{=}\frac{1}{n-1} \sum_{i=1}^{n}\left(q_i-\bar{q}\right)^2=\widehat S_q^2.$$ Now construct the CUSUM process as
\begin{eqnarray}
       U_{q}(k)&=&\frac{1}{\sqrt{n}~\widehat S_q }\left[ \sum_{i=1}^{k} q_i-k\Bar{q}   \right] \mbox{~~for all~~} k=1,\ldots,n \nonumber\\ 
       &=&\frac{ S_q}{\widehat S_q} \frac{1}{\sqrt{n}~ S_q }\left[ \sum_{i=1}^{k} q_i-k\Bar{q}   \right]\\ \nonumber 
       \label{cusum process zero}
\end{eqnarray}
Let us consider $u \in (0,1)$, and denote $k= \lfloor{nu}\rfloor$. Hence, from Equation-\ref{cusum process zero} we can write
\begin{equation}
  U_{q}^{(n)}(u)=  U_{q}(\lfloor{nu}\rfloor)= \frac{1}{\sqrt{n~}  S_{q}} \left[ \sum_{i=1}^{\lfloor{nu}\rfloor} q_i-u \sum_{i=1}^n q_i   \right].
    \label{concentration_asym_zero}
\end{equation}
Now using Donsker's theorem \cite[see][Ch. 16]{billingsley2013convergence}  and Slutsky's theorem \cite[see][Ch. 9]{athreya2006measure} we can write 
\begin{eqnarray}
    \frac{1}{\sqrt{n~}  S_{q}} \sum_{i=1}^{\lfloor{nu}\rfloor} q_i &\implies& W(u)   \text{~~~~~for all~~~~} 0<u\leq1,
\end{eqnarray}
where `$\implies$' represents week convergence, and $W(u)$ is the Winner processes on $[0,1]$. Hence, Equation-\ref{concentration_asym_zero} becomes
\begin{equation}
  U_{q}^{(n)}(u)\implies  W(u)-u~W(1)=B_0(u),
    \label{bbridge_zero}
\end{equation}
where, $B_0(u)$ is standard Brownian bridge. Noting that $\widehat S_{q}^2=\widehat S_{Q}^2$ and $U_{q}(k)=U_Q(k).$
It is immediate that
\begin{equation}
  U_{Q}^{(n)}(u)\implies B_0(u) \text{~~~~~~under~~} H_0.
\end{equation}
Hence, the lemma follows.
\end{proof}

 The next  theorem provides guarantees for consistency of
the proposed single change point estimate in Equation-\ref{est_loc}.  

\begin{thm}[Consistency of the Estimated Changepoint]
Let $k^* \in[0,n]$ be the location of the true single changepoint and $\hat{k}^*$ be the estimated location of the changepoint. Then $\hat{k}^* \overset{p}\rightarrow k^*$. Hence the estimated location of the changepoint is consistent.
\end{thm}
\begin{proof}
Assume that \( k \leq k^* \), where \( k^* \) represents the true but unknown location of the changepoint. Define \( m_1 = E(Q_i) \) for \( i \leq k^* \) and \( m_2 = E(Q_i) \) for \( i > k^* \). Additionally, let \( \Delta = m_1 - m_2 \) denote the difference in expected values before and after the changepoint. Denoting $ q_i= Q_i-E( Q_i)$ for $i= 1, \cdots,n,$ we can write the partial sum 
\begin{eqnarray}
\frac{1}{\sqrt{n}}\left[ \sum_{i=1}^{k}  Q_i - k\Bar{Q} \right] &=& \frac{1}{\sqrt{n}} \bigg[ \bigg( \sum_{i=1}^{k} ( Q_i - m_1) 
- \frac{k}{n} \bigg\{ \sum_{i=1}^{k^*}( Q_i - m_1) \nonumber \\
&& + \sum_{i=k^*+1}^{n} ( Q_i - m_2) \bigg\} \bigg) 
+ \frac{k(n-k^*)}{n} \Delta \bigg] \nonumber \\
&=& \frac{1}{\sqrt{n} }\left[ \sum_{i=1}^{k}  q_i-k\Bar{q}   \right]+ \frac{k(n-k^*)}{n^{3/2}} \Delta
\label{before_true_cp_est_loc}
\end{eqnarray}
Similarly, the following can be written for $k >k^*$
\begin{eqnarray}
 \frac{1}{\sqrt{n} }\left[ \sum_{i=1}^{k}  Q_i-k\Bar{Q} \right]
&=& \frac{1}{\sqrt{n}}\left[ \sum_{i=1}^{k}  q_i - k\bar{q} \right]+ \frac{k^*(n-k)}{n^{3/2}} \Delta
\label{after_true_cp_est_loc}
\end{eqnarray}
Let \( k = \lfloor n u \rfloor \) and \( k^* = \lfloor n u^* \rfloor \) for some \( u, u^* \in (0,1) \). Using Equations-\ref{before_true_cp_est_loc} and \ref{after_true_cp_est_loc}, we can express the following:
\begin{eqnarray}
   \frac{1}{\sqrt{n}~\widehat S_Q }\left[ \sum_{i=1}^{k}  Q_i-k\Bar{Q}   \right] &=& \begin{cases}
              \displaystyle  \frac{1}{\sqrt{n} \widehat S_Q}\left[ \sum_{i=1}^{k}  q_i-k\bar{q}   \right]+  \frac{1}{ \widehat S_Q} \frac{k(n-k^*)}{n^{3/2}} \Delta & \text{~if~} k\leq k^{*}\\
       \displaystyle   \frac{1}{\sqrt{n}\widehat S_Q}\left[ \sum_{i=1}^{k}  q_i - k\bar{q} \right]+  \frac{1}{ \widehat S_Q}  \frac{k^*(n-k)}{n^{3/2}} \Delta & \text{~if~} k >k^*
    \end{cases} \nonumber \\
    &=& \begin{cases}
              \displaystyle \left(\frac{ S_q}{\widehat S_Q}\right) \left( \frac{1}{\sqrt{n}  S_q}\left[ \sum_{i=1}^{\lfloor nu \rfloor}  q_i - u \sum_{i=1}^{ n}  q_i \right]+  \frac{1}{  S_q}  \sqrt{n}(1-u^*)u \Delta \right)& \text{~if~} u\leq u^{*} \vspace{0.2cm}\\ \nonumber 
       \displaystyle  \left(\frac{ S_q}{\widehat S_Q}\right) \left(  \frac{1}{\sqrt{n}  S_q}\left[ \sum_{i=1}^{\lfloor nu \rfloor}  q_i - u \sum_{i=1}^{ n}  q_i \right]+
        \frac{1}{  S_q} \sqrt{n}(1-u)u^* \Delta \right)& \text{~if~} u >u^* .
    \end{cases} \\ 
    &=& \displaystyle \left(\frac{ S_q}{\widehat S_Q}\right) \left\{ \frac{1}{\sqrt{n}  S_q}\left[ \sum_{i=1}^{\lfloor nu \rfloor}  q_i - u \sum_{i=1}^{ n}  q_i \right]+    \sqrt{n} c_* \right\},
    \label{consistency_proof_est_loc}
\end{eqnarray}
where $c_*=\displaystyle \frac{ \min\{u,u^{*}\}( 1-\max\{u,u^{*}\})\Delta}{ S_q}.$ Denoting $\text{Var}( q_1)= S_{q}^2$,
under $H_1$ it can be shown that 
$
    \widehat S_Q^2 \overset{p}{\longrightarrow}  S_{q}^2 + \left[u^{*}(1-u^{*})\Delta\right]^2.
    \label{est_variance_under_H_1}$ It can be easily shown that $ \left(\frac{ S_q}{\widehat S_Q}\right)=O_p(1)$, and the partial sum processes, $ U_{Q}^{(n)}(u)=\displaystyle\frac{1}{\sqrt{n}  S_q}\left[ \sum_{i=1}^{\lfloor nu \rfloor}  q_i - u \sum_{i=1}^{ n}  q_i \right]$ weakly converges to standard Brownian bridge on $[0,1]$, and hence $O_p(1)$. Therefore using the Equation-\ref{consistency_proof_est_loc} we have the following 
\begin{eqnarray}
    \widehat{k}^{*}&=& \displaystyle \arg \max_{1 \leq k < n} |U_Q(k)|=\displaystyle \arg \max_{1 \leq k < n} | U_{Q}^{(n)}(u)+\sqrt{n} c_*|\nonumber\\
\implies \widehat{u}^*    &=&\arg \max_{1 \leq u< 1} [O_p(1)+\sqrt{n} c_*]\nonumber\\
   % &=& \arg \max_{1 \leq u< 1}O_p(1)+ \arg \max_{1 \leq u< 1} [\sqrt{n} c_*]\nonumber\\
    &\overset{p}{\longrightarrow}& \arg \max_{1 \leq u< 1} [\sqrt{n} c_*]\nonumber\\
    &=& u^*~~~~~~~~~~[\mbox{since}~c^*~~ \mbox{is maximized at}~~ u=u^*]\nonumber\\
 \implies \widehat{k}^*&\overset{p}{\longrightarrow}& k^* \\ \nonumber
\end{eqnarray}
and hence, the theorem follows.
\end{proof}

\begin{cor}
     Under $H_0$, the test statistic, $ M_n(\hat{k}^*)$ we have the following:
$$\lim_{n \rightarrow\infty} \mathbb{P}_{H_0}\left(\mathcal{M}_n(\hat{k}^*)\geq k_\alpha  \right)\rightarrow \alpha.$$
\label{level_corr_torus}
\end{cor}
\begin{proof}
    As we obtained in Lemma-\ref{bbridge_lemma_torus}, the asymptotic distribution of the test statistic is the Kolmogorov distribution in general under the null hypothesis $H_{0t}$. So, the proof is straightforward.
\end{proof}

Under the alternative hypothesis, $H_1$, the proposed test is consistent, which is indicated by the following corollary.

\begin{cor}
    % Under $H_1$, the test statistic, $$ M_n= \max_{1\leq k<n} |T(k)|\rightarrow \infty.$$
For a fixed value of $\alpha \in (0,1)$ the probability of type II error   goes to zero exponentially, as the sample size increases to infinity, i.e.

$$\lim_{n \rightarrow\infty} \mathbb{P}_{H_1}\left(\mathcal{M}_n< k_\alpha  \right) \rightarrow  0.$$
\label{consistency_corr_torus}
\end{cor}
\begin{proof}
Introducing $k_{\alpha,n}=\left(\sqrt{n} c_*-k_\alpha \frac{\widehat S_Q}{  S_q }\right)$, and using Equation-\ref{consistency_proof_est_loc}  we obtain
\begin{eqnarray}
    \mathbb{P}(\text{type II error})&=& \mathbb{P}_{H_1}\left(\max_{1\leq k<n}|U_Q(k)|< k_\alpha  \right) \nonumber\\ 
    &=& \mathbb{P}\left(-\max_{1\leq k<n}\displaystyle  \left| \frac{1}{\sqrt{n}  S_q}\left[ \sum_{i=1}^{k}  q_i - k\bar{q} \right] \right|>  \left(\sqrt{n} c_*-k_\alpha \frac{\widehat S_Q}{  S_q }\right) \right) \nonumber\\ 
    &\leq& \mathbb{P}\left(\max_{1\leq k<n}\displaystyle  \left| \frac{1}{\sqrt{n}  S_q}\left[ \sum_{i=1}^{k}  q_i - k\bar{q} \right] \right|>  \left(\sqrt{n} c_*-k_\alpha \frac{\widehat S_Q}{  S_q }\right) \right).\nonumber\\ \nonumber
\text{Hence,~~} \lim_{n\rightarrow \infty}   \mathbb{P}(\text{type II error})  &=&  \lim_{n\rightarrow \infty} \mathbb{P}\left(\max_{1\leq u<1}\displaystyle  \left| B_0(u) \right|>  k_{\alpha,n} \right) \leq \lim_{n\rightarrow \infty}  2e^{-2(k_{\alpha,n})^2} \rightarrow 0.
% \\ \nonumber
%      &\leq& 2e^{-(k_{\alpha,n})^2} \longrightarrow 0, \text{~~as~~}n \rightarrow \infty.
\end{eqnarray}
\end{proof}

\section{Change in mean direction of spherical data }
\label{changepoint_mean_sphere}

%\subsection{Detection of mean change in spherical data }  
\label{section_cp_sphere_mean}

Let $\Psi_i=(\phi_i,\theta_i) \in [0,2\pi) \times [0,\pi), \mbox{~~for~~} i=1, \ldots,n$ be independent angular random vectors. We are interested in addressing  the following testing problem : 

\begin{eqnarray}
     H_{0s} &:& \Psi_{i}\overset{\mathrm{i.i.d.}}{\scalebox{1.5}{$\sim$}} F(\psi;\boldsymbol{\xi}_1,\boldsymbol{\eta}) 
     \mbox{~~for all~~} i =1,2,\ldots n, \nonumber\\
     H_{1s}&:&   \begin{cases}
               \Psi_{i} \overset{\mathrm{i.i.d.}}{\scalebox{1.5}{$\sim$}} F(\psi;\boldsymbol{\xi}_1,\boldsymbol{\eta}) & \text{, } 1 \leq i \leq k^*\\
         \Psi_{i}\overset{\mathrm{i.i.d.}}{\scalebox{1.5}{$\sim$}} F(\psi;\boldsymbol{\xi}_2,\boldsymbol{\eta})  & \text{, } (k^*+1) \leq i \leq n,
    \end{cases}
    \label{general test}
\end{eqnarray}
where $\boldsymbol{\xi}_1,\boldsymbol{\xi}_2 $ are  suitable  vector-valued  parameters representing the location (mean directions) of the distributions and $\boldsymbol{\xi}_1 \neq \boldsymbol{\xi}_2,$ 
under the alternative hypothesis $H_{1s}$. In both hypotheses, it is assumed that the concentration cum shape-parameter vector $\boldsymbol{\eta}$ remains unchanged for the entire sequence.

We consider the corresponding  mean shifted angles for $$\phi_i^c=[(\phi_i-\hat{\mu}_\phi) \mod 2\pi] \mbox{~and~}  \theta_i^c=[(\theta_i-\hat{\mu}_\theta) \mod \pi], $$ where   $\hat{\mu}_\phi$, $\hat{\mu}_\theta$ are the estimated circular mean direction of 
$\phi_i, \theta_i,$ respectively, $ \text{~for~} i=1, \cdots,n$, on the surface of a sphere.  
% Without loss of generality, the sign of the  circular random variables, $\phi \in [0, 2\pi)$ can be defined as   
% \begin{eqnarray}
%     sgn(\phi)&=& 2\left( \delta_{(\phi<\pi)}-0.5 \right) \in \{ 
%  -1,1\}.
%     \label{sign_function_sphere}
% \end{eqnarray}
Using the Definition-\ref{square_of_an_angle} (for sphere), we get the corresponding square areas as
$$\hat{c}_i=sgn^2(\phi_i^c)A_C^{(0)}(\phi_i^c) \mbox{~~~and~~} \hat{d}_i=A_C^{(0)}(\theta_i^c),$$ respectively, together with $\widehat{(cd)}_i=sgn(\phi_i^c) \sqrt{A_C^{(0)}(\phi_i^c) \cdot A_C^{(0)}(\theta_i^c)}$. Hence, calculate the curved variance and curved co-variance as

\begin{eqnarray}
    \hat{cv}_{\phi}^{(s)}=\frac{1}{n}\sum_{i=1}^n~ \hat{c}_i,\hspace{0.5cm} \hat{cv}_{\theta}^{(s)}=\frac{1}{n}\sum_{i=1}^n~ \hat{d}_i,~\mbox{and~~~}
    \widehat{ccv}_{\phi\theta}^{(s)}&=&\frac{1}{n}\sum_{i=1}^n~ \widehat{(cd)}_i,
\end{eqnarray}

Now, we obtain  the  \textit{curved dispersion matrix}  as 
\begin{eqnarray}
    \Sigma_{(s)}= \begin{bmatrix}
\hat{cv}_{\phi}^{(s)} & \widehat{ccv}_{\phi\theta}^{(s)}  \\
\widehat{ccv}_{\phi\theta}^{(s)} & cv_{\theta}^{(s)}
\end{bmatrix}
%\label{inv_matrix}
\end{eqnarray}
which also converges to its theoretical analog $ \Sigma_{(s)}$ similarly with probability $1$ following the Lemma-\ref{est_cd_matrix_lemma}. Hence, a similar convergence holds for the inverse which is given by
\begin{eqnarray}
    \hspace{0.4cm}\Sigma_{(s)}^{-1}= \frac{1}{\hat{cv}_{\phi}^{(s)} \cdot
\hat{cv}_{\theta}^{(s)}-\widehat{ccv}_{\phi\theta}^{(s)} \cdot \widehat{ccv}_{\phi\theta}^{(s)}}\begin{bmatrix}
\hat{cv}_{\theta}^{(s)} & -\widehat{ccv}_{\phi\theta}^{(s)}  \\
-\widehat{ccv}_{\phi\theta}^{(s)} & \hat{cv}_{\phi}^{(s)}
\end{bmatrix}~\mbox{denoted as} \begin{bmatrix}
\iota \hat{cv}_{\theta}^{(s)} & \iota \widehat{ccv}_{\phi\theta}^{(s)}  \\
\iota \widehat{ccv}_{\phi\theta}^{(s)} & \iota  \hat{cv}_{\phi}^{(s)}
\end{bmatrix} \nonumber\\
\label{inv_matrix sphere}
\end{eqnarray}

Let ${c}_i=sgn^2(\phi_i) A_C^{(0)}(\phi_i)$, and ${d}_i=A_C^{(0)}(\theta_i)$ be the square areas on the surface of a sphere, respectively, together with ${(cd)}_i= sgn(\phi_i)
 \sqrt{A_C^{(0)}(\phi_i) \cdot A_C^{(0)}(\theta_i)}$. We calculate an expression analogous to the quadratic form associated with the \textit{Mahalanobis distance } using the matrix $\Sigma_{(s)}^{-1}$ and the vector 
 $\left( sgn(\phi_i)
 \sqrt{A_C^{(0)}(\phi_i)},  \sqrt{A_C^{(0)}(\theta_i)} \right)^{T}$ to obtain

\begin{eqnarray}
    Q^{(s)}_i&=& \iota \hat{cv}_{\phi}^{(s)} \cdot {c}_i +\iota \hat{cv}_{\theta}^{(s)} \cdot {d}_i +2 \cdot\iota \widehat{ccv}_{\phi\theta}^{(s)}     \cdot{(cd)}_i, 
\label{qform_nc_sphere}
\end{eqnarray}
$  \mbox{~for~} i=1,2,\ldots,n$. Now, consider the estimated variance of the sequence, $\{ Q^{(s)}\}$ as
$$S_{Q^{(s)}}^2{=}\frac{1}{n-1} \sum_{i=1}^{n}\left(Q^{(s)}_i-\Bar{Q}^{(s)}\right)^2,$$ where $n~\bar{Q}^{s}=\displaystyle \sum_{i=1}^{n}Q^{(s)}_i,$ and we define a  CUSUM process

\begin{equation}
       Z(k)=\frac{1}{\sqrt{ n~S_{Q^{(s)}}^2}} \left[ \sum_{i=1}^{k} Q{_i}^{(s)}-k\Bar{Q}^{(s)}  \right]  \mbox{~~for all~~} k=1,\ldots,n 
       \label{sphere cusum process}
\end{equation}
to obtain the test statistic
\begin{equation}
    \mathcal{S}_n=\displaystyle \max_{1 \leq k < n}|Z(k)|   .
    \label{sphere_test_statistic}
\end{equation}
Following a similar argument as in Equation-\ref{concentration asym}, Equation-\ref{bbridge}, and Lemma-\ref{bbridge_lemma_torus}, it can be shown that \( \mathcal{S}_n \overset{d}{\to} K_\infty \).  
As a result, we reject the null hypothesis \( H_{0s} \) if \( \mathcal{S}_n > k_{\alpha} \), where \( k_{\alpha} \) denotes the upper \( \alpha \)-quantile of the distribution \( K_\infty \). Furthermore, analogous to Corollary-\ref{consistency_corr_torus}, it can be established that this test is also consistent.

\newpage
\section{Numerical Studies}
\label{simulation}
A comprehensive simulation study has been conducted to numerically evaluate the performances of the proposed test statistics identifying changepoint in the mean direction for toroidal and spherical distributions. The simulation for the well-known toroidal and spherical distributions.

\subsection{Measures of performance}
To measure the performance of the proposed changepoint detection method in the presence of a single changepoint, we conducted simulations under an alternative hypothesis framework. 
The power of the test statistic was evaluated across a grid of shift magnitudes in the two angular dimensions, resulting in a power surface. This surface represents the proportion of successful detections over repeated simulations for each combination of shift values. To provide an interpretable summary of the results, we also present contour plots of the power surface, where each contour level indicates constant power across the shift parameter space.

Beyond estimating and testing for the changepoint $k^*$, we also aim to provide a confidence interval that reflects uncertainty in the estimate $\hat{k}^*$. Since deriving the exact or even asymptotic distribution of $\hat{k}^*$ is complicated, we rely on a permutation-based strategy.
Rather than permuting the entire dataset as under the null hypothesis, we perform separate permutations on observations up to the estimated location of the changepoint and after the estimated changepoint $\hat{k}^*$. This is justified because, under the alternative hypothesis, observations within each segment (pre- and post-changepoint) are identically distributed.
Using this strategy, we generate $R$ permuted datasets, and from each one, we compute a new changepoint estimate $\hat{k^*}_1^*, \ldots, \hat{k^*}_R^*$. Let $\hat{k^*}^*_{(\alpha/2)}$ and $\hat{k^*}^*_{(1-\alpha/2)}$ denote the empirical $\alpha/2$ and $1 - \alpha/2$ quantiles of these $R$ estimates. Then a two-sided $100(1 - \alpha)\%$ confidence interval for the true changepoint location $k^*$ is given by:
$$
\left( 2\hat{k}^* - \hat{k^*}^*_{(1-\alpha/2)}, \; 2\hat{k}^* - \hat{k^*}^*_{(\alpha/2)} \right).
$$
This construction centers the interval around $\hat{k}^*$, adjusting for the variability seen across the permuted samples, and captures the uncertainty without requiring strong distributional assumptions. Construction of confidence intervals in this way is well known in the literature, for example \cite[see][]{drikvandi2025distribution}.

Now, for the multiple changepoint scenario, let $\mathcal{C}$  and $\hat{\mathcal{C}}$ be the set of true and estimated changepoints, respectively, or in other words, it is the set of corresponding segment boundaries, where each segment has a different mean direction. Then the total number of segmentations are $|\mathcal{C}|-1$, $|\hat{\mathcal{C}}|-1$ , respectively.
Now, to evaluate segmentation accuracy, we primarily adopt the adjusted Rand index (ARI), as introduced by \cite{hubert1985comparing}. This widely recognized metric compares clustering results to assess how closely the estimated segmentation matches the true segmentation.
We interpret an adjusted Rand index exceeding 0.95 as evidence of near-perfect segmentation. An ARI around 0.9 indicates good segmentation quality, although minor oversegmentation or omission of nearby changepoints might still be present. Conversely, ARI values falling below 0.8 suggest unreliable segmentation performance and reduced practical relevance. In this study, segmentation results are presented along with their ARI scores in Table~4 in the Appendix.

In addition to the ARI, we compute the Hausdorff distance to quantify the estimated changepoint location. Considering any two segmentations $\mathcal{C}$ and $\mathcal{C}'$, the  distance $d(\mathcal{C}, \mathcal{C}')$ is defined as:
\[
d(\mathcal{C}, \mathcal{C}') = \frac{1}{n} \max_{c \in \mathcal{C}} \min_{c' \in \mathcal{C}'} |c - c'|.
\]
 Then $d(\mathcal{C}, \hat{\mathcal{C}})$ represents the maximum distance from each true changepoint in $\mathcal{C}$ to its closest corresponding changepoint in $\hat{\mathcal{C}}$. This measure penalizes missed detections (undersegmentation). Similarly, $d(\hat{\mathcal{C}}, \mathcal{C})$ penalizes oversegmentation by evaluating the proximity of estimated changepoints to the true ones. The Hausdorff distance between $d(\mathcal{C}, \hat{\mathcal{C}})$ is then defined as:
\[
d_H(\mathcal{C}, \hat{\mathcal{C}}) = \max\{ d(\mathcal{C}, \hat{\mathcal{C}}), d(\hat{\mathcal{C}}, \mathcal{C}) \}.
\]
 The use of ARI and Hausdorff distancein nonparametric changepoint detection is rich in literature, for example, see \cite{londschien2023random}.

The effectiveness of change point estimation methods is influenced by the interplay between the sample size, the magnitude of distributional shifts between segments, and the number of change points. As shown in Figure-\ref{sin_model_power_plot} and Table-\ref{Ari_hd_tble}, for the single changepoint scenario, the power reaches 1, and for the multiple changepoint scenario, it attains average adjusted Rand indices (ARIs) exceeding 0.975 as the sample size increases. This suggests that these datasets have a high signal-to-noise ratio, making change point detection relatively straightforward. We have presented two simulation studies, one for a single another for a multiple changepoint scenario, to study the effect of signal length on the power (for a single changepoint) and ARI and Hausedroff distances (for multiple).

\subsection{Toroidal distributions}
\label{simulation_torus}
\begin{figure}[h!]
    \centering
        {\includegraphics[trim= 20 20 20 20, clip, width=0.6\textwidth, height=0.35\textwidth]{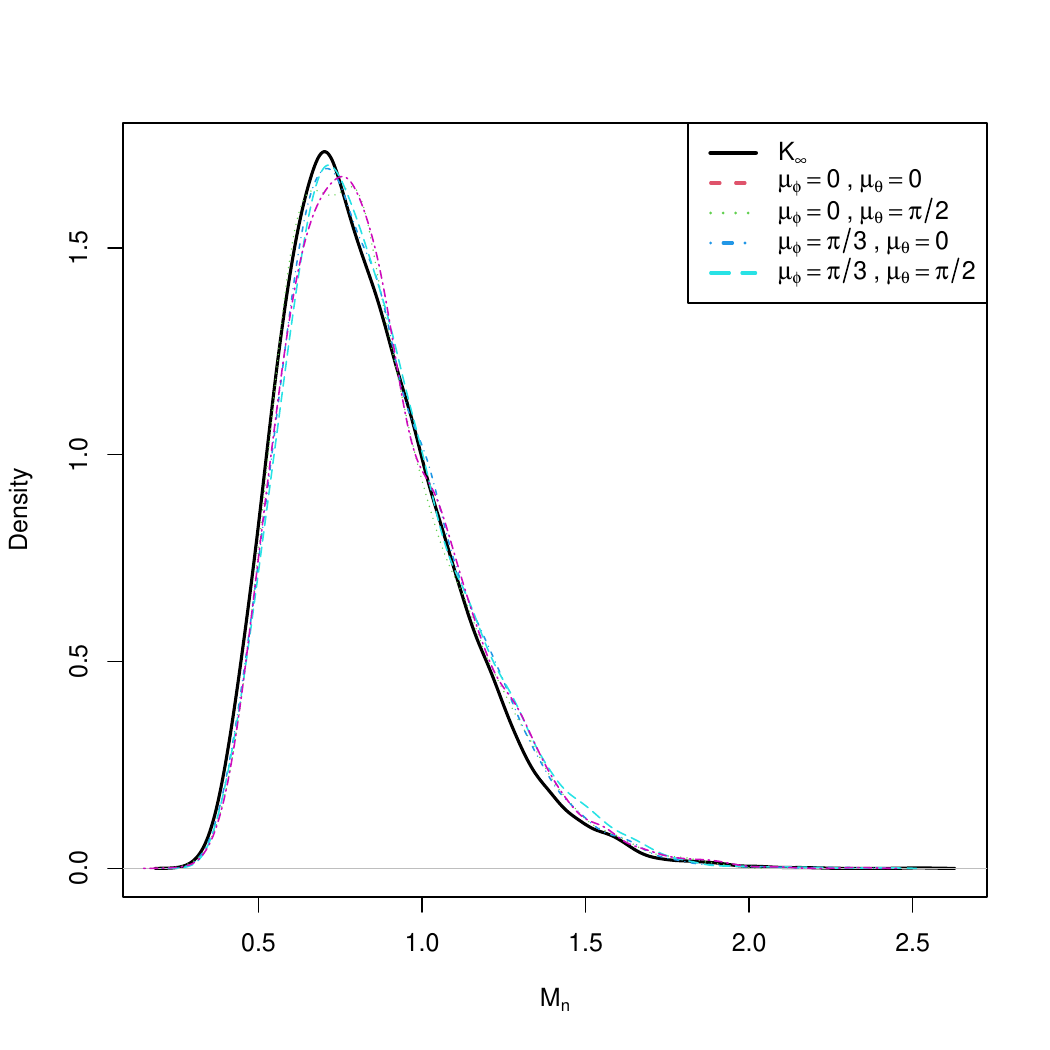}}

       % \subfloat[]{%
       %  {\includegraphics[trim= 20 20 20 20, clip, width=0.5\textwidth, height=0.35\textwidth]{ torus_null_density_rvmcos_mean.pdf }}}
       %    \subfloat[]{%
       %  {\includegraphics[trim= 20 20 20 20, clip, width=0.5\textwidth, height=0.35\textwidth]{ sphere_null_density_kent_pos_mean.pdf}}} \hspace{5pt}

       %    \subfloat[von cos]{%
       %  {\includegraphics[trim= 20 20 20 20, clip, width=0.5\textwidth, height=0.35\textwidth]{torus_null_density_voncos_mean.pdf}}}
       %    \subfloat[]{%
       %  {\includegraphics[trim= 20 20 20 20, clip, width=0.5\textwidth, height=0.35\textwidth]{sphere_null_density_kent_mean.pdf}}} 
        \caption{The density plots of the test statistic, $\mathcal{M}_n$ under $H_{0t}$ with a sample of size $n=1000$ from von Mises sine model.}

    % \caption{(a)is the density plots of the test statistic, $\mathcal{M}_n$ under $H_{0t}$ with a sample of size $n=1000$ from von Mises sine model. (b) is the density plots of the test statistic, $\mathcal{S}_n$ under $H_{0s}$ with a sample of size $n=1000$ from  Fisher distribution.}

    % \caption{(a), (c), and (e) are the density plots of the test statistic, $\mathcal{M}_n$ under $H_{0t}$ with a sample of size $n=1000$ from von Mises sine model, von Mises cos model, and a new distribution proposed by \cite{biswas2024changepoint}, respectively. (b), (d), and (f) are the density plots of the test statistic, $\mathcal{S}_n$ under $H_{0s}$ with a sample of size $n=1000$ from  Fisher distribution, Kent distribution with positive ovalness parameter, and Kent distribution with negative ovalness parameter, respectively.}
    \label{torus_null_density_plot}
\end{figure}

Here, we have considered one of the widely used toroidal distributions,
the bivariate  von Mises sine-model, due to \cite{singh2002probabilistic}, with the probability density function 
\begin{equation}
    f_{\mbox{vMsine}}(\phi,\theta)=C_{\mbox{vMsine}} \exp{ \{\kappa_1\cos(\phi-\mu_{\phi})+\kappa_2\cos(\theta-\mu_{\theta})+\kappa_3 \sin(\phi-\mu_{\phi})\sin(\theta-\mu_{\theta})    \}},
    \label{sine model bv}
\end{equation}
where, $(\mu_{\phi},\mu_{\theta})\in [0, 2\pi)$, $\kappa_1, \kappa_2>0$, $\kappa_3 \in \mathbb{R}$, and $C$, the normalizing constant, is given by 

$$C_{\mbox{vMsine}}=4\pi^2\displaystyle \sum_{m=0}^{\infty}  \binom{2m}{m} \bigg(  \frac{\lambda^2}{4\kappa_1\kappa_2} \bigg)^m I_m(\kappa_1)I_m(\kappa_2),$$  and
$I_m(\kappa)$ denotes the modified Bassel function of the first kind of order $m.$  
%\subsubsection{Performance under null }
\paragraph{Performance under null:}
We have considered a sample size of $n=1000$ and the number of iterations being $10^4$ to study the null distribution. Figure-\ref{torus_null_density_plot} displays a density plot of the distribution of the test statistic, $\mathcal{M}_n$ under $H_{0t}$  with  $\kappa_1=2, \kappa_2=2,$ and $\kappa_3=0$, and different mean direction vectors, $(\mu_\phi, \mu_\theta)=(0,0), (\mu_\phi, \mu_\theta)=(0,\frac{\pi}{2}), (\mu_\phi, \mu_\theta)=(\frac{\pi}{3},0) $, and $(\mu_\phi, \mu_\theta)=(\frac{\pi}{3},\frac{\pi}{2})$.
It is observed that the densities of the test statistics are nearly identical irrespective of the different mean direction vectors. This property is further supported by the results in Table-\ref{table:cut-off_tru_est_mean}, where the empirical cut-off values and empirical sizes remain stable across different dependency parameter settings under the null hypothesis, $H_{0t}$. 

\begin{table}[h!]
\begin{center}
{\renewcommand{\arraystretch}{1}
 \scalebox{.75}{ \begin{tabular}{|c|c|cc|cc|c|}
\cline{1-7}
\textbf{Sample Size} & 
\begin{tabular}{c}
\textbf{Concentration Parameters:} \\
$\kappa_1, \kappa_2$ \\
\textbf{Dependency Parameter:} 
$\kappa_3$
\end{tabular} 
& \multicolumn{2}{c|}{$(\mu_\phi, \mu_\theta)$} 
& \multicolumn{2}{c|}{$(\hat{\mu}_\phi,\hat{\mu}_\theta)$} 
& $K_\infty$ \\ \cline{3-7}
 & & \textbf{0.95th qtl.}   & \textbf{E.S} & \textbf{0.95th qtl.}  &\textbf{E.S}  & \textbf{0.95th qtl.}\\ \cline{1-7}
$n=50$ & 
\begin{tabular}{c}
$\kappa_1 = \kappa_2 = 2$ \\
$\kappa_3 = 0$
\end{tabular}
& 1.2599 & 0.049 & 1.2577 & 0.048 & 1.2537 \\ \cline{1-7}
$n=150$ & 
\begin{tabular}{c}
$\kappa_1 = \kappa_2 = 2$ \\
$\kappa_3 = 1$
\end{tabular}
& 1.2937 & 0.051 & 1.3257 & 0.052 & 1.3120 \\ \cline{1-7}
$n=500$ & 
\begin{tabular}{c}
$\kappa_1 = \kappa_2 = 2$ \\
$\kappa_3 = -1$
\end{tabular}
& 1.3410 & 0.054 & 1.3479 & 0.051 & 1.3391 \\ \cline{1-7}
$n=1000$ & 
\begin{tabular}{c}
$\kappa_1 = \kappa_2 = 2$ \\
$\kappa_3 = 0$
\end{tabular}
& 1.3441 & 0.049 & 1.3391 & 0.053 & 1.3445 \\ \cline{1-7}
\end{tabular}}}
\end{center}
\caption{Cut-off values (0.95th quantile) of the test statistic $\mathcal{M}_n$ under the null hypothesis $H_{0t}$, computed using both the true mean direction $(\mu_\phi, \mu_\theta)$ and its estimate $(\hat{\mu}_\phi,\hat{\mu}_\theta)$. For each case, the corresponding empirical size (E.S) is also reported. Samples are drawn from the von Mises cosine model with varying dependency parameters $\kappa_3$ and fixed concentration parameters $\kappa_1 = \kappa_2 = 2$, across sample sizes $n = 50$, $150$, $500$, and $1000$. The table additionally includes cut-off values obtained from the limiting distribution $K_\infty$ with the grid sizes are same with the sample sizes. (Equation~\ref{bbridge}).}
\label{table:cut-off_tru_est_mean}
\end{table}

\noindent
\textbf{False positive rate:}
To assess the reliability of the proposed changepoint detection method in scenarios with no structural changes, we conducted a Monte Carlo simulation study under the null hypothesis. Specifically, for each chosen sample size $n \in \{50, 250, 500, 1000\}$, we generated 1000 independent realizations of bivariate toroidal data using the wrapped von Mises sine distribution with parameters $\kappa_1 = \kappa_2 = 4$, $\kappa_3 = 0$, and mean directions fixed at $(0, 0)$. Since these data contain no true changepoints, any detection is considered a false positive. 
The recursive binary segmentation (RBS) algorithm was applied to each realization, and the false positive rate (FPR) was estimated as the proportion of simulations in which at least one changepoint was falsely detected. A detection threshold of 1.314 was used, corresponding to a nominal significance level of $\alpha = 0.05$. The simulation results, summarized in Table-\ref{tab:fpr}, demonstrate that the proposed method maintains control over the false discovery rate. For moderate to large sample sizes, the estimated FPR remained close to or slightly below the 5\% nominal level, confirming the conservativeness and reliability of the proposed test statistic under the null condition.
\begin{table}[h!]
\centering
\begin{tabular}{|c|c|}
\hline
\textbf{Sample Size ($n$)} & \textbf{Estimated FPR (\%)} \\
\hline
50   & 3.23 \\
250  & 4.20 \\
500  & 5.55 \\
1000 & 5.50 \\
\hline
\end{tabular}
\caption{Estimated False Positive Rates (FPR) under the Null hypothesis, $H_{0t}$ for Different Sample Sizes}
\label{tab:fpr}
\end{table}

To evaluate the performance of the proposed test statistic $\mathcal{M}_n$, in Table-\ref{table:cut-off_tru_est_mean}, we compute its empirical 0.95th quantile cut-off values under the null hypothesis $H_{0t}$ using both the true mean direction $(\mu_\phi, \mu_\theta)$ and its estimate $(\hat{\mu}_\phi, \hat{\mu}_\theta)$. In addition to the cut-off values, the corresponding empirical sizes (E.S) are reported to assess size accuracy. Simulated samples of sizes $n = 50$, $150$, $500$, and $1000$ are generated from the von Mises sine model with concentration parameters $\kappa_1 = \kappa_2 = 2$ and varying dependency parameters $\kappa_3 \in \{0, 1, -1\}$. Each configuration is replicated 1000 times for reliable estimation. For reference, the table also includes cut-off values obtained from the limiting distribution $K_\infty$ (Equation~\ref{bbridge}).

\begin{figure}[h!]
\centering
\subfloat[]{%
{\includegraphics[trim= 0 0 0 0, clip,width=0.3\textwidth,height=0.3\textwidth]{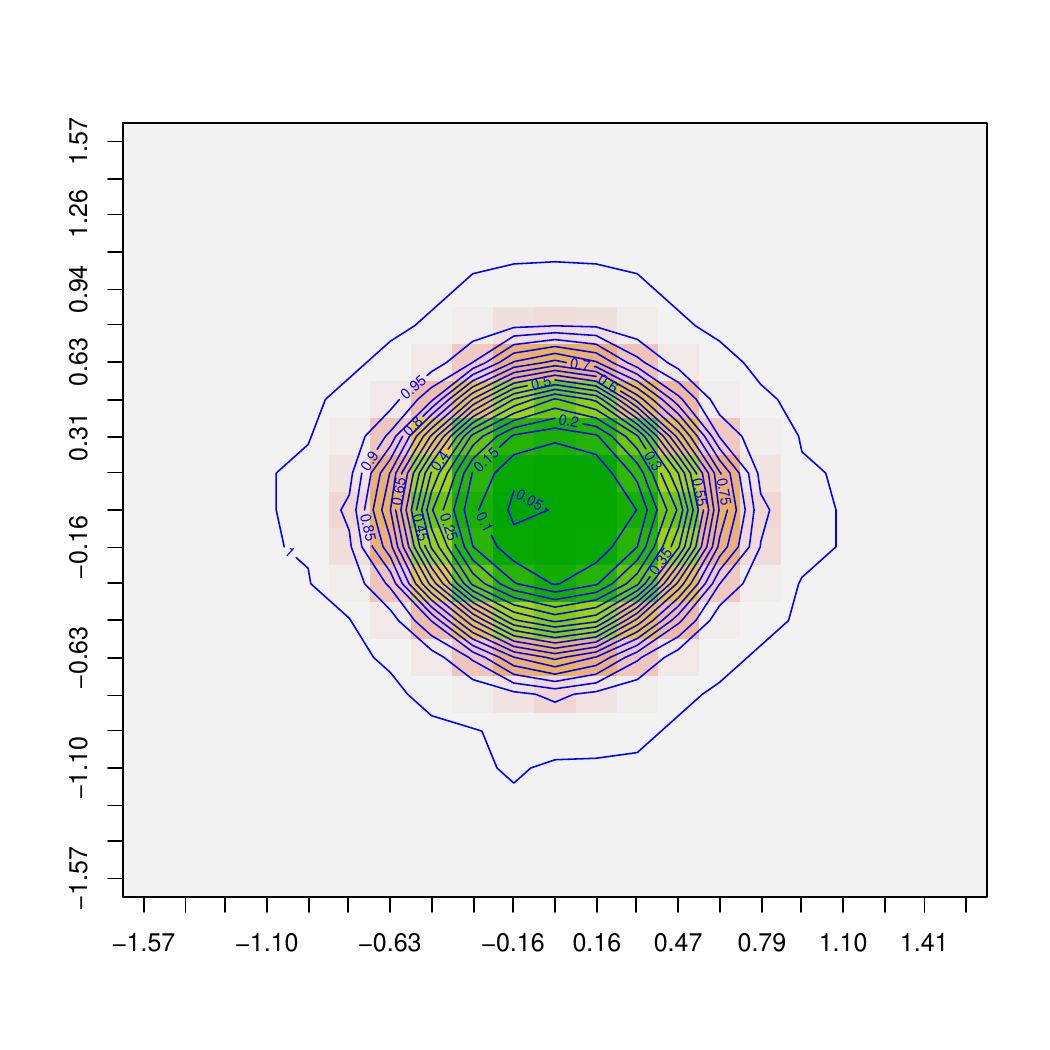}}} 
\subfloat[]{%
{\includegraphics[trim= 0 0 0 0, clip,width=0.3\textwidth,height=0.3\textwidth]{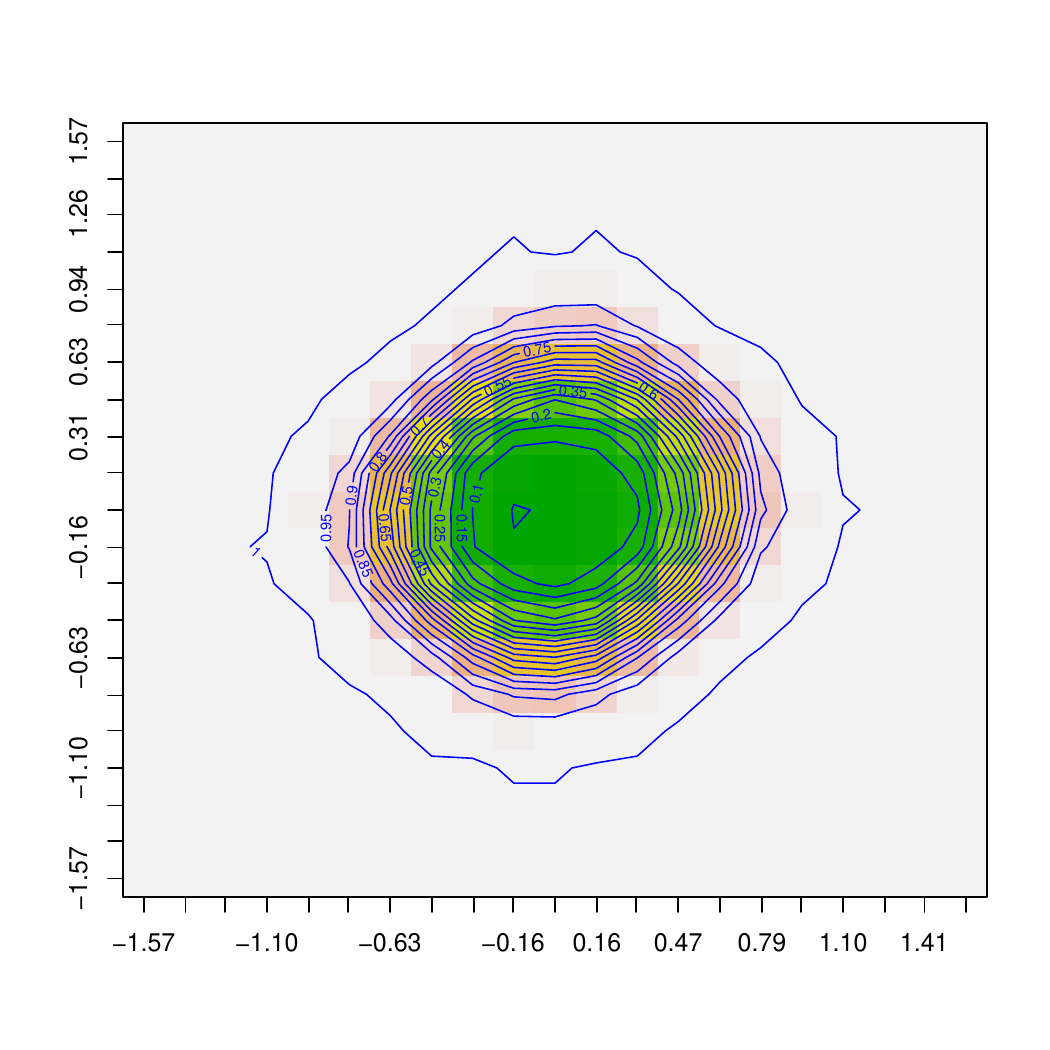}}} 
\subfloat[]{%
{\includegraphics[trim= 0 0 0 0, clip,width=0.3\textwidth,height=0.3\textwidth]{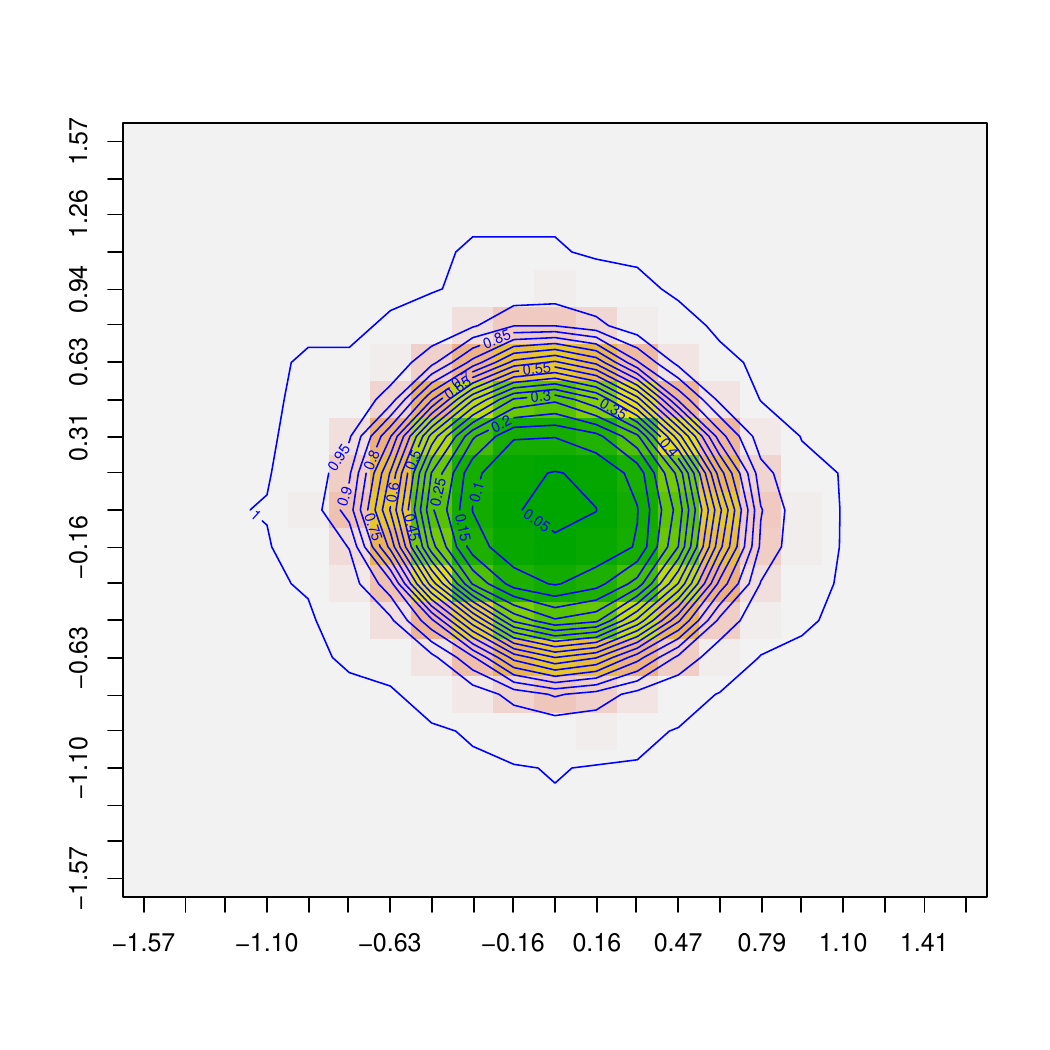}}} \hspace{5pt}
\subfloat[ ]{%
{\includegraphics[trim= 20 20 20 20, clip,width=0.3\textwidth,height=0.3\textwidth]{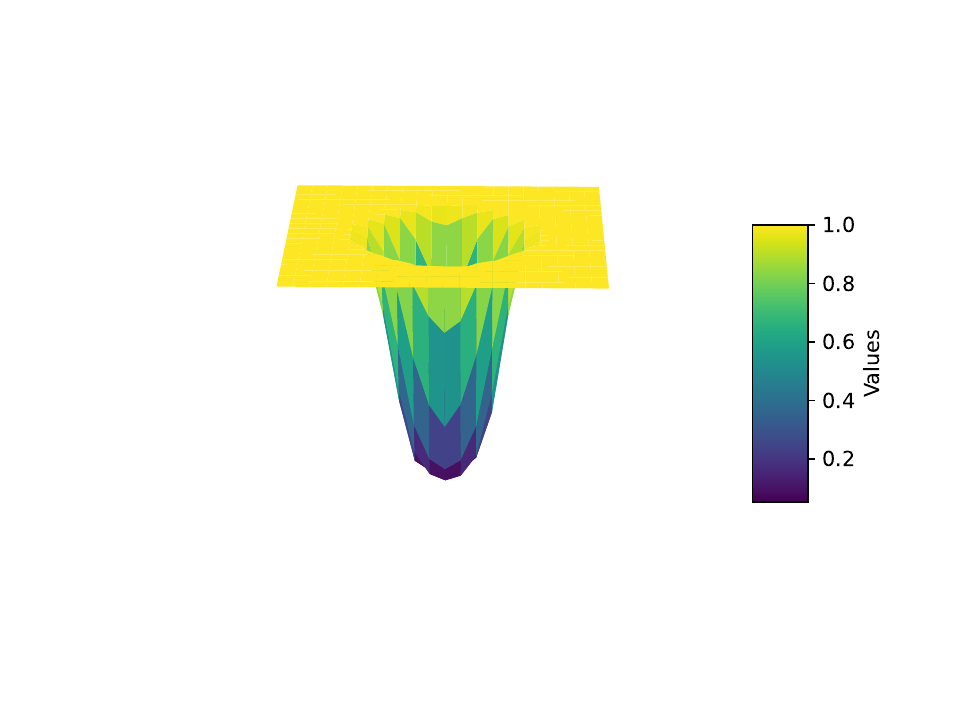}}}
\subfloat[ ]{%
{\includegraphics[trim= 20 20 20 20, clip,width=0.3\textwidth,height=0.3\textwidth]{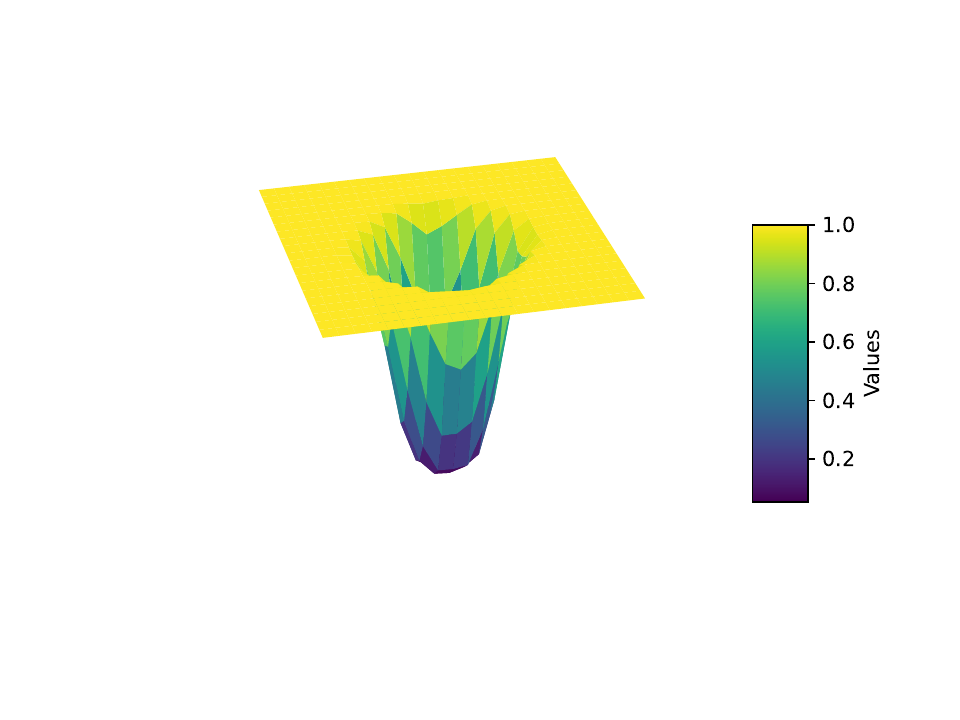}}}
\subfloat[ ]{%
{\includegraphics[trim= 20 20 20 20, clip,width=0.3\textwidth,height=0.3\textwidth]{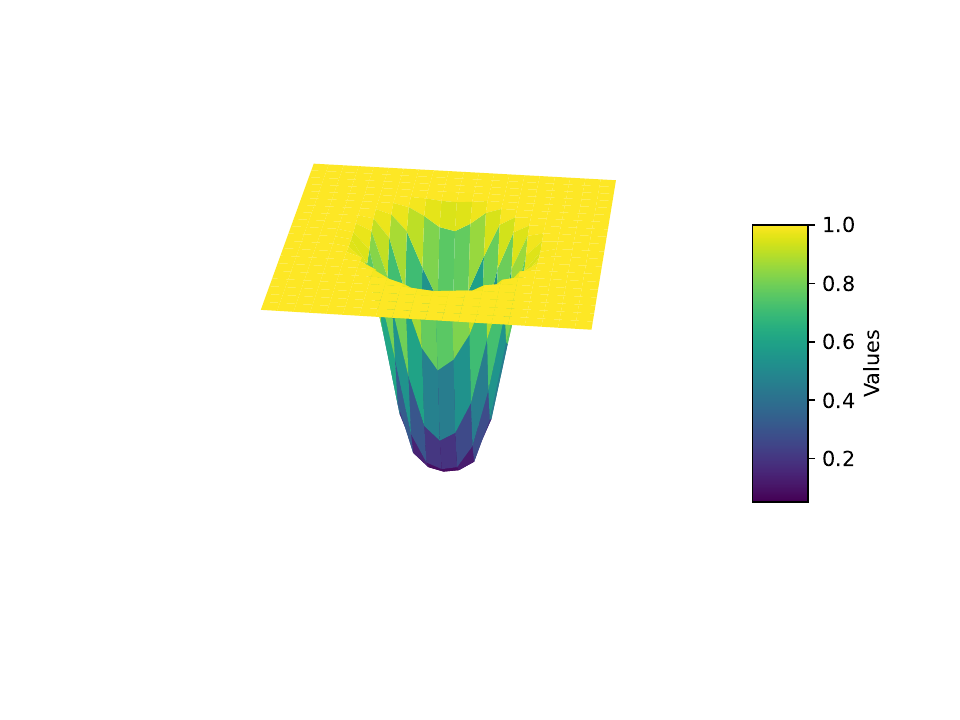}}}

\caption{(a),(b), (c) are the contour plots; (d), (e), (f) are the corresponding surface plots of power under $H_{1t}$ when the location of the changepoint is considered at $k^{*}=\frac{n}{2},$ for von mises sine model with zero association, positive association, and negative association, respectively.} 
\label{sin_model_power_plot}
\end{figure}

%\subsubsection{Performance under alternative}
\paragraph{Performance under $ \boldsymbol{H_{1t}}$:}
The performance of the proposed changepoint detection method is measured under single and multiple changepoint scenarios.\\
\textbf{Single changepoint:}
In the following we discuss the performance of the under alternative hypothesis for the single changepoint scenario.

\noindent
\textbf{Power surface and contour:}
To generate the power surface and the corresponding contour, the location of the changepoint is considered at $k^{*}=\frac{n}{2},$ and the mean direction vector before the change is $(\mu_\phi, \mu_\theta)=(0,0).$  After the change, a shift of $(\delta_\phi,\delta_\theta)$ in the mean direction vector is added to the initial one. Both $\delta_\phi,\delta_\theta$ take $21$ equispaced values in $[-\frac{\pi}{2},\frac{\pi}{2}].$ We performed  $10^4$ iterations to compute the power of the test statistic, $\mathcal{M}_n$ in Equation-\ref{torus_test_statistic} for sample size of $n=500$  at the level of $5\%.$
The plots are reported for three types of dependent data for this model. Since the value $\kappa_3$ decides the dependency between the random angles in this model, we keep $\kappa_1 = \kappa_2 = 2$ throughout and vary $\kappa_3$. 
Figure~\ref{sin_model_power_plot}(a), (b), and (c) depict the contour plots for data generated from the model in Equation-\ref{sine model bv} under three scenarios: independence (\(\kappa_3 = 0\)), right-tilted association (positive dependence, \(\kappa_3 = 2\)), and left-tilted association (negative dependence, \(\kappa_3 = -2\)), respectively. The corresponding surface plots are shown in Figure~\ref{sin_model_power_plot}(d), (e), and (f). These surface and contour plots clearly demonstrate that the power of the test converges to one. 
 In addition, we evaluate the empirical power of $\mathcal{M}_n$ under the alternative hypothesis $H_{1t}$. In Table-\ref{table:sacc_power_diff_loc} the powers are reported for change-point locations at $\frac{n}{3}$, $\frac{n}{2}$, and $\frac{2n}{3}$ for sample sizes $n = 60$, $150$, and $600$. For this simulation the samples are drawn from the von Mises sine model with concentration parameters $\kappa_1 = \kappa_2 = 2.5$ and a non-zero dependency parameter $\kappa_3 = 1$. The mean direction before the change point is set to $(\mu_\phi, \mu_\theta) = (0, 0)$, and a shift of $\left(\frac{\pi}{6}, \frac{\pi}{6}\right)$ is applied after the change point.

 \begin{table}[h!]
\begin{center}
\renewcommand{\arraystretch}{1.5}
\begin{tabular}{|c|c|c|c|}
\hline
\textbf{Sample Size} & Change Point at $\boldsymbol{\frac{n}{3}}$ & Change Point at $\boldsymbol{\frac{n}{2}}$ & Change Point at $\boldsymbol{\frac{2n}{3}}$ \\ 
\hline
$n = 60$  & 0.104 & 0.173 & 0.137 \\ 
$n = 150$ & 0.248 & 0.377 & 0.344 \\ 
$n = 600$ & 0.783 & 0.952 & 0.888 \\ 
\hline
\end{tabular}
\end{center}
\vspace{0.4cm}
\caption{Empirical power of the test statistic $\mathcal{M}_n$  under the alternative hypothesis $H_{1t}$, based on samples drawn from the von Mises cosine model with concentration parameters $\kappa_1 = \kappa_2 = 2.5$ and dependency parameter, $\kappa_3 = 1$. Power values are reported for changepoint locations at $\frac{n}{3}$, $\frac{n}{2}$, and $\frac{2n}{3}$, corresponding to sample sizes $n = 60$, $150$, and $600$. The mean direction before the changepoint is $(\mu_\phi, \mu_\theta) = (0, 0)$, and after the changepoint a shift of $(\delta_\phi, \delta_\theta) = \left(\frac{\pi}{6}, \frac{\pi}{6}\right)$ is applied.}
\label{table:sacc_power_diff_loc}
\end{table}

\noindent
\textbf{Signal strength:}
\begin{figure}[h!]
\centering
\subfloat[]{%
{\includegraphics[width=0.5\textwidth,height=0.5\textwidth]{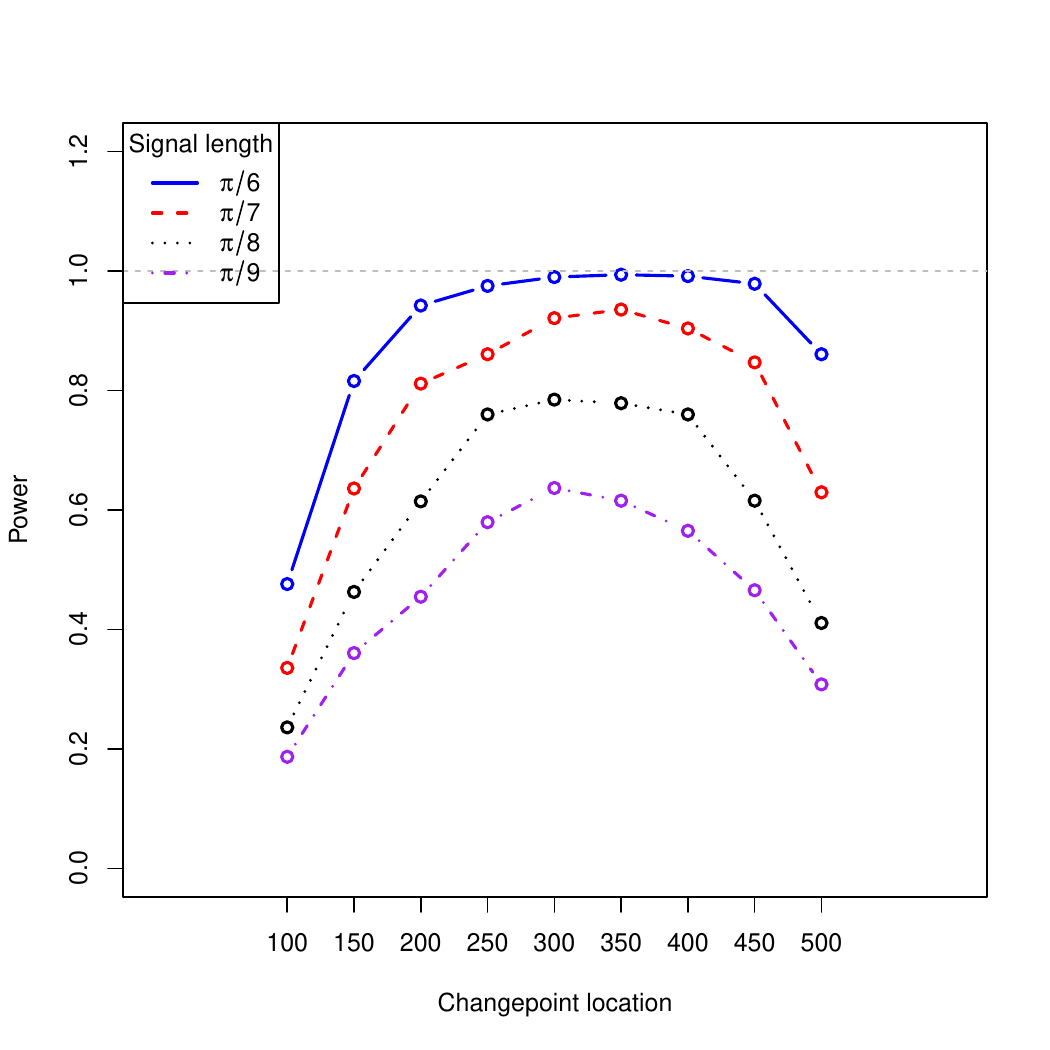}}} 
\subfloat[ ]{%
{\includegraphics[width=0.5\textwidth,height=0.5\textwidth]{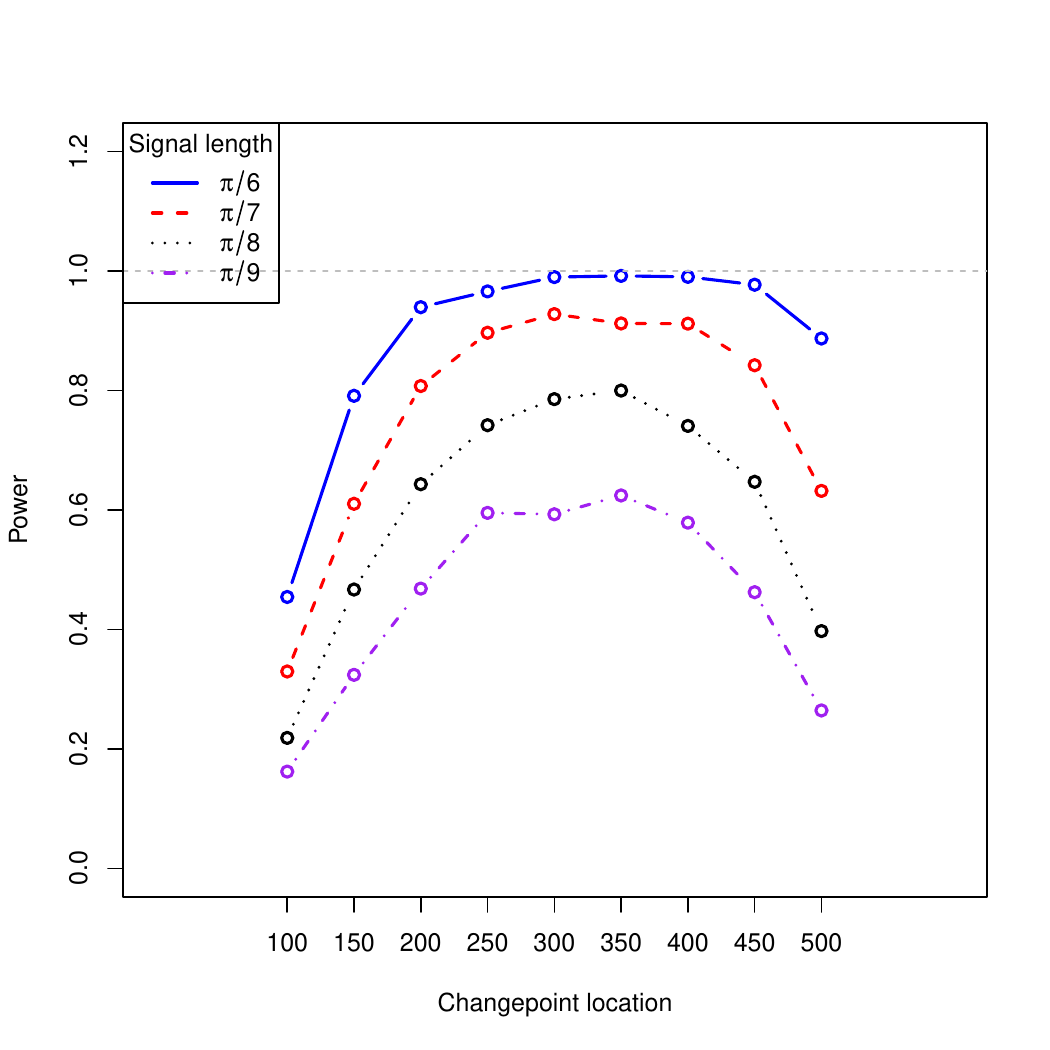}}}\hspace{5pt}
\caption{
Empirical power curves of the proposed test statistic $\mathcal{M}_n$ as a function of signal length for different change point locations under a single change point scenario. Data are generated from the bivariate von Mises sine model with fixed concentration parameters and sample size $n = 600$. The first segment has mean direction $(0, 0)$, and the second segment has a shifted mean direction of $(\delta, \delta)$, where $\delta \in \{\pi/9, \pi/8, \pi/7, \pi/6\}$. Change points are placed at nine different locations ranging from early to late in the sequence: 100, 150, 200, 250, 300, 350, 400, 450, and 500. Panels (a) and (b) correspond to dependency parameters $\kappa_3 = 0$ and $\kappa_3 = -1$, respectively. Power is estimated over 1000 simulations at a nominal level of 0.05. The results show that power increases with signal strength and is higher when the change occurs closer to the center of the sequence.}
\label{signal_length}
\end{figure}
To examine how signal strength and change point location affect the performance of our method, we simulate data from a two-segment model with a single change point. A total of 600 observations are generated from the von Mises sine model with fixed concentration parameters. Observations in the first segment have mean direction $(0, 0)$, while those in the second segment share the same concentration but have mean direction shifted by $(\delta, \delta)$, where $\delta \in \{\pi/9, \pi/8, \pi/7, \pi/6\}$ denotes the signal length. The change point is placed at varying locations within the sequence: 100, 150, 200, 250, 300, 350, 400, 450, and 500. For each setting, power is computed over 1000 simulations as the proportion of instances in which the proposed test correctly detects a change at the nominal level of 0.05.
Figure-\ref{signal_length}(a) and (b) display the power curves as functions of signal length for different change point locations, under dependency parameter values $\kappa_3 = 0$ and $\kappa_3 = -1$, respectively. A similar trend is observed when $\kappa_3 = 1$ (not shown). As expected, the power increases with signal length across all change point locations. For example, when the change occurs near the center of the sequence, the test achieves higher power under a shift of $\pi/6$ compared to a shift of $\pi/9$. This pattern holds across all change point positions.
However, the position of the change point also has a notable effect on detection performance. Specifically, change points located near the sequence boundaries result in lower power, particularly when the signal strength is moderate. In such cases, the shorter segment provides insufficient information for stable estimation of the change point. These findings are consistent with theoretical insights in change point analysis: detection becomes more difficult when the signal is weak and the change point is positioned asymmetrically, due to increased variability in parameter estimation.

\noindent
\textbf{Confidence interval:}
We next assessed inference for a single changepoint under the bivariate von~Mises sine model for varying sample sizes $n \in \{60,150,300,600\}$. For each run, observations up to the changepoint were generated with mean direction $(0,0)$, and those after the changepoint with mean direction $(\delta,\delta)$, where $\delta = \pi/3$ was fixed. Concentration parameters were set to $\kappa_1 = \kappa_2 = 4$ and $\kappa_3 = 0$. Three changepoint locations $n/3$, $n/2$, and $2n/3$-were analyzed separately. For each setting, 1,000 Monte Carlo replications were conducted. Changepoints were estimated using the recursive detection procedure with a threshold of 1.314087. For each estimated changepoint, we constructed a permutation-based confidence interval by permuting observations within each segment (pre- and post-estimate), recalculating the test statistic 100 times, and computing symmetric normal-approximation intervals based on the empirical distribution of argmax locations.
Table-\ref{singal_cp_CI} reports the median and mode confidence intervals across settings. While the absolute interval widths (in index units) increase slightly with $n$, the relative widths (as a percentage of the sequence length) decrease, confirming improved precision at larger sample sizes. For example, at location $n/3$, the median interval for $n=60$ is $[16,27]$ (of width $=11$, i.e., $11/60\approx 18.3\%$), whereas for $n=600$ it is $[192,209]$ (of width $=17$, i.e., $17/600\approx 2.8\%$).

\noindent
Similarly, for central changepoints at $n/2$, the interval for $n=150$ is $[70,80]$ (of width $=10$, $10/150\approx 6.7\%$) and for $n=600$ it is $[294,306]$ (of width $=12$, $12/600=2.0\%$). A comparable reduction is seen at $2n/3$: for $n=150$, $[94,105]$ (of width $=11$, $11/150\approx 7.3\%$) versus $[392,407]$ for $n=600$ (of width $=15$, $15/600=2.5\%$). Across all locations, intervals are tightest (in relative terms) when the changepoint is central, reflecting the informational advantage of balanced segments; boundary‑proximal changepoints yield wider intervals due to shorter segment lengths. These trends align with the power results and demonstrate that, although absolute widths can grow with $n$, the relative uncertainty shrinks markedly as sample size increases.

\begin{table}[]
\centering
\label{tab:ci_single_grouped}
\renewcommand{\arraystretch}{1.82}
\scalebox{.76}{\begin{tabular}{c|cc|cc|cc}
\toprule
\multirow{2}{*}{\textbf{Sample Size ($n$)}} &
\multicolumn{2}{c|}{\textbf{$n/3=(20,50,100,200)$}} &
\multicolumn{2}{c|}{\textbf{$n/2=(30,75,150,300)$}} &
\multicolumn{2}{c}{\textbf{$2n/3=(40,100,200,400)$}} \\
\cline{2-7}
 & \textbf{Median CI} & \textbf{Mode CI} & \textbf{Median CI} & \textbf{Mode CI} & \textbf{Median CI} & \textbf{Mode CI} \\
\midrule
60 & [16, 27] & [16, 26] & [26, 34] & [27, 34] & [35, 44] & [36, 44] \\
150 & [44, 58] & [44, 57] & [70, 80] & [70, 80] & [94, 105] & [94, 106] \\
300 & [93, 109] & [93, 109] & [145, 155] & [145, 155] & [193, 206] & [193, 205] \\
600 & [192, 209] & [191, 209] & [294, 306] & [295, 306] & [392, 407] & [393, 406] \\
\bottomrule
\end{tabular}}
\caption{
Permutation-based confidence intervals for single changepoint detection under the bivariate von~Mises sine model. Each cell reports the interval as [Lower, Upper], with separate sub-columns for median and mode values of the bounds, based on 1,000 Monte Carlo replications.
}
\label{singal_cp_CI}
\end{table}

In the following we discuss the performance under the alternative hypothesis, $H_{1t}$ for the multiple changepoint scenario.\\
\textbf{Multiple Changepoints:}
To evaluate the performance and robustness of the proposed changepoint detection method under multiple changepoint scenarios, we performed simulations across varying sample sizes: $n \in \{60, 150, 300, 600\}$. Each dataset was synthetically generated as bivariate angular data on the torus using the wrapped von Mises sine distribution via the \texttt{rvmsin} function in the \texttt{BAMBI} package. Every sequence contained three true changepoints located approximately at $n/3$, $n/2$, and $2n/3$. These changepoints induced abrupt shifts in the mean directions of the distribution. Specifically, the data were segmented into four homogeneous regions with centers at $(0,0)$, $(\pi, \pi)$, $(2\pi, 2\pi)$, and back to $(\pi, \pi)$, implemented via angular shifts of $0$, $2\pi/2$, $4\pi/2$, and $2\pi/2$ radians respectively. The concentration parameters were fixed at $\kappa_1 = \kappa_2 = 4$ and $\kappa_3 = 0$ to ensure moderate dispersion and independence between the two angular components.
For each sample size, 1000 independent realizations were generated. The recursive binary segmentation (RBS) algorithm was applied to detect changepoints. Performance metrics including the average number of detected changepoints, the adjusted Rand index (ARI), and the median Hausdorff distance, were computed. Table-\ref{Ari_hd_tble} summarizes these results.
As the sample size increased, the detection performance improved substantially. At $n = 600$, the method accurately identified nearly all changepoints, yielding an average of 3.48 detected changepoints (SE = 0.02), an ARI of 0.9803 (SE = 0.0014), and a median Hausdorff distance of only 0.2\%. Even at moderate sample sizes ($n = 150$ and $n = 300$), the method maintained high ARI scores ($> 0.96$) and small localization errors. For the smallest sample size ($n = 60$), performance slightly degraded, but the method remained statistically informative with an ARI of 0.8752 and a median Hausdorff distance of 3.3\%. These findings confirm the method's effectiveness in accurately identifying multiple changepoints in toroidal sequences across a range of sample sizes.

\begin{table}[h!]
\centering
\label{tab:performance_summary}
\renewcommand{\arraystretch}{1.8}
\scalebox{.68}{\begin{tabular}{|c|c|c|c|}
\hline
\textbf{Sample Size ($n$)} & \textbf{Avg. \# of CPs (SE)} & \textbf{Adjusted Rand Index (SE)} & \textbf{Median Hausdorff Distance (\%)} \\
\hline
60   & 2.72 (0.02) & 0.8752 (0.0053) & 3.3 (3.3) \\
150  & 3.33 (0.02) & 0.9696 (0.0014) & 0.7 (0.7) \\
300  & 3.45 (0.02) & 0.9744 (0.0015) & 0.3 (0.3) \\
600  & 3.48 (0.02) & 0.9803 (0.0014) & 0.2 (0.2) \\
\hline
\end{tabular}}
\caption{
Performance summary of the proposed changepoint detection method across varying sample sizes. 
Each result is averaged over 1000 independent simulations. The table reports the average number of detected changepoints (with standard errors), the adjusted Rand index (ARI) with standard errors, and the median Hausdorff distance (with median absolute deviations in parentheses), all measuring segmentation accuracy and localization precision.
}
\label{Ari_hd_tble}
\end{table}

\begin{figure}[h!]
\centering
\subfloat[ ]{%
{\includegraphics[trim= 0 0 0 0, clip, width=0.35\textwidth,height=0.35\textwidth]{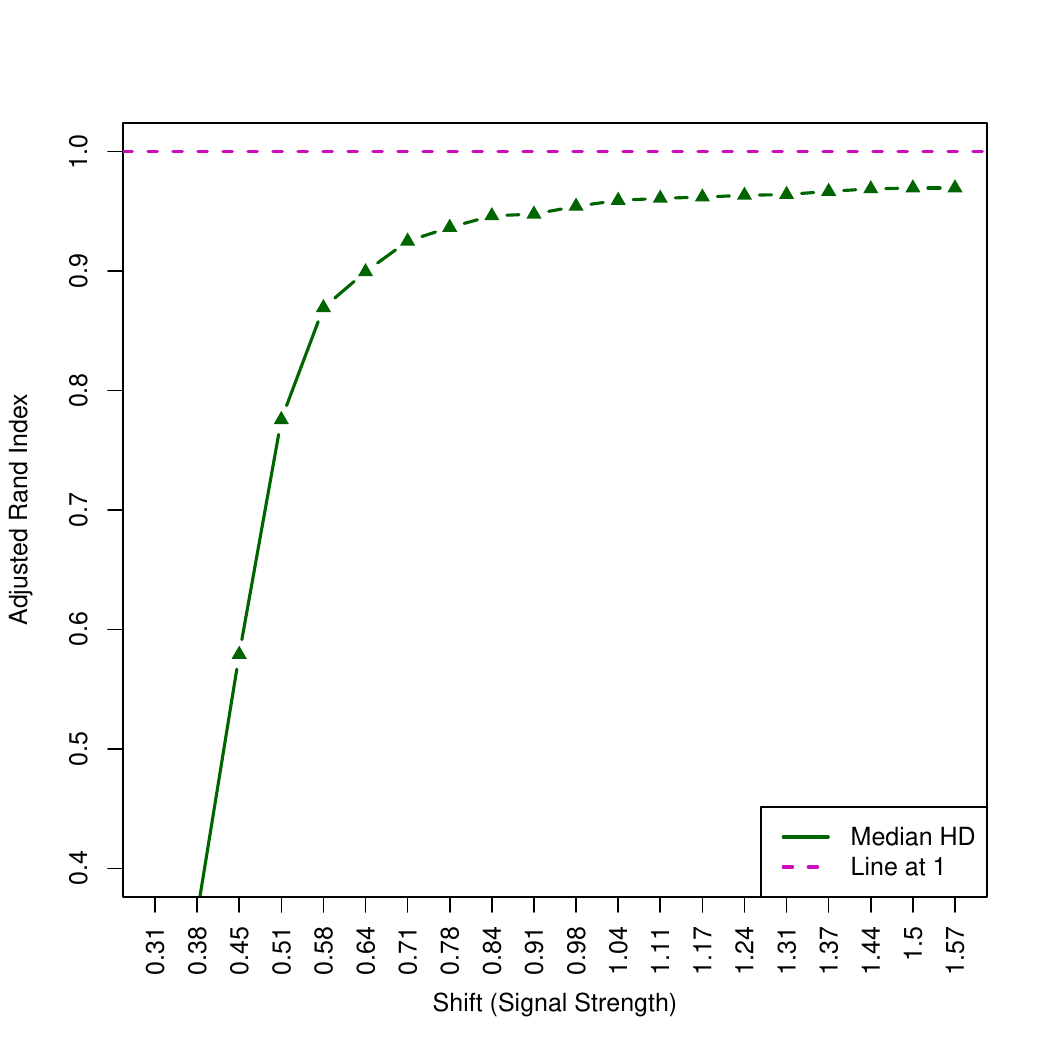}}}
\subfloat[]{%
{\includegraphics[width=0.35\textwidth,height=0.35\textwidth]{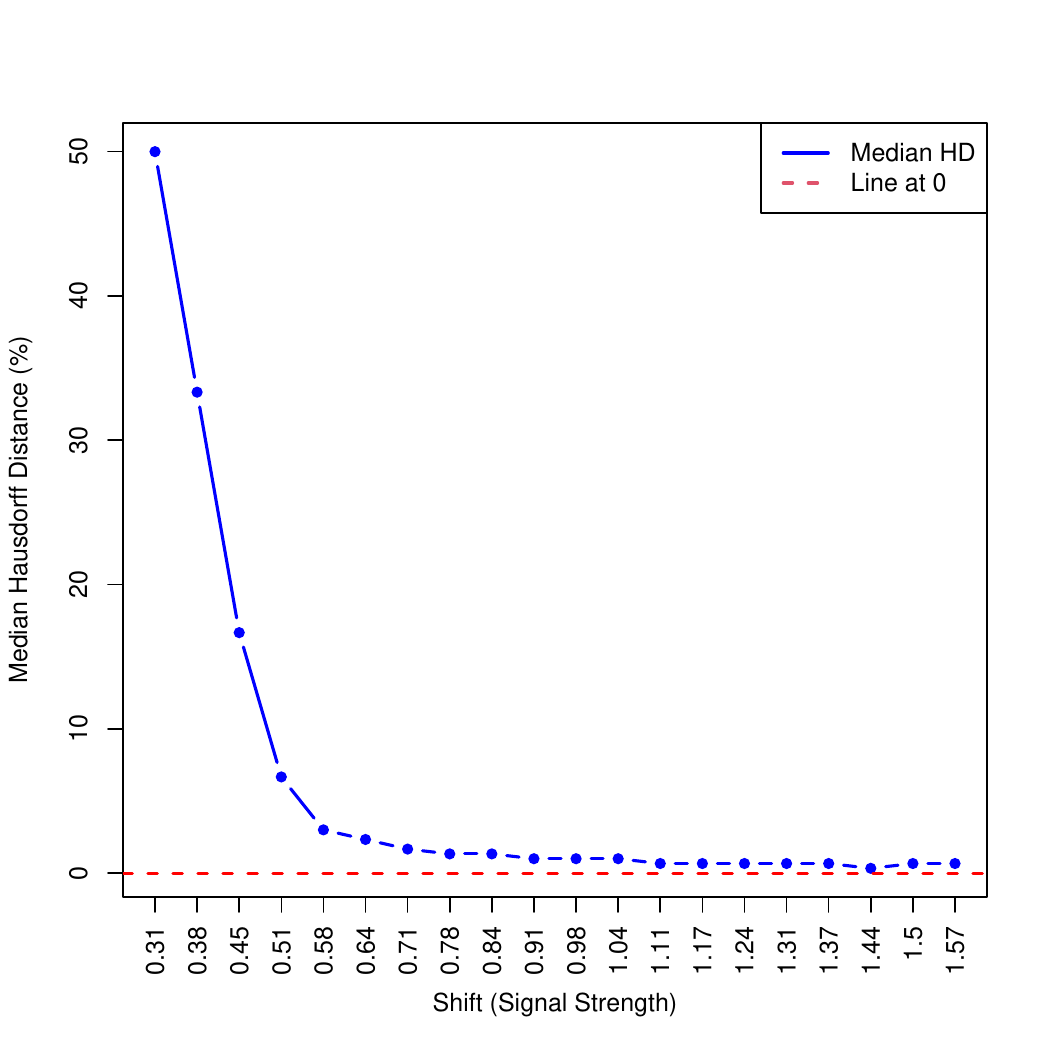}}}
\caption{
Performance of the proposed changepoint detection method as a function of signal strength $\delta \in [\pi/10, \pi/2]$ in the multi-changepoint toroidal setting. The left panel shows the median symmetric Hausdorff distance (as a percentage of sequence length) between the estimated and true changepoint sets. The right panel displays the average Adjusted Rand Index (ARI) comparing true and estimated segmentations. Each point is based on 1000 simulations. As the signal strength increases from $\pi/10$ to $\pi/2$.}
\label{signal_length_multi}
\end{figure}
\noindent
\textbf{Signal strength:} To investigate the effect of signal strength on changepoint detection performance, we simulate data from a four-segment model with three true changepoints located at positions $n/3$, $n/2$, and $2n/3$ in a sequence of length $n = 300$. Each segment consists of independent samples from a bivariate von Mises sine model with concentration parameters fixed at $\kappa_1 = \kappa_2 = 4$ and no toroidal dependence ($\kappa_3 = 0$). The mean directions vary across segments to introduce location-based changes: the first segment has mean direction $(0, 0)$, the second segment is shifted to $(1.5\delta, 1.5\delta)$, the third returns to $(0, 0)$, and the fourth shifts to $(\delta, \delta)$, where $\delta$ varies across a grid from $\pi/10$ to $\pi/2$ and represents the signal strength. For each value of $\delta$, we simulate 1000 independent datasets and apply the proposed recursive changepoint detection algorithm. Performance is evaluated using the adjusted Rand index (ARI) and the  Hausdorff distance between the estimated and true changepoints. As shown in Figure-\ref{signal_length_multi}(a) and (b), detection performance improves significantly with increasing signal strength. When the shift is weak (e.g., $\delta = 0.31$), the  ARI is low, approximately $0.17$ with a large Hausdorff distance of approxmitely $50\%$. As $\delta$ increases, detection accuracy improves rapidly; for $\delta \geq 0.9$, the method reliably estimated all changepoints, achieving an ARI above $0.95$ and median Hausdorff distances below $1\%$. These results highlight the critical role of signal strength in reliable changepoint estimation for directional data, and demonstrate that the proposed method achieves near-perfect segmentation even under moderate shift magnitudes.\\

The simulations for the null density, power surface, and the corresponding contour can be shown for another well-known toroidal distribution, the
bivariate angular von Mises cosine model due to  \cite{mardia2007protein}  with the probability density function 
\begin{equation}
    f_{\mbox{vMcos}}(\phi,\theta)=C_{\mbox{vMcos}} \exp{ \{\kappa_1\cos(\phi-\mu_{\phi})+\kappa_2\cos(\theta-\mu_{\theta})+\kappa_3 \cos(\phi-\mu_{\phi}-\theta+\mu_{\theta})    \}},
    \label{cosine model bv}
\end{equation}
where, $(\mu_{\phi},\mu_{\theta})\in [0, 2\pi)$, $\kappa_1, \kappa_2>0$, $\kappa_3 \in \mathbb{R}$, and $C$, the normalizing constant, is given by 

$$C_{\mbox{vMcos}}=4\pi^2\displaystyle  \bigg[ I_0(\kappa_1)I_0(\kappa_2)I_0(\kappa_3)+ 2 \displaystyle \sum_{m=0}^{\infty}  I_m(\kappa_1)I_m(\kappa_2)I_m(\kappa_3)\bigg],$$  and
$I_r(\kappa)$ denotes the modified Bassel function of the first kind of order $r.$

\subsection{Spherical distributions}
\label{simulation_sphere}
Here, we have considered the well-known spherical distribution, the bivariate  Fisher distribution on the sphere due to \cite{fisher1953dispersion},  and the probability density function given by
$$
f_{\mbox{vMF}}(\boldsymbol{x;\mu^T },\kappa)= \frac{\kappa}{\sinh{\kappa}}\exp\{\kappa \boldsymbol{\mu^T x} \},
$$
where $\kappa\geq 0,||\boldsymbol{\mu} ||=1$. Now,  with the spherical polar coordinate transformations, $\boldsymbol{ x}=(\cos\theta,\sin\theta\cos\phi,\sin\theta\sin\phi)^T,\boldsymbol{ \mu}=(\cos\alpha,\sin\alpha\cos\beta,\sin\alpha\sin\beta)^T $ the immediate above Fisher density can be written as 
\begin{eqnarray}
    f_{\mbox{vMF}}(\theta, \phi)= \frac{\kappa \sin \theta}{4\pi \sinh{\kappa}}\exp\{\kappa[\cos \theta \cos\alpha+ \sin\theta\sin\alpha \cos(\phi-\beta)   ] \},
\end{eqnarray}
where $ \phi,\beta \in (0,2\pi];~~~ \theta,\alpha\in [0,\pi].$ More details about the spherical distributions can be found in \citeauthor{Mardia_2000}, \citeyearpar[Ch. 9, pp. 168]{Mardia_2000}.

\begin{figure}[h!]
    \centering
        {\includegraphics[trim= 20 20 20 20, clip, width=0.6\textwidth, height=0.35\textwidth]{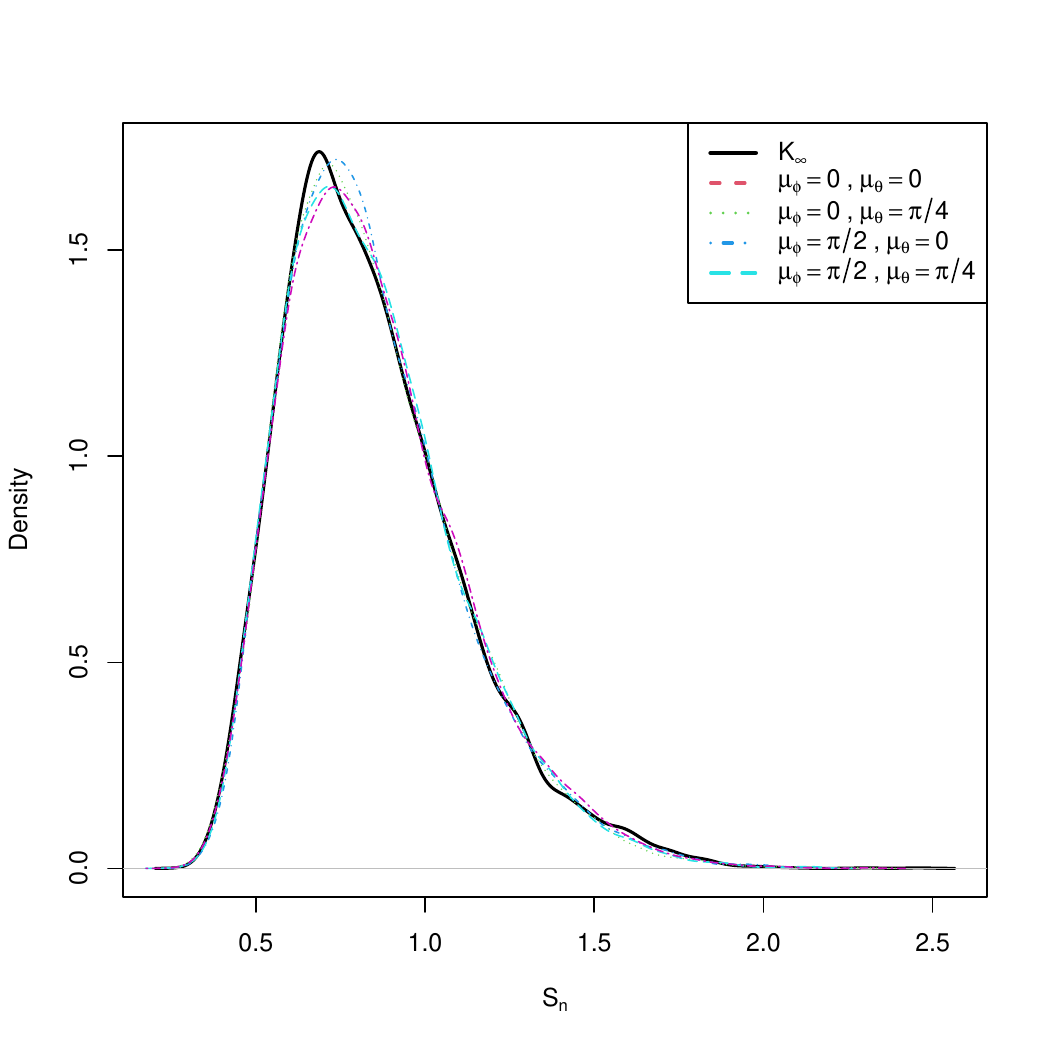}}

       % \subfloat[]{%
       %  {\includegraphics[trim= 20 20 20 20, clip, width=0.5\textwidth, height=0.35\textwidth]{ torus_null_density_rvmcos_mean.pdf }}}
       %    \subfloat[]{%
       %  {\includegraphics[trim= 20 20 20 20, clip, width=0.5\textwidth, height=0.35\textwidth]{ sphere_null_density_kent_pos_mean.pdf}}} \hspace{5pt}

       %    \subfloat[von cos]{%
       %  {\includegraphics[trim= 20 20 20 20, clip, width=0.5\textwidth, height=0.35\textwidth]{torus_null_density_voncos_mean.pdf}}}
       %    \subfloat[]{%
       %  {\includegraphics[trim= 20 20 20 20, clip, width=0.5\textwidth, height=0.35\textwidth]{sphere_null_density_kent_mean.pdf}}} 

\caption{The density plots of the test statistic, $\mathcal{S}_n$ under $H_{0s}$ with a sample of size $n=1000$ from  Fisher distribution.}
    % \caption{(a) is the density plots of the test statistic, $\mathcal{M}_n$ under $H_{0t}$ with a sample of size $n=1000$ from von Mises sine model. (b) is the density plots of the test statistic, $\mathcal{S}_n$ under $H_{0s}$ with a sample of size $n=1000$ from  Fisher distribution.}

    % \caption{(a), (c), and (e) are the density plots of the test statistic, $\mathcal{M}_n$ under $H_{0t}$ with a sample of size $n=1000$ from von Mises sine model, von Mises cos model, and a new distribution proposed by \cite{biswas2024changepoint}, respectively. (b), (d), and (f) are the density plots of the test statistic, $\mathcal{S}_n$ under $H_{0s}$ with a sample of size $n=1000$ from  Fisher distribution, Kent distribution with positive ovalness parameter, and Kent distribution with negative ovalness parameter, respectively.}
    \label{sphere_null_density_plot}
\end{figure}

  Figure-\ref{sphere_null_density_plot} displays a density plot of the distribution of the test statistic, $\mathcal{S}_n$ under $H_{0s}$  with the sample size of $n=1000$ from the bivariate  Fisher distribution with  $\kappa=2$, and different mean direction vectors such as $(\mu_\phi, \mu_\theta)=(0,0), (\mu_\phi, \mu_\theta)=(0,\frac{\pi}{4}), (\mu_\phi, \mu_\theta)=(\frac{\pi}{2},0) $, and $(\mu_\phi, \mu_\theta)=(\frac{\pi}{2},\frac{\pi}{4})$. 
 % Although the Fisher distribution is widely used, some datasets do not fit well with the Fisher distribution and instead seem to originate from distributions with oval-shaped density contours. To address this, \cite{kent1982fisher} proposed a model, with the density detailed in \citeauthor{Mardia_2000}, \citeyearpar[Ch. 9, pp. 176]{Mardia_2000}.  For $\kappa=5$ Figure-\ref{null_density_plot}(d) and \ref{null_density_plot}(f) display the density plots of the distribution of the test statistic, $\mathcal{S}_n$ under $H_{0s}$  with the sample size of $n=1000$ from the independent bivariate  Kent distribution  with  $ \beta=3$, different mean direction vectors such as, $(\mu_\phi, \mu_\theta)=(0,0), (\mu_\phi, \mu_\theta)=(0,\frac{\pi}{6}), (\mu_\phi, \mu_\theta)=(\frac{2\pi}{3},0) $, and $(\mu_\phi, \mu_\theta)=(\frac{2\pi}{3},\frac{\pi}{6})$, and  $ \beta=-3$,  different mean direction vectors such as, $(\mu_\phi, \mu_\theta)=(0,0), (\mu_\phi, \mu_\theta)=(0,\frac{\pi}{3}), (\mu_\phi, \mu_\theta)=(\frac{2\pi}{3},0) $, and $(\mu_\phi, \mu_\theta)=(\frac{2\pi}{3},\frac{\pi}{3})$, respectively.
 In total $10^4$  iterations have been conducted for each of the above specifications. 
It is evident from Figure-\ref{sphere_null_density_plot} that the densities of the distribution of the test statistics are nearly identical irrespective of the different mean direction vectors of Fisher distribution. Moreover, these are close to the density of the limiting distribution of the random variable $K_{\infty}$ (Equation-\ref{bbridge}). 

% The same is observed in Figures \ref{null_density_plot}(d) and \ref{null_density_plot}(f) when the samples are drawn from the Kent distribution.

\begin{figure}[h!]
\centering
\subfloat[]{%
{\includegraphics[width=0.3\textwidth,height=0.3\textwidth]{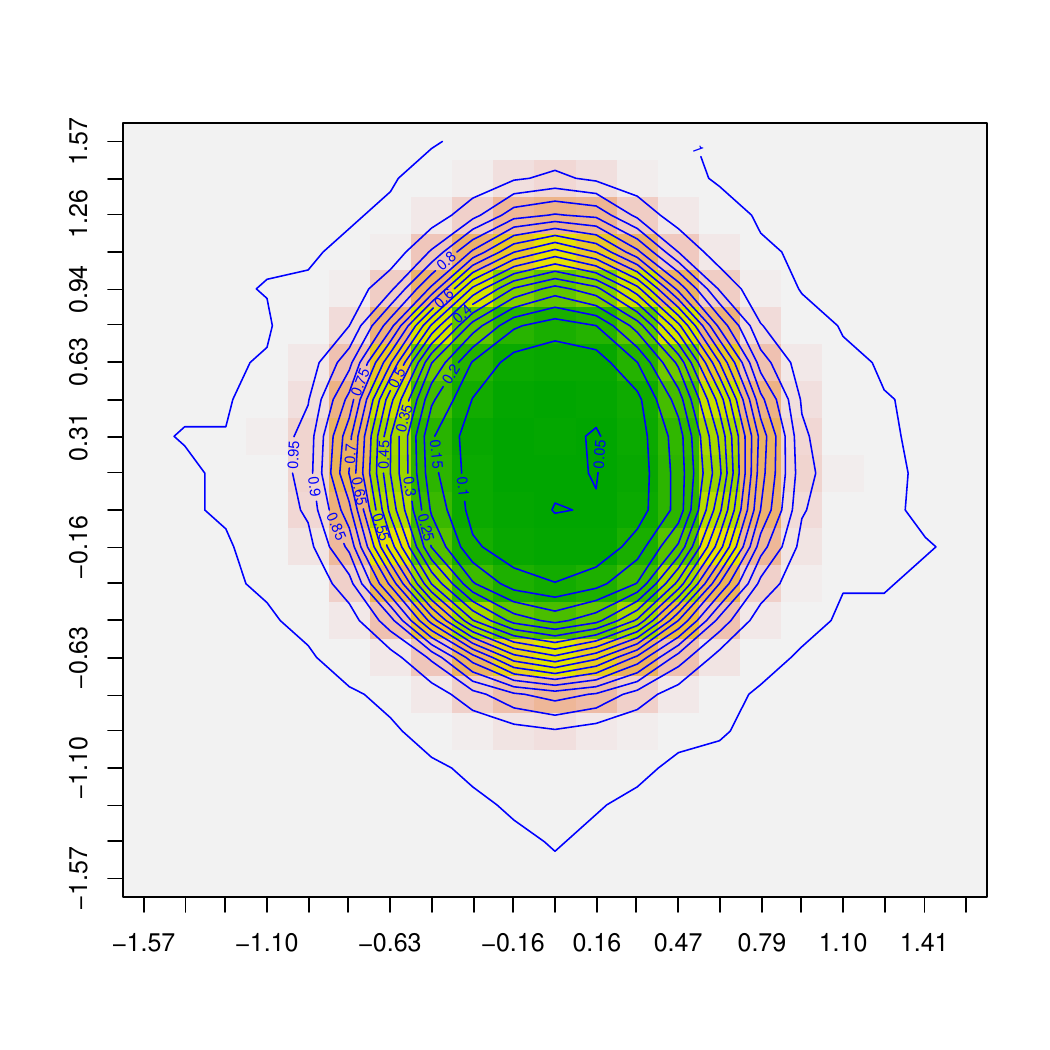}}} 
\subfloat[ ]{%
{\includegraphics[width=0.3\textwidth,height=0.3\textwidth]{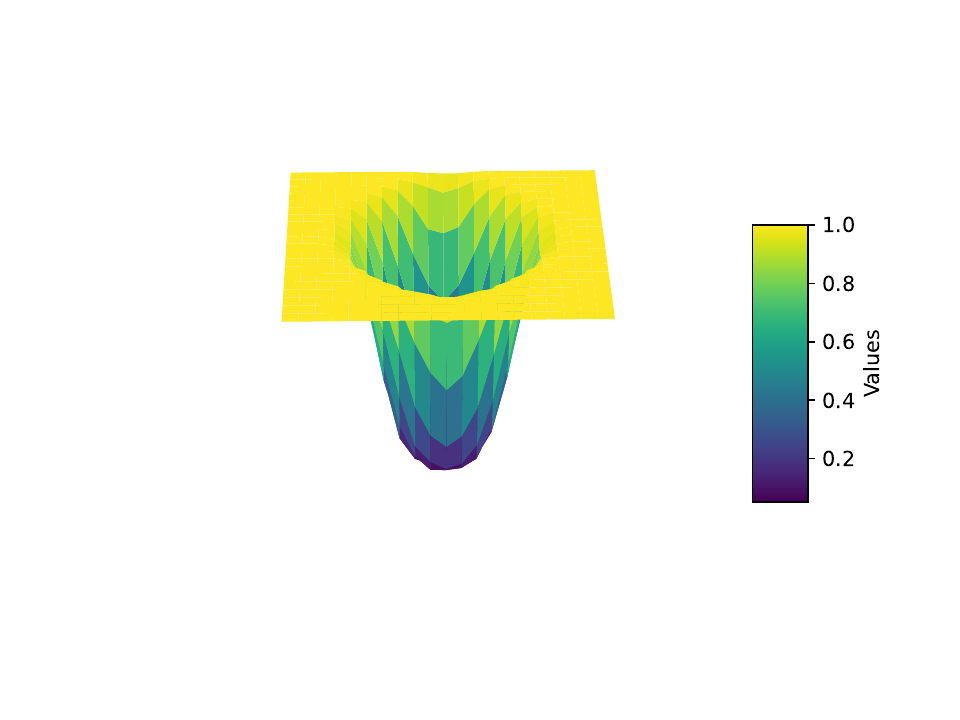}}}\hspace{5pt}

% \subfloat[]{%
% {\includegraphics[width=0.4\textwidth,height=0.4\textwidth]{kent_pos_sphere_contour.pdf}}} 
% \subfloat[ ]{%
% {\includegraphics[width=0.4\textwidth,height=0.4\textwidth]{kent_power_splot_beta_positive.pdf}}}\hspace{5pt}

% \subfloat[]{%
% {\includegraphics[width=0.4\textwidth,height=0.4\textwidth]{kent_neg_sphere_contour.pdf}}} 
% \subfloat[ ]{%
% {\includegraphics[width=0.4\textwidth,height=0.4\textwidth]{kent_power_splot_beta_negative.pdf}}}

\caption{(a), the contour plot; (b) is the corresponding surface plots of power under $H_{1s}$ when the location of the changepoint is considered at $k^{*}=\frac{n}{2},$  for Fisher distribution.} 
% \caption{(a), the contour plot; (b) is the corresponding surface plots of power under $H_{1s}$ when the location of the changepoint is considered at $k^{*}=\frac{n}{2},$  for Fisher distribution, Kent distribution with positive ovalness parameter, and Kent distribution with negative ovalness parameter, respectively.} 
\label{sphere_splot_contour_plot}
\end{figure}

To generate the power surface and the corresponding contour, the location of the changepoint is considered at $k^{*}=\frac{n}{2},$ and the mean direction vector before the change is $(\mu_\phi, \mu_\theta)=(0,0).$  After the change, a shift of $(\delta_\phi,\delta_\theta)$ in the mean direction vector is added to the initial one. Here, $\delta_\phi$ take $21$ equispaced values in $[-\frac{\pi}{2},\frac{\pi}{2}],$ and $\delta_\theta$ take $21$ equispaced values in $[-\frac{\pi}{2.5},\frac{\pi}{2.5}].$ We performed  $10^4$ iterations to compute the power of the test statistic, $\mathcal{S}_n$ in Equation-\ref{sphere_test_statistic} for sample size of $n=500$  at the level of $5\%.$ Figure-\ref{sphere_splot_contour_plot}(a) and (b) depict the power surface and the corresponding contour plot for the concentration parameter $\kappa=2$. The surface and contour plots distinctly show that the power of the test approaches one.
% Next, for the concentration parameter, $\kappa=5$ and ovalness parameters $\beta=3$ to obtain Figures-\ref{sphere_splot_contour_plot}(c), \ref{sphere_splot_contour_plot}(d). Similarly Figures-\ref{sphere_splot_contour_plot}(e), \ref{sphere_splot_contour_plot}(f) show the power surface and the corresponding contour plot with ovalness parameters $\beta=-3$. 

The simulations for the null density, power surface, and the corresponding contour can be shown for another well-known spherical distribution, the
 Kent distribution due to  \cite{kent1982fisher}, with the density detailed in \citeauthor{Mardia_2000}, \citeyearpar[Ch. 9, pp. 176]{Mardia_2000}. 

\subsection{Comparison:}
To the best of our knowledge, there are currently no dedicated changepoint detection methods developed specifically for toroidal data. As suggested by the referees, for comparative purposes, we have employed  Graph-based change-point (GCP) detection due to \cite{chen2015graph}, which uses the RBS scheme, Multivariate Nonparametric Change Point detection (MNCP) due to \cite{padilla2021optimal} that uses the multivariate nonparametric Wild Binary Segmentation scheme (MNWBS), and the Changeforest method due to \cite{londschien2023random}, which employed the  RBS scheme as a representative baseline, despite its methodological mismatch for the toroidal data framework considered in this study. The R-package, \texttt{gSeg} is utilized for GCP. The code for implementing MNCP can be found at \texttt{\url{https://github.com/hernanmp/RMNCP/tree/master/R}}, and for Changeforest, we have used the \texttt{changeforest} software package available in Python. 
The comparison of GCP, MNCP, and Changeforest with the proposed Recursive Binary Segmentation (RBS) approach was conducted under identical simulation settings. Specifically, synthetic bivariate toroidal data were generated using von Mises sine models, producing sequences of length $n = 300$ with three true changepoints positioned at approximately 100, 150, and 200. Four homogeneous segments were simulated: the first segment centered at $(0,0)$; the second segment introduced a shift of $0.9 \times \pi/2$ radians in both angular components; the third segment applied a further shift of $1.1 \times \pi/2$ radians; and the final segment applied a shift of $2 \times \pi/2$ radians. Concentration parameters were fixed at $\kappa_1 = \kappa_2 = 4$ and $\kappa_3 = 0$, ensuring independence between angular components and isolating the location shifts as the sole source of structural change.
\begin{table}[h!]
\begin{center}
\renewcommand{\arraystretch}{1.5}
\scalebox{.9}{
\begin{tabular}{|c|c|c|c|}
\hline
Methods & Median HD Distance (\%) &ARI & Average CPs \\ 
\hline
\textbf{MNCP (MNWBS)}  & 16.7 (0.0) & 0.8801 (0.0023) & 2.54 (0.02) \\ 
\textbf{Changrforest (RBS)}  & 25 (0.0) & 0.8847 (0.0320) & 2.43 (0.02) \\
\textbf{GCP (RBS)} &  16.7 (0.0)    & 0.8747 (0.0016) & 2.35 (0.02) \\
\textbf{Proposed (RBS)} & \textbf{1.7 (1.3)} & \textbf{0.9425 (0.0019)} & \textbf{3.25 (0.02)}\\ 
\hline
\end{tabular}}
\end{center}
\caption{Comparison of changepoint detection methods on simulated bivariate toroidal data over 1,000 Monte Carlo replications with true changepoints at indices 100, 150, and 200 in sequences of length 300. Reported metrics include median symmetric Hausdorff distance (HD) in \% (with median absolute deviation in parentheses), Adjusted Rand Index (ARI), and average number of detected changepoints (with standard errors in parentheses).}
\label{table:comparison_table}
\end{table}
Detection outcomes were evaluated using both the proposed method, embedded within a recursive binary segmentation (RBS) framework, and a baseline existing method designed for multivariate nonparametric changepoint detection in real-valued data. Across 1,000 Monte Carlo replications, we recorded the proportion of detections at each candidate changepoint location.  A system with a processor $Intel \circledR ~ Xeon(R)~ CPU ~E5-2630~ v3~ @ ~2.40GHz \times 16$, RAM $64.0$ GB, graphics$NVS~ 315/PCIe/SSE2$, Ubuntu 22.04.2 LTS, $64$-bit OS has been used for the above simulation.
Table-\ref{table:comparison_table} summarizes the comparative performance of changepoint detection methods on simulated bivariate toroidal data. The proposed method significantly outperforms existing approaches across all evaluation metrics. It achieves the lowest median symmetric Hausdorff distance (1.7\%) and the highest Adjusted Rand Index (0.9425), indicating precise localization and strong segmentation agreement. Moreover, it detects an average of 3.25 changepoints, closely matching the true number of structural changes, which is three. In contrast, the baseline methods MNCP, Changeforest, and GCP exhibit higher localization errors and tend to underestimate the number of changepoints. These results highlight the superior accuracy and reliability of the proposed method in handling the intrinsic geometry of toroidal data, particularly in detecting subtle and curvature-driven changes that standard methods fail to capture. Also, in terms of computation time, the proposed method, GCP, and the Changeforest method take approximately $30$ seconds, whereas the MNCP method takes approximately two hours and nineteen minutes to complete the task.
In summary, neither MNCP nor Changeforest is specifically designed for non-Euclidean data, and even though GCP addresses non-Euclidean data generally, none of these existing methods takes the intrinsic geometry of the torus into account. In this experimental setup, the second segment’s mean lies in a region of positive curvature, while the mean of the third segment shifts to a region of negative curvature on the curved torus. The actual change in mean direction is intentionally subtle, only about $0.2 \times \pi/2$ and corresponds to a transition across different curvatures.
Such small shifts and curvature-driven changes are not reliably detected by standard methods like MNCP, Changeforest, or GCP, as these approaches do not properly accommodate the unique non-linear structure and phase behavior in toroidal data.
The proposed method successfully identifies all three changepoints even in the presence of mild and curvature-based shifts. This highlights the clear advantage of the method in accurately detecting complex and subtle multiple changepoints in toroidal sequences, where existing techniques typically fail to capture the true structural changes due to their inability to address non-linear phase shifts inherent to toroidal manifolds.

\section{Data Analysis}
\label{data_analysis_biporjoy}
\begin{figure}[t]
\centering
\subfloat[]{%
{\includegraphics[width=0.3\textwidth,height=0.3\textwidth]{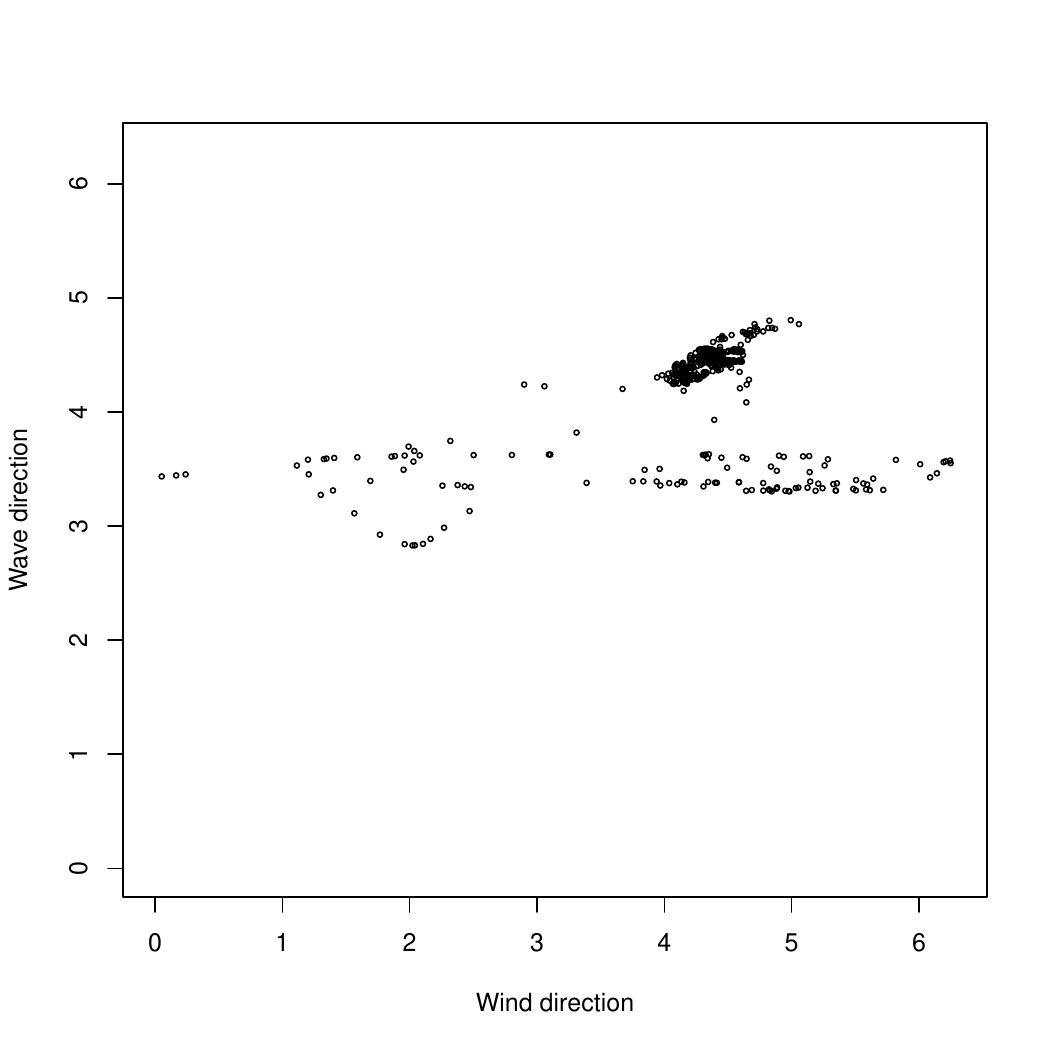}}}
\subfloat[ ]{%
{\includegraphics[trim= 100 100 100 100, clip, width=0.3\textwidth,height=0.3\textwidth]{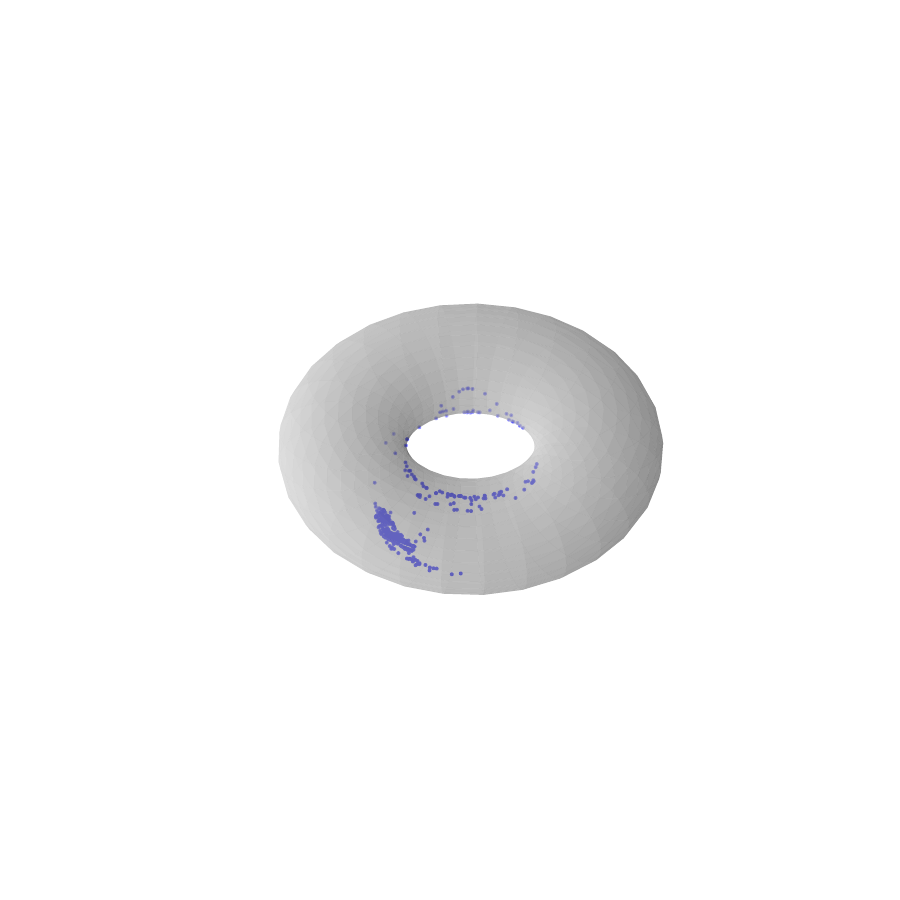}}}
% \subfloat[ ]{%
% {\includegraphics[width=0.5\textwidth,height=0.5\textwidth]{biporjoy_temoral_plot_wave.pdf}}}\hspace{5pt}

% \subfloat[]{%
% {\includegraphics[width=0.5\textwidth,height=0.5\textwidth]{biporjoy_temoral_plot_wind.pdf}}} 

% \caption{(a) scatter plot of the wind-wave directions on the flat torus. (b)  a circular temporal plot of the direction of the wave. (c)  circular temporal plot of the wind direction, and (d) scatter plot of the wind-wave directions on the curved torus.} 
\caption{(a) scatter plot of the wind-wave directions on the flat torus. (b)  scatter plot of the wind-wave directions on the curved torus.} 
\label{Biporjoy_plot_torus}
\end{figure}

Intending to study the possible association of the wind direction and wave direction at a chosen location with the meteorological events described above, we collected the 10-meter-above-the-sea-level wind direction and mean wave direction data \cite[see][]{data2023} at the location with coordinates $17.3^{\circ} N$, and $67.3^{\circ}E$, which is about 1200 km from the location of the landfall.  The hourly data spans from June 6, 2023, 0000 UTC to June 20, 2023, 0630 UTC. This resulted in a total of 348 observations reported in degrees. Figure-\ref{Biporjoy_plot_torus}(a) and (b) represent the planner and the curved torus plot of the data, respectively.
As discussed above, several significant meteorological events happened during the period 6th - 19th June 2023, which indicates the possibility of changepoints being present in the data. Since the mean direction is unknown, we executed the test for the toroidal mean direction to determine the existence of changepoints for the mean direction in this data set. Subsequently, using the method developed in Section-\ref{section_cp_torus_mean} and the binary segmentation procedure, we found the presence of multiple changepoints. Table-\ref{table:data_biporjoy_cp_table} reports some of the significant changepoints along with the $95\%$ confidence interval that closely correspond to meteorological phenomena associated

\begin{figure}[t]
\centering
\subfloat[ ]{%
{\includegraphics[trim= 20 20 20 20, clip, width=0.3\textwidth, height=0.3\textwidth]{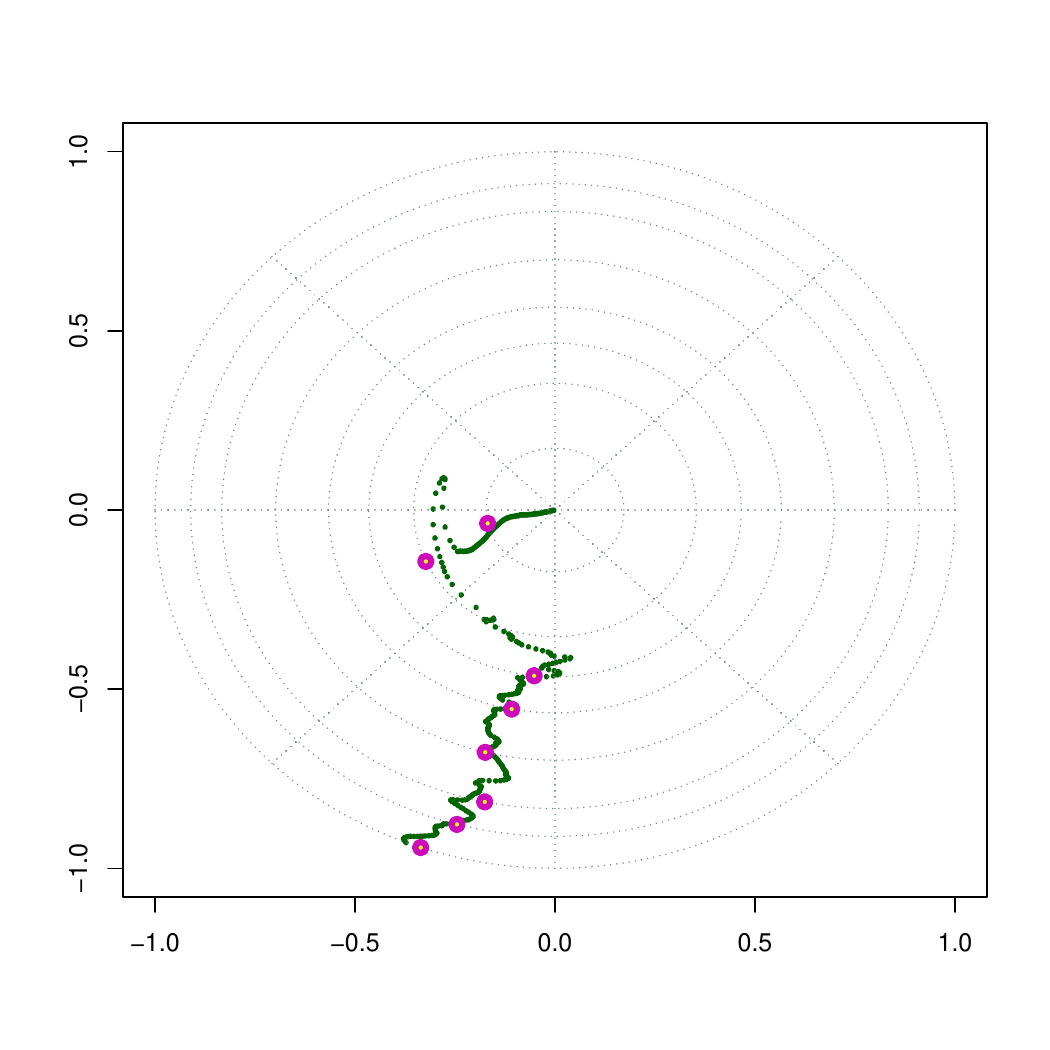}}}
\subfloat[]{%
{\includegraphics[trim= 20 20 20 20, clip, width=0.3\textwidth, height=0.3\textwidth]{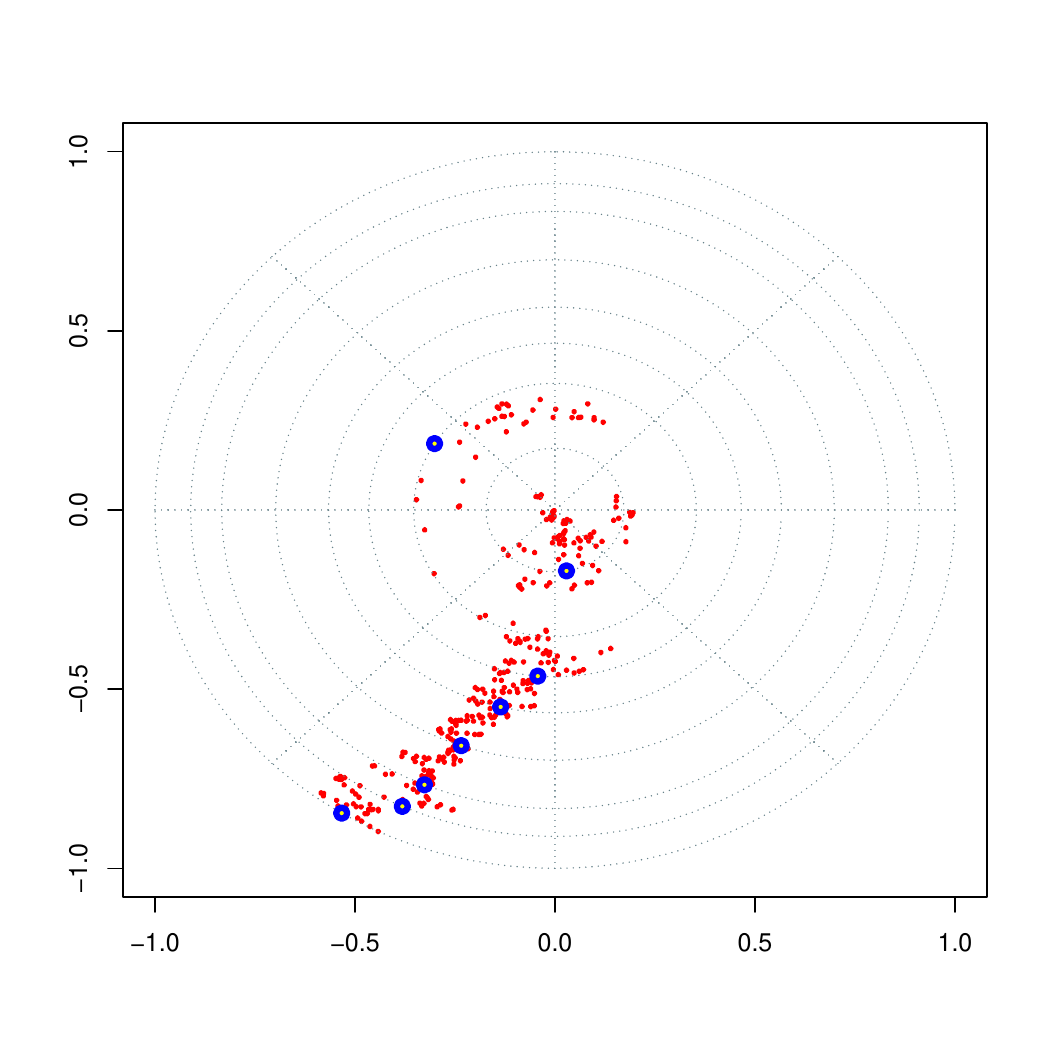}}} 
\subfloat[ ]{%
{\includegraphics[trim= 0 0 0 0, clip, width=0.31\textwidth,height=0.29\textwidth]{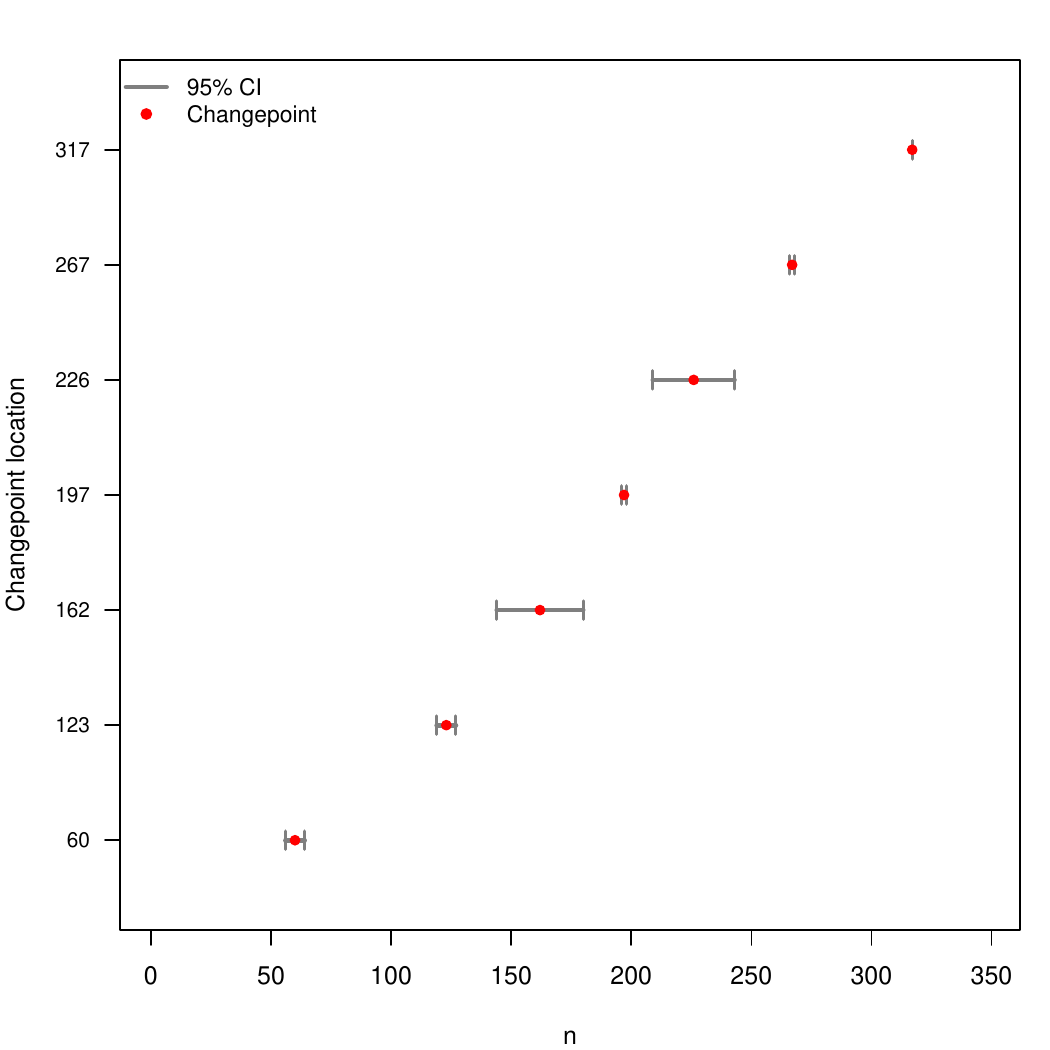}}}

\caption{The circular temporal plot of  (a) wave direction, (b) wind direction, and (c) is the $95 \%$ confidence intervals for the detected changepoints and estimated changepoints are in red dots of the Super Cyclonic Storm (SuCS) ``BIPORJOY'' }
\label{biporjoy_wind_wave_temporal_plot}

\end{figure}

with the Super Cyclonic Storm, ``Biporjoy''. Table-\ref{table:biporjoy_wind_wave_segment_mean_table}  shows the estimated values of the mean direction of the wind and wave directions for each segment. It may be noted that we used the limiting distributions $K_{\infty}$ in Equation-\ref{bbridge} to obtain the corresponding p-values. We also represent the data using a circular temporal plot \cite[see][]{biswas2024changepoint} in Figure-\ref{biporjoy_wind_wave_temporal_plot}(a) and (b), where eight annular circles from the center to outward represent the corresponding estimated changepoints in the mean direction of the wind and wave direction, respectively. The segment-wise estimated mean is represented by blue, and magenta bubble plots at the outer end of the corresponding segment. The Figure-\ref{biporjoy_wind_wave_temporal_plot}(c) is the $95 \%$ confidence intervals for the detected changepoints and estimated changepoints are in red dots of the wind and wave directions of the ``Biporjoy'' cyclone.

\begin{table}[t]
\centering
\renewcommand{\arraystretch}{1.1} % Adjust the cell height
\begin{tabular}{ |>{\centering\arraybackslash}p{3.08cm}|>{\centering\arraybackslash}p{2.05cm}|>{\centering\arraybackslash}p{2.05cm}|>{\centering\arraybackslash}p{4.05cm}| }
  \hline
Data Segment & Estimated CP & P-value & 95\% CI  \\
  \hline
1--348      & 123 & 0.0000 & [122, 124] \\
1--123      & 60  & 0.0000 & [54, 66] \\
124--348    & 197 & 0.0000 & [195, 199] \\
124--197    & 162 & 0.0008 & [155, 169] \\
198--348    & 226 & 0.0000 & [217, 235] \\
227--348    & 267 & 0.0000 & [266, 268] \\
268--348    & 317 & 0.0000 & [317, 317] \\
  \hline
\end{tabular}
\vspace{0.3cm}
\caption{Significant changepoints using the binary segmentation scheme for the wind-wave direction data of the Biporjoy cyclone.}
\label{table:data_biporjoy_cp_table}
\end{table}

% \begin{table}[h!]
% \centering
% \renewcommand{\arraystretch}{1.1} % Adjust the cell height
% \begin{tabular}{ |>{\centering\arraybackslash}p{2.5cm}|>{\centering\arraybackslash}p{2.5cm}|>{\centering\arraybackslash}p{2.5cm}| }
%   \hline
%   Homogeneous segments& Mean direction for wind in radians (degrees) & Mean direction for wave in radians (degrees) \\

%   \hline
%  1-60	&	4.88 (279.60)	&	3.36 (192.51)	\\
%  %\cline{1-3}
% 61-123	&	2.59 (148.39)	&	3.56 (203.97)	\\
% %\cline{1-3}
% 124-162	&	4.62 (264.70)	&	4.60 (263.56)	\\
% %\cline{1-3}
% 163-197	&	4.47 (256.11)	&	4.52 (258.97)	\\
% %\cline{1-3}
% 198-243	&	4.37 (250.38)	&	4.46 (255.53)	\\
% %\cline{1-3}
% 244-290	&	4.31 (246.94)	&	4.50 (257.83)	\\
% %\cline{1-3}
% 291-317	&	4.28 (245.22)	&	4.44 (254.39)	\\
% %\cline{1-3}
% 318-348	&	4.15 (237.77)	&	4.37 (250.38)	\\
%   \hline
% \end{tabular}
% \vspace{0.3cm}
% \caption{Values of the mean direction in radian (degree) for each segment for  Biporjoy cyclone data.}
% \label{table:biporjoy_wind_wave_segment_mean_table}
% \end{table}

\begin{table}[h!]
\centering
\renewcommand{\arraystretch}{1.1} % Adjust the cell height
\begin{tabular}{ |>{\centering\arraybackslash}p{2.5cm}|>{\centering\arraybackslash}p{2.5cm}|>{\centering\arraybackslash}p{2.5cm}| }
  \hline
  Homogeneous segments& Mean direction for wind in radians (degrees) & Mean direction for wave in radians (degrees) \\

  \hline
 1-60	&	4.88 (279.60)	&	3.36 (192.51)	\\
 %\cline{1-3}
61-123	&	2.59 (148.39)	&	3.56 (203.97)	\\
%\cline{1-3}
124-162	&	4.62 (264.70)	&	4.60 (263.56)	\\
%\cline{1-3}
163-197	&	4.47 (256.11)	&	4.52 (258.97)	\\
%\cline{1-3}
198-226	&	4.37 (250.38)	&	4.46 (255.53)	\\
%\cline{1-3}
227-267	&	4.33 (248.56)	&	4.51 (258.51)	\\
%\cline{1-3}
268-317	&	4.28 (245.22)	&	4.44 (254.39)	\\
%\cline{1-3}
318-348	&	4.15 (237.77)	&	4.37 (250.38)	\\
  \hline
\end{tabular}
\vspace{0.3cm}
\caption{Values of the mean direction in radian (degree) for each segment for  Biporjoy cyclone data.}
\label{table:biporjoy_wind_wave_segment_mean_table}
\end{table}

% \begin{algorithm}
% \caption{Recursive Binary Segmentation for Changepoint Detection in Angular Data}
% \label{algo:binary_segmentation}

% \textbf{Input:} Sequence of bivariate angular observations $\{(\phi_t, \theta_t)\}_{t=1}^n$ with corresponding time indices, significance level $\alpha$, and minimum segment length $h$.

% \textbf{Output:} Estimated set of changepoint locations $\widehat{\mathcal{C}}$.

% \begin{algorithmic}[1]
% \State Initialize $\widehat{\mathcal{C}} \gets \emptyset$
% \State Initialize segment list $\mathcal{S} \gets \{(1, n)\}$
% \While{$\mathcal{S} \neq \emptyset$}
%     \State Select and remove a segment $(s, e)$ from $\mathcal{S}$
%     \If{time gap between $t = s$ and $t = e$ $>$ 24 hours}
%         \State \textbf{continue} to next segment
%     \EndIf
%     \If{$e - s + 1 < h$}
%         \State \textbf{continue} to next segment
%     \EndIf
%     \State Apply the test statistic $\mathcal{M}_n$ to the segment $(s, e)$
%     \State Compute $p$-value based on empirical or asymptotic null distribution
%     \If{$p < \alpha$}
%         \State Let $\widehat{k}$ be the estimated changepoint location within $(s, e)$
%         \State Add $\widehat{k}$ to $\widehat{\mathcal{C}}$
%         \State Add new subsegments $(s, \widehat{k})$ and $(\widehat{k}+1, e)$ to $\mathcal{S}$
%     \EndIf
% \EndWhile
% \State Sort $\widehat{\mathcal{C}}$ in increasing order
% \State \Return $\widehat{\mathcal{C}}$
% \end{algorithmic}
% \end{algorithm}

\begin{figure}[h!]
\centering
\subfloat[]{%
{\includegraphics[width=0.3\textwidth,height=0.3\textwidth]{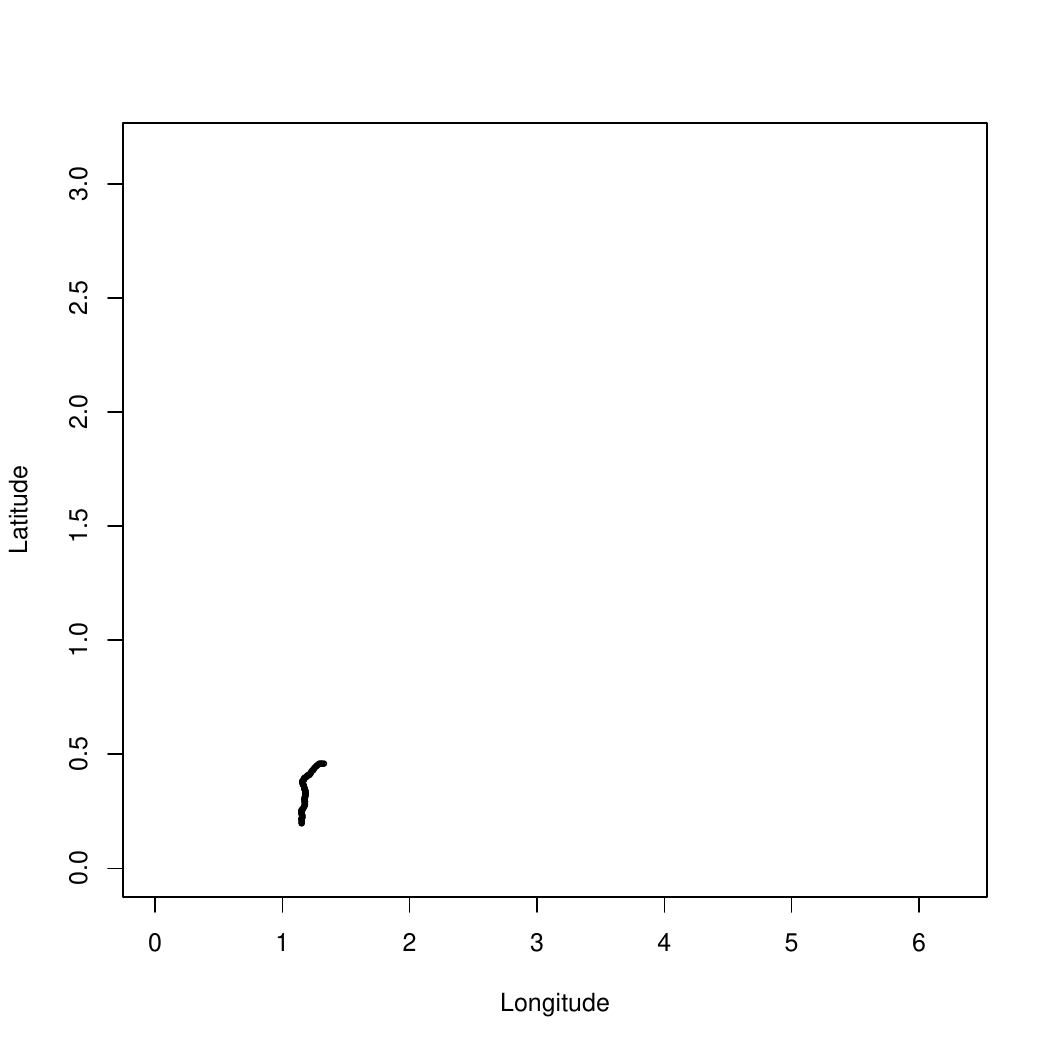}}} 
\subfloat[ ]{%
\reflectbox{\includegraphics[trim= 100 100 100 100, clip, width=0.3\textwidth,height=0.3\textwidth,angle=0]{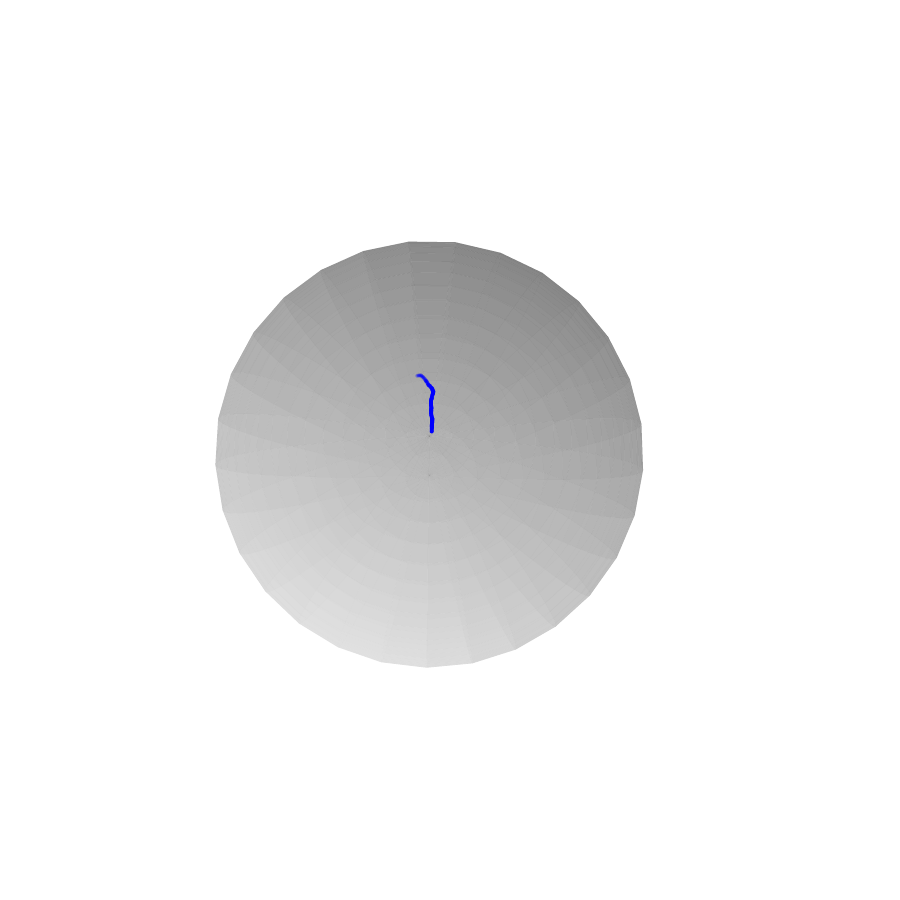}}}
% \subfloat[ ]{%
% {\includegraphics[width=0.45\textwidth,height=0.45\textwidth]{biporjoy_temoral_plot_lat.pdf}}}\hspace{5pt}

% \subfloat[]{%
% {\includegraphics[trim= 20 20 20 20, clip,width=0.45\textwidth,height=0.45\textwidth]{biporjoy_temoral_plot_long.pdf}}} 
% \caption{(a) scatter plot of the latitude-longitude directions on the flat sphere. (b) circular temporal plot of the latitude. (c)  circular temporal plot of the longitude, and (d)  scatter plot of the latitude-longitude directions on the sphere.} 

\caption{(a) scatter plot of the latitude-longitude directions on the flat sphere. (b) scatter plot of the latitude-longitude directions on the sphere.} 
\label{Biporjoy_plot_sphere}
\end{figure}

\begin{figure}[h!]
\centering
\subfloat[]{%
{\includegraphics[trim= 20 20 20 20, clip, width=0.3\textwidth, height=0.3\textwidth]{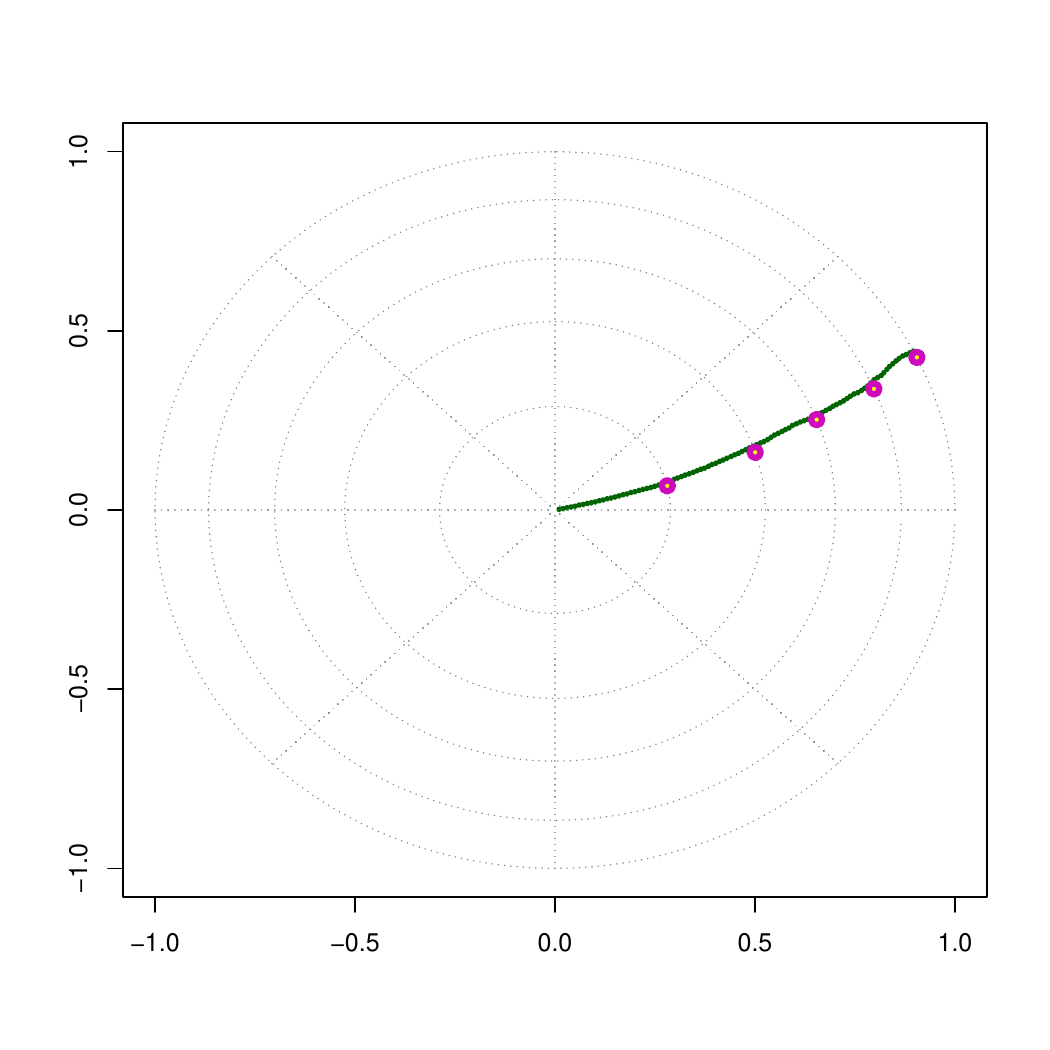}}} 
\subfloat[ ]{%
{\includegraphics[trim= 20 20 20 20, clip, width=0.3\textwidth, height=0.3\textwidth]{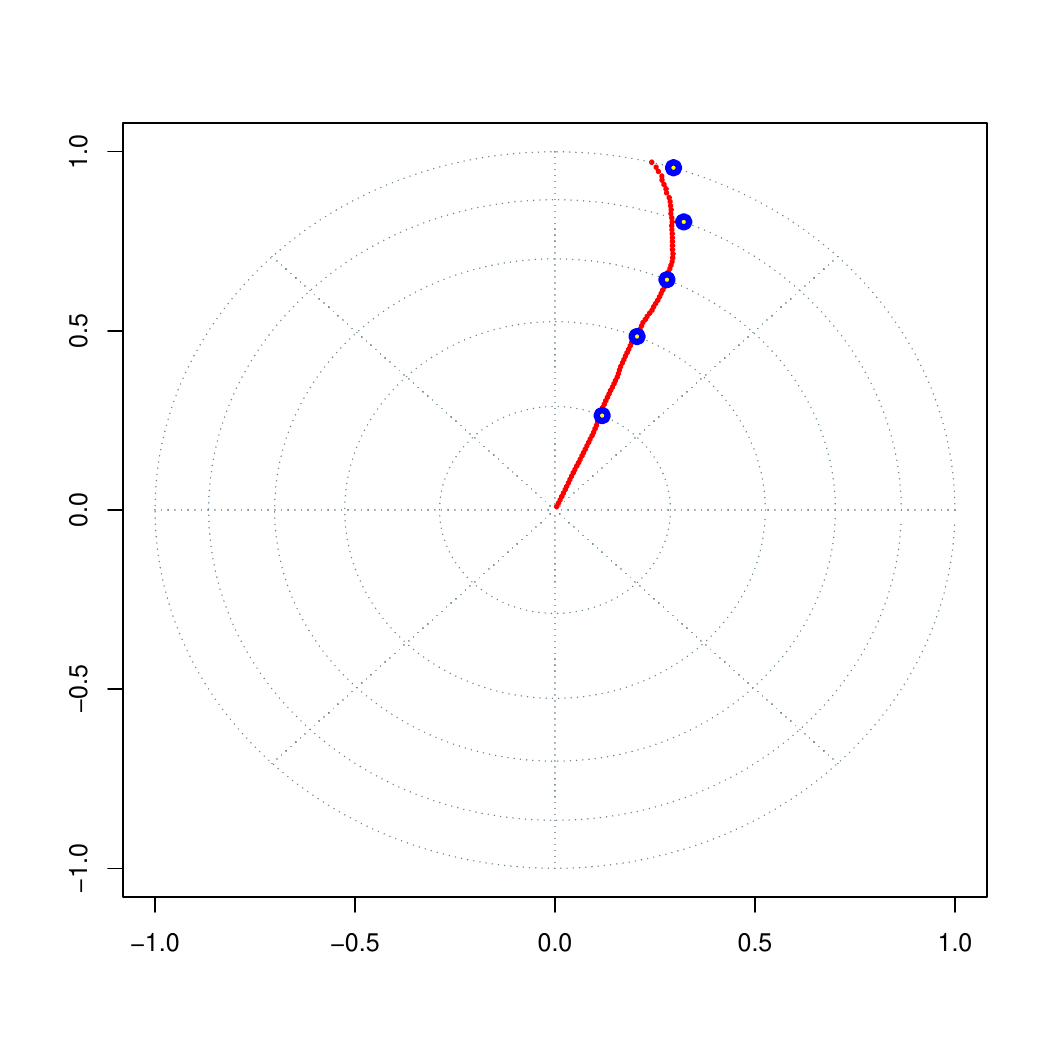}}}
\subfloat[]{%
{\includegraphics[width=0.31\textwidth,height=0.29\textwidth]{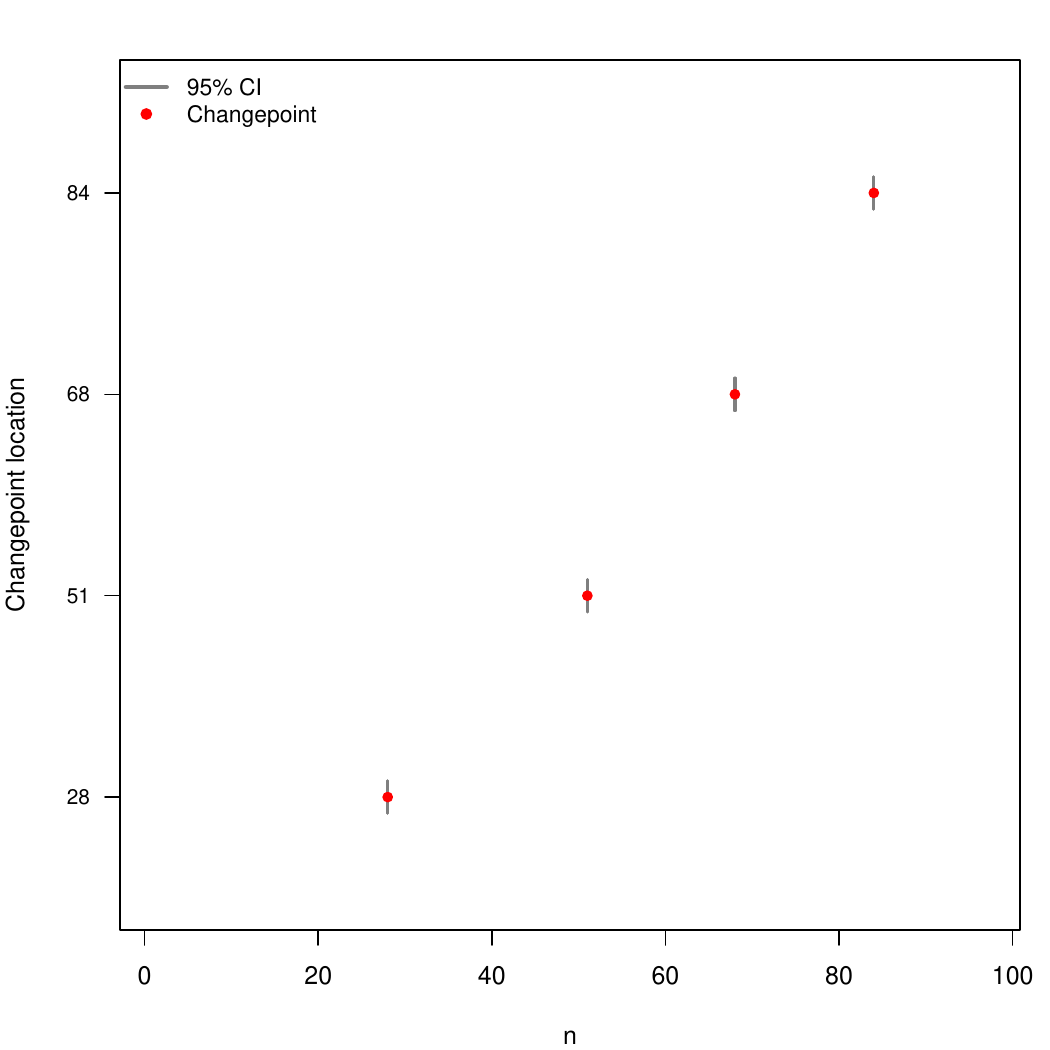}}}

\caption{ The circular temporal plot of (a) latitude, (b) longitude, and (c) is the $95 \%$ confidence intervals for the detected changepoints and estimated changepoints are in red dots of the path of the Super Cyclonic Storm (SuCS) ``BIPORJOY''.}
\label{biporjoy_path_temporal_plot}

\end{figure}

We have collected latitude and longitude data for the path of the Biporjoy cyclone, considering the spherical nature typical of cyclone paths. The data spans from June 6, 2023, 0000 UTC, to June 18, 2023, 0000 UTC, totaling 97 observations in degrees. The observations are not hourly observations.  Figure-\ref{Biporjoy_plot_sphere}(a) and (b) represent the planner plot and the plot on the sphere of the data, respectively. The cyclone altered its course many times during its journey, traversing a long distance. This frequent change in direction made it particularly challenging to forecast the path of the cyclone. Given the significant meteorological events between June 6-19, 2023, suggesting potential changepoints, we conducted tests on the spherical mean direction to detect these changepoints. Subsequently, using the method developed in Section-\ref{section_cp_sphere_mean} and the binary segmentation procedure, we identified multiple changepoints along with the $95\%$ confidence interval for the path of the cyclone, as detailed in Table-\ref{table:data_biporjoy_lat_long_table}. Table-\ref{table:biporjoy_path_segment_mean_table} consists of estimated values of the mean direction of the latitudes and longitudes of the cyclone path for each segment.
Here, we again utilized the limiting distributions \( K_{\infty} \) to derive corresponding p-values for our analysis. Additionally, the data is represented using a circular temporal plot in Figure-\ref{biporjoy_path_temporal_plot}(a) and (b), where five annular circles from the center to outward represent the corresponding estimated changepoints in the mean direction of the latitude and longitude, respectively. The segment-wise estimated mean is represented by blue, and magenta bubble plots at the outer end of the corresponding segment. Figure-\ref{biporjoy_path_temporal_plot}(c) is the $95 \%$ confidence intervals for the detected changepoints and estimated changepoints are in red dots for the path of the ``Biporjoy'' cyclone.

% \begin{table}[h!]
% \centering
% \renewcommand{\arraystretch}{1.1} % Adjust the cell height
% \begin{tabular}{ |>{\centering\arraybackslash}p{4.5cm}|>{\centering\arraybackslash}p{4.5cm}|>{\centering\arraybackslash}p{4.5cm}| }
%   \hline
% Data  segment & Estimated changepoint & P-value \\
%   \hline
%   1-97 & 68 & 0.0000 \\
%   %\cline{1-3}
%   1-68 & 28 & 0.0000\\
%   %\cline{1-3}
%   29-68 & 51 & 0.0000 \\
%   %\cline{1-3}
%   69-97 & 84 & 0.0004 \\
%     \hline
% \end{tabular}
% \vspace{0.3cm}
% \caption{ Some significant changepoints using the binary segmentation scheme for the  direction of the path of the  Biporjoy cyclone }
% \label{table:data_biporjoy_lat_long_table}
% \end{table}    

\begin{table}[h!]
\centering
\renewcommand{\arraystretch}{1.1} % Adjust the cell height
\begin{tabular}{ |>{\centering\arraybackslash}p{3.05cm}|>{\centering\arraybackslash}p{2.08cm}|>{\centering\arraybackslash}p{2.05cm}|>{\centering\arraybackslash}p{4.05cm}| }
  \hline
Data Segment & Estimated CP & P-value & 95\% CI \\
  \hline
1--97   & 68 & 0.0000 & [68, 68] \\
1--68   & 28 & 0.0000 & [28, 28] \\
29--68  & 51 & 0.0000 & [51, 51] \\
69--97  & 84 & 0.0004 & [84, 84] \\
  \hline
\end{tabular}
\vspace{0.3cm}
\caption{Significant changepoints using the binary segmentation scheme for the direction of the path of the Biporjoy cyclone.}
\label{table:data_biporjoy_lat_long_table}
\end{table}

\begin{table}[h!]
\centering
\renewcommand{\arraystretch}{1.1} % Adjust the cell height
\begin{tabular}{ |>{\centering\arraybackslash}p{2.5cm}|>{\centering\arraybackslash}p{2.5cm}|>{\centering\arraybackslash}p{2.5cm}| }
  \hline
  Homogeneous segments& Mean direction for latitude in radians (degrees) & Mean direction for longitude in radians (degrees) \\

  \hline
 1-28	&	0.23 (13.53)	&	1.15 (65.89)	\\
 %\cline{1-3}
29-51	&	0.31 (17.81)	&	1.17 ( 67.03)	\\
%\cline{1-3}
52-68	&	0.36 (21.08)	&	1.16  ( 66.46)	\\
%\cline{1-3}
69-84	&	0.40 (22.97)	&	1.19 (68.18)	\\
%\cline{1-3}
85-97	&	0.44 (25.21)	&	1.27 (72.76)	\\
  \hline
\end{tabular}
\vspace{0.3cm}
\caption{Values of the mean direction in radian (degree) for each segment for path of the Biporjoy cyclone data.}
\label{table:biporjoy_path_segment_mean_table}
\end{table}

\newpage
\section{Conclusion}
\label{conclusion}

It has been discovered that the characteristics of the underlying distribution abruptly change at unknown instances in many temporally ordered data sets. Finding this kind of instance is crucial for a lot of applications. Although this subject has been extensively researched for linear data, bivariate angular data have received no attention so far. The changepoint problems for the mean direction of bivariate angular data (spherical and toroidal) are examined for the first time in the literature. The concept of the ``square of an angle'' has been introduced using the intrinsic geometry of a curved torus. 
Analogous to the dispersion matrix for bivariate linear random variables, the ``curved dispersion matrix" for bivariate angular random variables is introduced.  Using this analogous measure of the ``Mahalanobis distance,'' we develop two new non-parametric tests to identify changes in the mean direction parameters for toroidal and spherical distributions. The limiting distributions of the test statistics are derived to follow the Kolmogorov distribution under the null hypothesis. The consistency of the proposed tests has been demonstrated under the alternative hypothesis.
The proposed methods have been put into practice to identify changepoints in mean direction for hourly wind-wave direction observations and to identify changepoints in the path of the cyclonic storm ``Biporjoy."

\bibliographystyle{apalike} 
\bibliography{buddha_bib}   
\end{document}